\documentclass[12pt,a4paper]{article}
\usepackage{graphicx}
\usepackage{epsfig}
\usepackage{enumitem}
\usepackage{subfigure}
\def\ab{\ifm{\sim}}
 \def\up#1{\ifm{^{#1}}}
\def\ff{$\phi$--factory}
\def\ifm#1{\relax\ifmmode#1\else$#1$\fi}
\def\DAF{DA\char8NE}
\def\f{\ifm{\phi}}
\def\epm{\ifm{e^+e^-}}
\def\L{\ifm{{\cal L}}}
\def\dt{\ifm{{\rm d}\,t}}
\def\deg{\ifm{^\circ}}
   
\def\ks{\ifm{K_S}} \def\kl{\ifm{K_L}}
  
\def\kpm{\ifm{K^\pm}}  
\def\figb#1;#2;{\parbox{#2cm}{\epsfig{file=#1.eps,width=#2cm}}}
\def\saltaunrigo{\vspace{6pt}}
\def\figbc#1;#2;{\cl{\figb #1;#2;}}
\let\cl=\centerline
\begin{document}

\begin{flushright}
    LNF - 10/3(P) \\
    February 10, 2010
\end{flushright}

\begin{center}
  \vspace{1.3cm}
    {\Large \bf Technical Design Report of the Inner Tracker for the KLOE-2 experiment \\} 
  \vspace{1cm}   
    {\bf The KLOE-2 Collaboration \\ }
 \saltaunrigo
    G.~De Robertis, O.~Erriquez, F.~Loddo, A.~Ranieri, \\
 {\it Dipartimento di Fisica, Universit\`a di Bari and INFN sezione di
   Bari, Bari, Italy} 
 \saltaunrigo

 G.~Morello, M.~Schioppa \\
 {\it Dipartimento di Fisica, Universit\`a della Calabria and INFN gruppo
   collegato di Cosenza, Cosenza, Italy}
 \saltaunrigo

 E.~Czerwinski, P.~Moskal, M.~Silarski, J.~Zdebik \\
 {\it Institute of Physics, Jagellonian University, Cracow, Poland}
 \saltaunrigo

 D.~Babusci, G.~Bencivenni, C.~Bloise, F.~Bossi, P.~Campana,
 G.~Capon, P.~Ciambrone, E.~Dan\`e, E.~De~Lucia, D.~Domenici,
 M.~Dreucci, G.~Felici, S.~Giovannella, F.~Happacher, E.~Iarocci,
 M.~Jacewicz, J.~Lee~Franzini, M.~Martini, S.~Miscetti, L.~Quintieri,
 V.~Patera, P.~Santangelo, I.~Sarra, B.~Sciascia, A.~Sciubba,
 G.~Venanzoni, R.~Versaci \\
 {\it Laboratori Nazionali di Frascati dell' INFN, Frascati, Italy}

 \saltaunrigo
 S.~A.~Bulychjev, V.~V.~Kulikov, M.~A.~Martemianov, M.~A.~Matsyuk \\
 {\it Institute for Theoretical and Experimental Physics (ITEP), Moscow, Russia}

 \saltaunrigo
 C.~Di~Donato \\ 
 {\it Dipartimento di Scienze Fisiche, Universit\`a di Napoli ``Federico II'' and INFN sezione di
   Napoli, Napoli, Italy}

 \saltaunrigo
 C.~Bini, V.~Bocci, A.~De~Santis, G.~De~Zorzi,
 A.~Di~Domenico, S.~Fiore, P.~Franzini, P.~Gauzzi \\
% C.~Bini, V.~Bocci, M.~Capodiferro, A.~De~Santis, G.~De~Zorzi,\\
% A.~Di~Domenico, S.~Fiore, P.~Franzini, P.~Gauzzi, A.~Pelosi \\
 {\it Dipartimento di Fisica, ``Sapienza'' Universit\`a di Roma 
   and INFN sezione di Roma, Roma, Italy}

 \saltaunrigo
 F.~Archilli, D.~Badoni, F.~Gonnella, R.~Messi, D.~Moricciani \\
 {\it Dipartimento di Fisica, Universit\`a di Roma ``Tor Vergata''
   and INFN sezione di Roma 2, Roma, Italy}

 \saltaunrigo
 P.~Branchini, A.~Budano, F.~Ceradini, B.~Di~Micco,
 E.~Graziani, F.~Nguyen, A.~Passeri, C.~Taccini, L.~Tortora \\
 {\it Dipartimento di Fisica, Universit\`a Roma Tre and INFN sezione di Roma
   Tre, Roma, Italy}

 \saltaunrigo
 L.~Kurdadze, D.~Mchedlishvili, M.~Tabidze \\
{\it Nuclear Physics Department and High Energy Physics Institute of Tbilisi State Univesity, Georgia}

 \saltaunrigo
 B.~Hoistad, T.~Johansson, A.~Kupsc, M.~Wolke \\
 {\it Department of Nuclear and Particle Physics, Uppsala University,
   Uppsala, Sweden}

 \saltaunrigo
 W.~Wislicki \\
 {\it A. Soltan Institute for Nuclear Studies, Warsaw, Poland}

 \vspace{1.0cm}
         {\bf and} 
 \vspace{1.0cm}

    N.~Lacalamita, R.~Liuzzi, M.~Mongelli, V.~Valentino\\
    {\it INFN sezione di Bari, Bari, Italy} 

    \saltaunrigo
    A.~Balla, S.~Cerioni, M.~Gatta, S.~Lauciani, M.~Pistilli\\
    {\it Laboratori Nazionali di Frascati dell' INFN, Frascati, Italy}

    \saltaunrigo
    A.~Pelosi \\
    {\it INFN sezione di Roma, Roma, Italy}

\end{center}
%%%%%%%%%%%%%%%%%%%%%%%%%%%%%%%%%%%%%%%%%%%%%%%%%%%%%%%%%%%%%%%%%%%%%%%%%%
\newpage
\tableofcontents
\newpage
\section{Introduction}
%\section*{Introduction}
%\addcontentsline{toc}{section}{Introduction}
The KLOE experiment collected an integrated luminosity $\int\!\L\dt\,$\ab2.5
fb\up{-1} at the Frascati \ff\  \DAF , an \epm\ collider operated at the energy of 1020 MeV, the mass of the
\f\ meson.

%\noindent
The experiment achieved several precision physics
results~\cite{bib:Rivista} both in
\begin{itemize}
\item Kaon physics, thanks to the unique
availability of pure \ks\ , \kl\ , \kpm\ beams, with the measurement of all
significant branching ratios and the unitarity and universality tests of
the weak interactions in the Standard Model, and also with several stringent
tests of $CPT$ symmetry
and quantum mechanics, and in
\item Hadronic physics,
with the study of the properties of scalar and pseudoscalar mesons with
unprecedented accuracy and the measurement of the $e^+e^-\to\pi^+\pi^-$
cross section representing the main hadronic contribution to the muon
anomaly.
\end{itemize}
%
% Equal energy positron and electron beams collide at an
%angle of ($\pi-$0.025) radians producing \f-mesons with a transverse momentum of \ab13 MeV.
%
\begin{figure}[!h]
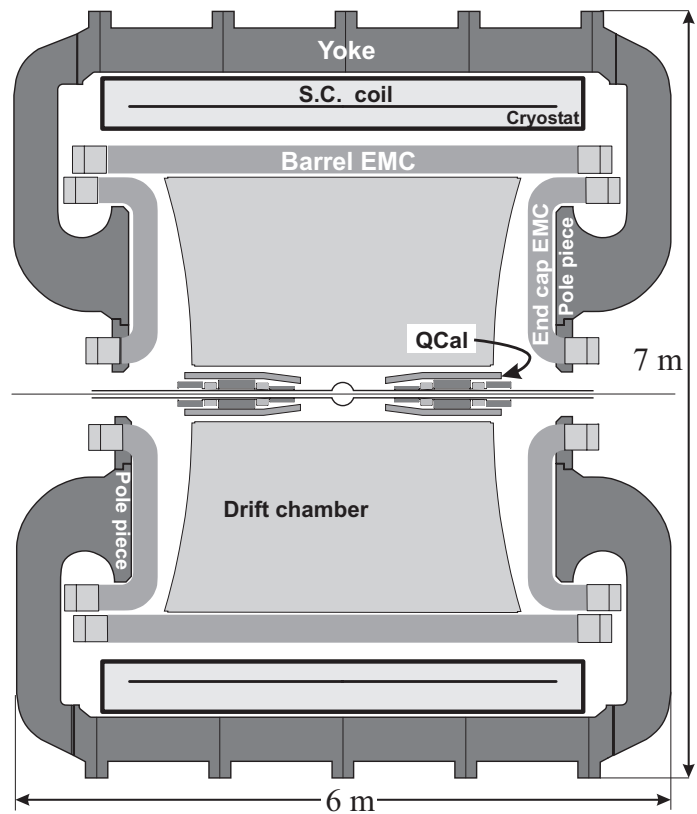

\centering
\figb DETpics/KloeDetector;9.;
\caption{The KLOE detector}
\label{fig:kloe}
\end{figure}
The detector (fig.~\ref{fig:kloe}) consists of a large volume drift chamber surrounded by an
electromagnetic sampling calorimeter and it is entirely
immersed in an axial magnetic field $B = 0.52$ T.
The drift chamber (DC) \cite{DC}, 3.3 m long and 4 m in
diameter, has a stereo geometry with 12,582 drift cells arranged in 58 layers
and operates with a 90\% helium-10\%
isobutane gas mixture. Tracking in the DC provides measurements of the
 momentum of charged particles with
$\sigma(p_{\perp})/p_{\perp}\leq0.4\%$ for polar angles larger than 45$^{\deg}$. The spatial resolution is
\ab150 $\mu$m in the bending plane, \ab2 mm on the $z$ coordinate and \ab3 mm on decay vertices.
%%%%%%%%%%%%%%%%%%%%%%%%%%%%%%%%%%%%%%%%%%%%%%%%%%%%%
The first track hit measured by the KLOE drift chamber
 is at a radius of 28 cm from the interaction point (IP).
After track extrapolation,
the vertex of a $K_S \to \pi^+\pi^-$ decay close to the IP
is reconstructed with a resolution of $\sim 1~\tau_S \simeq 6$ mm.
%, while the
%invariant mass from the momenta of the two pion tracks is reconstructed
%with a resolution of $\sim 1$ MeV.
%%%%%%%%%%%%%%%%%%%%%%%%%%%%%%%%%%%%%%%%%%%%%%%%%%%%%
The electromagnetic calorimeter (EMC) \cite{EMC} consists of a cylindrical
barrel and two endcaps, covering a solid angle of 98\% of 4$\pi$.
Particles crossing the lead-scintillator-fiber structure of the EMC,
segmented into five planes in depth, are detected as local energy deposits.
Deposits close in time and space are grouped into clusters.
The energy and time resolution for electromagnetic showers
are $\sigma_{E}/E = 5.7\%/\sqrt{E(\mbox{GeV})}$ and
$\sigma_{t} = 57\mbox{ ps}/\sqrt{E(\mbox{GeV})} \oplus 100 \mbox{ ps}$, respectively.
The trigger \cite{TRIG} requires two isolated energy
deposits in the EMC with: E~$>$~50 MeV in the barrel and E~$>$~150 MeV in the endcaps.
Cosmic-ray muons are identified as events with two energy
deposits with E~$>$~30 MeV in the outermost EMC planes and are vetoed at the trigger level.
A software filter, based on the topology and multiplicity
of EMC clusters and DC hits, is applied
to reject machine background.

% The KLOE-2 physics program is focused on events produced close to the interaction point(IP) such as
% \ks\ ,\kl\, for interferometry studies, \kpm\ and $\eta$ decays \cite{k2eoi}.
% The optimization of low momentum tracks detection is
% therefore of fundamental importance to improve the acceptance for this
% class of events, the resolution and tails of the distribution of their
% tracking and vertexing parameters, and the robustness with respect to
% machine background. To this extent an upgrade of the KLOE experiment is foreseen
% with the introduction of a new detector between the beam pipe and the
% DC inner wall: the Inner Tracker (IT).
% The detector requirements are:
% \begin{itemize}
%   \item $\sigma_{r\phi} = 200\mu m$ and $\sigma_Z = 500\mu m$ spatial resolutions;
%   \item 5~kHz/cm$^2$ rate capability;
%   \item 2\% X$_0$ overall material budget.\\
% \end{itemize}
% The KLOE-2 physics program includes the study of $K_S$, $\eta$ and $\eta^\prime$ rare decays, as well as hadron physics and interferometry.
% For the fulfilment of such a program the reconstruction performance for the decay products coming from the interaction region must be improved.
% One the main hardware upgrades of the detector is the insertion of a Inner Tracker (IT) inside the present Drift Chamber.
% The motivation for the addition of an Inner Tracker in the KLOE apparatus is the optimization for the physics coming from the interaction region, namely for a fine reconstruction of the $K_S$, $\eta$ and $\eta^\prime$ decay products.

\par
After the completion of the KLOE data taking, a proposal \cite{k2eoi,k2rollin}
has been presented for a physics program to be carried out with an
upgraded KLOE detector, KLOE-2~\cite{bib:let,bib:het,bib:cals}, at an upgraded \DAF\ machine,
%upgraded in luminosity.
 which has been assumed to deliver an integrated luminosity
${\cal O} (20)$
fb$ ^{-1} $.

\par
The KLOE physics program required a detector capable of reconstructing
with good accuracy a large fraction of \kl\
decays, in a big fiducial volume, and \ks\ decays coming from the
interaction region.
At KLOE-2 an enhanced interest will be focused on physics coming from the
interaction region: \ks\ decays, \ks\-\kl\
interference, $\eta$, $\eta'$ and \kpm\ decays, multi-lepton events. Therefore an
improvement of the detection performance and capabilities close to the
interaction point (IP) would be extremely beneficial for the KLOE-2 physics
program.

%The first track hit measured by the KLOE drift chamber
% is at a radius of 28 cm from the interaction point (IP).

%The improvement of the decay vertex reconstruction capability
%and the optimization of low momentum tracks detection are
%therefore of fundamental importance.
%%{\bf write something on specific decays }
%To this extent an upgrade of the KLOE experiment is proposed
%with the introduction of a new detector between the beam pipe and the
%DC inner wall: the Inner Tracker (IT).
In this document an upgrade of the KLOE setup with a new detector, the Inner
Tracker (IT), is discussed.
The IT will be placed between the beam pipe and the DC inner wall in order to:
\begin{itemize}
\item
reduce the track extrapolation length and improve the decay vertex reconstruction capability;
\item
increase the geometrical acceptance
for low momentum tracks, presently limited by the KLOE magnetic field and by the distance of the DC first layer, and optimize their detection;
\item
improve the track momentum resolution.
\end{itemize}

\section{Physics issues}
\subsection{Neutral Kaon Interferometry}
A $\phi$-factory is a unique place to investigate the evolution of the
entangled
kaons, and to search for decoherence and CPT violation effects,
possibly induced by underlying quantum gravitational phenomena.
\par As discussed in Ref.\cite{k2eoi,didohand}, from the analysis of $CP$
violating decay mode $\phi \rightarrow K_S K_L  \rightarrow \pi^+\pi^-,
\pi^+\pi^-$ several important tests of fundamental physics can be done. 
Quantum interference effects in this channel
have been observed for the first time by the KLOE collaboration \cite{kloeqm},
measuring the distribution
$I(\pi^+\pi^-,\pi^+\pi^-;|\Delta t|)$, with $\Delta t$ the time difference
of the two $\pi^+\pi^-$ decays.
The analysis of the full KLOE data set \cite{didovalencia} provided the most
stringent
limits on the
%$\alpha$, $\beta$,
$\gamma$, $\Re \omega$, and $\Im \omega$
parameters related to
possible decoherence and $CPT$ violation effects in the neutral kaon system~\cite{bib:gamma_omega}.
Moreover, the most precise test of quantum coherence on an
entangled two-particle system has been performed, putting very stringent limits on
the decoherence parameters $\zeta_{00}$ and $\zeta_{SL}$~\cite{bib:deco}. All these
measurements are dominated by statistical uncertainties.
\par
\begin{figure}[t]
\vspace{9.0cm}
%\special{psfile=PHYSpics/plot_work.eps voffset=-70 hoffset=0
\includegraphics{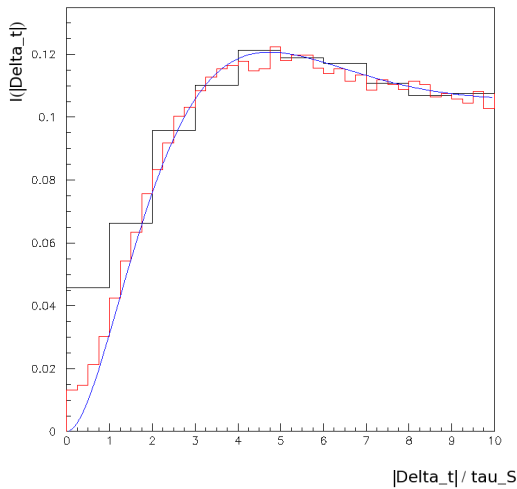}
\caption{
 The $I(\pi^+\pi^-,\pi^+\pi^-;|\Delta t|)$ distribution as a function of
$|\Delta t|$ (in $\tau_S$ units) with the
present KLOE resolution $\sigma_{|\Delta t|}\approx  \tau_S$
(histogram with wide bins), with an improved resolution
$\sigma_{|\Delta t|}\sim 0.3\,\tau_S$ (histogram with narrow bins), and
in the ideal case (solid line). Simulation results.
\label{fig:resol} }
\end{figure}
\begin{figure}[t]
\vspace{10.0cm}
%\special{psfile=PHYSpics/rollin3_1.eps voffset=-70 hoffset=0
\includegraphics{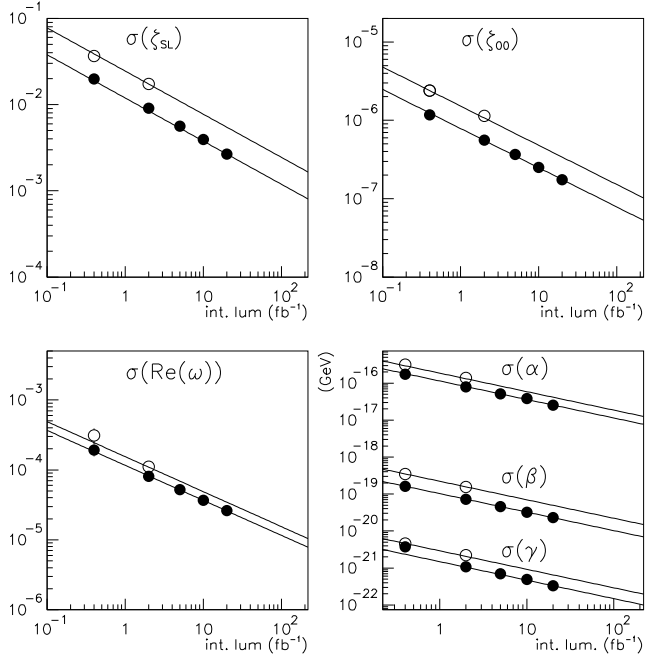}
 \caption{
 The statistical sensitivity to the parameters $\zeta_{SL}$ (top-left),
 $\zeta_{0\bar{0}}$ (top-right),
 $\Re \omega$ (bottom-left) and 
$\alpha$, $\beta$, $\gamma$ (bottom-right) with the present KLOE resolution
$\sigma_{|\Delta t|}\approx \tau_S$
(open circles), and with an improved resolution
$\sigma_{|\Delta t|}\sim 0.3\,\tau_S$ (full circles);
solid lines represent 1/$sqrt(\L)$ function.
    \label{fig:resol2} }
\end{figure}

In general, all decoherence effects (including $CPT$ violation effects related to decoherence phenomena)
should manifest as a deviation from the quantum mechanical
prediction $I(\pi^+\pi^-, \pi^+\pi^-;|\Delta t|=0)=0$. Hence the
reconstruction of events in the region at $\Delta t \approx 0$, i.e. with
two vertices close to each other and both near the IP, is crucial for a
precise determination of the parameters related to 
$CPT$ violation and decoherence. The vertex resolution,
mainly due to track extrapolation, affects the
$I(\pi^+\pi^-, \pi^+\pi^-;|\Delta t|)$ distribution precisely in this region, as shown
in fig.~\ref{fig:resol}. 
The resolution has been taken into account by properly folding and fitting
the original distribution with the smearing matrix and the efficiency curve
obtained with the Monte Carlo simulation used for the KLOE data analysis
\cite{kloeqm}. The effect of an improved resolution
$\sigma_{|\Delta t |}\approx 0.3\,\tau_S$ has simply been obtained by
taking into account the corresponding $|\Delta t |$ scale factor in the 
smearing
matrix and in the efficiency curve. In this way all the peculiar asymmetries
and behaviours of the $|\Delta t|$ resolution around $\Delta t =0$ are 
maintained, even though they are shrinked.
The impact of the resolution on the decoherence parameter
measurements has to be carefully evaluated. Indeed, the resolution has two main effects:
(1) it contributes to the statistical sensitivity on the measurement of the
decoherence (and $CPT$ violation) parameters, which are
obtained from a fit to the measured $I(\pi^+\pi^-, \pi^+\pi^-;|\Delta t|)$ distribution;
(2) it introduces a source of systematic uncertainty, due to its folding in
the fit procedure.
In fig.~\ref{fig:resol2} the statistical uncertainty on
several decoherence and $CPT$-violating parameters is shown as a function of the integrated
luminosity for $\sigma_{|\Delta t|}\approx \tau_S$ (present KLOE resolution)
and  $\sigma_{|\Delta t|} \sim 0.3\,\tau_S$.
As it can be seen in the last case an improvement
%in the uncertainties
of about a factor two in statistical sensitivity
could be achieved. This improvement would be equivalent to an increase in
luminosity of a factor of four, and it is necessary, with a target
integrated luminosity of \L ~20 fb-1, to reduce the statistical
uncertainties below or approximately at the same level of the expected 
systematic ones.
Therefore the insertion of an inner tracker between the spherical beam pipe and
the drift chamber is mandatory for a significant improvement of the KLOE
results, and in general for the neutral kaon interferometry program of KLOE-2.
\\ In fact also the $\Delta t$ distribution of other decay modes like
%\begin{itemize}
%\item
$\phi \rightarrow K_S K_L  \rightarrow \pi^+\ell^-\bar{\nu} ,
\pi^-\ell^+\nu$ and
%\item
$\phi \rightarrow K_S K_L  \rightarrow \pi\pi , \pi\ell\nu$
%\end{itemize}
are interesting
in the interference region (corresponding to kaon decays
close to the IP), being sensitive to $CP$, $CPT$ and/or $\Delta S=\Delta Q$
rule violation effects \cite{didohand}.
%to the imaginary part of
%$CPT$ and $\Delta S = \Delta Q$ rule violating parameters, or the phase $\phi_{\pi\pi}$ of the $\eta_{\pi\pi}$ parameter
In analogy with the measurement of the decoherence parameters, the statistical sensitivity and the systematic
uncertainties in the measurement of these parameters could be largely improved
with the use of an inner tracker.
%, allowing KLOE-2 to perform the
%best direct measurement of such parameters.

\subsection{Rare \ks\ decays}
Significant results in kaon physics can be obtained at KLOE-2 improving
the present knowledge on \ks\ rare decays.
The measurement of \ks\ decays at the $\phi$-factory has the unique
feature to rely on \ks\ pure beams, tagged by the reconstruction of
\kl\ decays and \kl\ interactions in the calorimeter. Background sources
are limited to the dominant \ks\ decay channels, $K_S \to \pi \pi$,
well under control using the constraint of the closed kinematics of the decay.
%
%severely controlled by the kinematic constraints obtained
%from \kl\ 4-momentum reconstruction.
\\
In general the insertion of an inner tracker would greatly improve the quality of tracking
and vertexing  for all charged $K_S$ decay modes;
it would also optimize low momentum tracks detection, which is
of fundamental importance to improve the acceptance for some rare $K_S$
decays.
%classes of events.
%specific decays as $K_S\to\pi e\nu$, $K_S\to\pi \mu\nu$,
%$K_S\to\pi^0 e^+e^-$, $K_S\to\pi^+\pi^- e^+e^-$.

%, the resolution and tails of the distribution of their
% tracking and vertexing parameters, and the robustness with respect to
% machine background.
The study of $K_S\to\pi e\nu$ decays is very important
to test $CPT$ symmetry, the $\Delta S=\Delta Q$ rule and to
measure $V_{us}$. The most precise measurement of this branching ratio (BR) has been
obtained by KLOE with a total uncertainty
of $\sim$1.3\% using an integrated luminosity
of 410 pb$^{-1}$ \cite{Ambrosino:2006si}. This accuracy is the combination of a 1.1\%
statistical and a 0.7\% systematic fractional uncertainties, largely dominated
by the precision of the knowledge of the distributions of tracking-related
quantities.
In fact the systematic uncertainty on the combined fit of these
distributions, like the difference between missing energy and momentum,
mainly comes from the background of $K_S\to\pi^+\pi^-$ decays not correctly
reconstructed, e.g. cases in which one pion decays to a muon before
entering the DC tracking volume.

%Using this result KLOE has measured for the first time the
%$K_S$ semileptonic charged asymmetry
%$A_S=(1.5\pm 9.6_{stat}\pm 2.9_{sys})\times 10^{-3}$, of great relevance
%for $CPT$ tests, and whose measurement benefits of a reduction in the ratio
%of some systematic effects.
With a target integrated luminosity of
%${\cal O} (10)$
10-20 fb$ ^{-1} $ and the detector upgraded with an inner tracker,
%With an integrated luminosity of 7.5 fb$^{-1}$,
KLOE-2 could measure the BR with a total uncertainty
dominated by systematic effects.
 These systematic effects could be reduced
at few per mil level with the IT,
thanks to a better quality of the tracking, a more
powerful rejection of the background from pion decays close to the IP,
and a larger acceptance for low momentum tracks.
%A possible improvement in
%the acceptance for semileptonic decays could be largely improved
%by reducing the KLOE magnetic field; the corresponding worsening in the
%transverse momentum resolution could be compensated by the presence
%of the inner tracker.
% could be accompained for and a larger acceptance.
%of the background due to
%obtained with the use of an inner tracker.
%for the BR and permil level for $A_S$
%with the help of the inner tracker.
\par
Similar considerations apply to other  $K_S$ rare decays that can be measured
at KLOE-2 as
%\begin{itemize}
%\item
(i) $K_S\to\pi \mu\nu$, with similar physics interest as the
$K_S\to\pi e\nu$ decay, but with a lower BR and larger background;
(ii)
%\item
$K_S\to\pi^+\pi^- e^+e^-$, important to perform tests of
Chiral Perturbation Theory and possible new $CP$ violation mechanisms;
%\item
(iii) $K_S\to\pi^0 e^+e^-$ very important to assess indirect $CP$ violation
contribution in $K_L$ decays
and to extract the direct $CP$ violating contribution.
% sensitive to
%$\Im(V_{td} V_{ts}^*)$.
%\end{itemize}

%of 0.8\% and
%the asymmetry $A_S$ at the 0.3\% level.
%
% Measurement of this decay greatly improve
%at step-1 inserting inner tracker.

\subsection{$\eta$ decays in four charged particles}
The radiative decays of the $\eta$ meson where the virtual photon converts
into a lepton pair allow to study the $\eta$ internal structure, and to
search for unconventional possible $CP$ violation mechanisms.
The BR of the $\eta \to \pi^+\pi^- e^+e^-$ decay has been recently measured
at KLOE \cite{versaci}
with a 3.4\% statistical uncertainty and
a 2.6\% systematic one.
%The first uncertainty is dominated by
%The transverse momentum distribution of the four tracks peaked at low
%values, $\sim 10$ MeV/c, strongly reduces
The detection efficiency for these decays is strongly limited by
the minimum detectable transverse momentum $\sim 23$ MeV/c with the KLOE DC,
 being the transverse momentum distribution of the four tracks peaked at low
values, $\sim 10$ MeV/c.
Moreover
the largest contribution (70\%) to systematic uncertainty is due to
the cut to reject $\eta \to \pi\pi \gamma$ events
with $\gamma$ conversion on the beam pipe.
The insertion of an inner tracker between the beam pipe and the KLOE DC
is expected to significantly improve both the acceptance of low momentum tracks
and the vertex and momentum resolution to reject much more efficiently the
background.
Similar benefits are expected for the $\eta \to e^+e^-e^+e^-$ and
$\eta \to e^+e^-\mu^+\mu^-$  decays with a BR
in the range $10^{-5}$ and  $10^{-7}$, respectively, well accessible at KLOE-2.
%
%the systematic uncertainty is dominated by the cut on background
%rejection

%out of the geometrical acceptance of the KLOE DC.
%The systematic uncertainty

\subsection{Multi-lepton events}
\begin{figure}[t]
\vspace{8.0cm}
\includegraphics{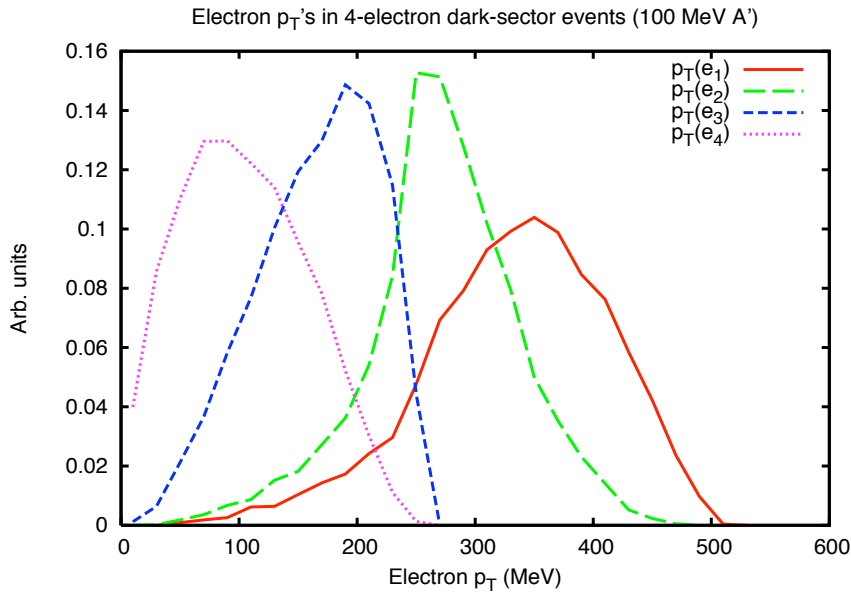}
 \caption{\it
Transverse momentum distribution
of the four electrons
produced in events involving a neutral vector boson
with a mass of 100 MeV connected with a secluded gauge sector. Electrons
are sorted by decreasing transverse momentum. 
\label{fig:4eDM} }
\end{figure}
Several puzzling astrophysical observations (PAMELA,
ATIC, INTEGRAL, DAMA) have been recently interpreted by postulating
the
existence of some secluded gauge sector weakly coupled with
the SM particles.
The typical energy scale for this new physics is 1 GeV or less, and
can therefore induce observable signals at \DAF .
One of the possible signatures for the secluded sector is the production
of multi-lepton events, whose energy spectrum is highly
dependent on the lepton flavor/multiplicity and on the actual
values of the parameters of the theory
(the mass of the vector boson, for instance).
As an example,
%with a rich
%phenomenology at low (${\cal O} (1 \hbox{ GeV})$) energies \cite{toro}.
%This possibility of detecting New Physics signals at \DAF\ seems very attractive.
%The typical signature will be the production of multi-lepton events,
%that must be fully reconstructed by the KLOE detector.
in fig.~\ref{fig:4eDM} the transverse momentum distribution
of the four electrons
produced in events involving the neutral vector boson mediating
the new interaction with a mass of 100 MeV, is shown.
The low momenta of the leptons in this kind of events
(about 30\% of the events have at least an electron below 50 MeV/c;
this fraction might become larger for
a different choice of the vector boson's mass or if muons
are present instead of electrons)
translates into a low geometrical acceptance
due to the KLOE magnetic field. The need to fully reconstruct the event
in addition to the necessity to reject the background due to
$\gamma$ conversions on the beam pipe makes, in this respect,
the insertion of an inner tracker
%can be proven to be
extremely beneficial.

\section{Detector design requirements.}
\begin{figure}[!h]
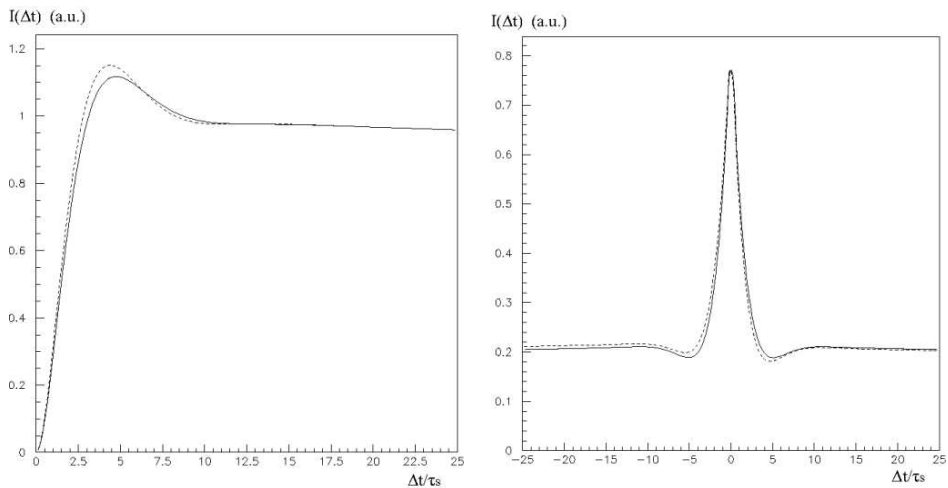

\centering
\figb PHYSpics/fullregen9_check14_handbook;6.; \kern0.2cm \figb PHYSpics/fullregen9_check13_handbook;6.;
\caption{(Left) Distribution of the decay time difference $|\Delta t|$ for
  $ \ks\kl \to \pi^+\pi^- , \pi^+\pi^- $ events
 (Right) Distribution of the decay time difference $\Delta$t for
  $\ks\kl \to \pi^- l^+ \nu , \pi^+  l^- \bar{\nu}$ events}
\label{fig:inter}
\end{figure}

The design of the IT is driven by physics requirements and space
constraints due to limited detector clearance.
The inner radius must preserve the \ks-\kl\ quantum
interference region (e.g. from \ks\ regeneration)
shown in fig.~\ref{fig:inter} in units of \ks\ lifetime ($\tau_S \simeq$ 0.6cm at a $\phi-factory$).
It follows that the inner radius of the IT should be at least at a distance of
$\simeq 20 \tau_S \simeq$ 12 cm from the IP.
The outer radius of the inner tracker is constrained from the
presence of the drift chamber inner wall at 25 cm.

\par
A crucial design parameter is the resolution on the \ks\ decay point,
occurring within few cm from the IP. An accurate study of
quantum interferometry requires an improvement on this resolution
of a factor 3-4 with respect to the present value ($\simeq$0.6 cm).
\par
In order to reconstruct the decay vertex nearby the IP, the charged decay tracks must be
extrapolated from the hit on the IT innermost layer
to the decay point. % corresponding to a is of the order of 10 cm.
This in turn implies that both the spatial resolution of the IT
and the multiple scattering contribution must be carefully taken into account
in the evaluation of the track momentum resolution.
To this extent and in order to minimize photon conversion,  the IT should
give a maximum contribution to the overall material budget of $\sim 2\%
X_0$.

\par
The event rate from $\phi-$decays at O($10^{33}$) cm$^{-2}$s$^{-1}$  luminosity is of the order of 10 KHz.
However a critical parameter for the inner tracker design is given by the occupancy
due to the machine background events.
The composition (electrons and/or photons), the rate, the spatial and momentum distribution of these events are difficult to estimate since they are highly dependent on the machine optics,
operating conditions, and on the configuration of screens and absorbers.
A first estimate of the background occupancy can be obtained
from the counting rate of the silicon detector of the FINUDA
experiment that collected data at the second \DAF\ interaction region: at a radius of 6 cm from IP
this detector collected 7-8 hits/plane in $2 \mu s$  of integration time
at $10^{32}$cm$^{-2}$s$^{-1}$ luminosity.
Scaling with the luminosity and detector's geometry, the occupancy for the IT
is 10-20 hits /layer, in one $\mu s$ of integration time.
An alternative estimate can be obtained by properly scaling the single counting
rate of the innermost layer of the KLOE DC positioned at 28 cm from the
IP. The result is found to be consistent with the previous estimate.
In order to minimize the possible effect of combinatorial for an
independent pattern recognition based on IT hits, the detector should have the
shortest possible integration time.
As a reference figure for the integration time of the IT,
we have considered $200$ ns, equal to the maximum delay between the $e^+e^-$ collision and the KLOE trigger signal formation.
%{\bf Misc: Simulation of Touschek effect from Caterina Bloise?}

\par
Summarizing, the detector requirements for the IT are:
\begin{itemize}
\item $\sigma_{r\phi}\sim 200\mu m$ and $\sigma_Z \sim 400\mu m$ spatial
  resolutions, to improve present resolution on the \ks\ decay point by a
  factor of $\sim$3;
\item 2\% X$_0$ overall material budget;
\item 5~kHz/cm$^2$ rate capability.
\end{itemize}
%%%%%%%%%%%%%%%%%%%%%%%%%%%%%%%%%%%%%%%%%%%%%%%%%%%%%%
\subsection{Detector layout}
\label{sec:detector_layout}
\begin{figure}[!h]
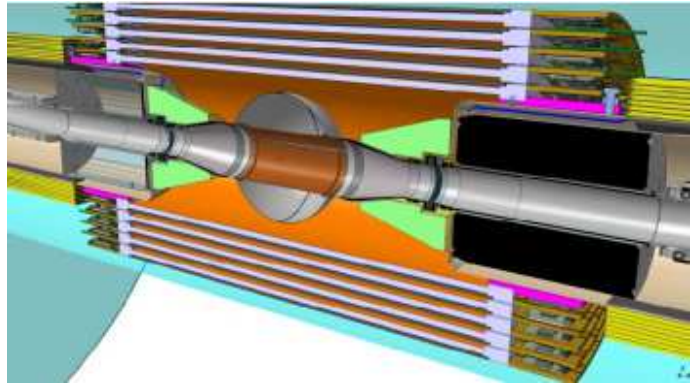

\centering
\figb ITpics/a2;9.;
\caption{Cross-section of the inner tracker at the interaction point.}
\label{fig:layout}
\end{figure}

% Items to be covered:
% \begin{itemize}
% \item Short description of the GEMs' principle of operation (e.g. from LHCb
%   Muon System TDR in 3.1) with refernces
% \item Comparison between the GEM solution and the other possible choices,
% namely Silicon VTX and Straw Tubes, poiting out the GEM advantages: low
% budget material (X$_0$), fast response and low occupancy.
% \item Write everything for a detector with 5 layers and then if
%   necessary/possible write a comment on having 4 layers instead.
% \end{itemize}
%
The Inner Tracker will be inserted in the available space inside the
Drift Chamber. The proposed solution consist of five independent tracking layers (L1-L5),
each providing a 3-D reconstruction of space points along the track with a
2-D readout. The innermost layer is placed at 12.7~cm from the beam line,
corresponding to 20~$\tau_S$ avoiding to spoil the $K_S K_L$ interference.
The outermost layer will be placed at 23.0~cm from the beam line, just
inside the internal wall of the Drift Chamber. The active length for all
layers is 700 mm (fig.~\ref{fig:layout}).

\par
\begin{figure}[!h]
\centering
\includegraphics[width=0.6\textwidth]{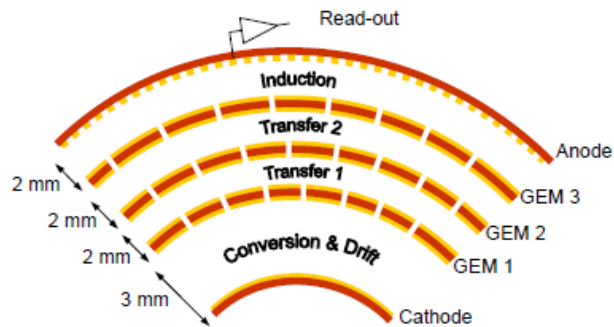}
\caption{Cross-section of the triple GEM detector.}
\label{fig:gemscheme}
\end{figure}

We have chosen to realize each layer as a cylindrical-GEM detector (CGEM).
The CGEM is a triple-GEM detector composed by concentric
cylindrical electrodes (fig.~\ref{fig:gemscheme}): from the cathode (the innermost electrode),
through the three GEM foils, to the anode readout (the outermost
electrode). The anode readout of each CGEM is segmented  with 650 $\mu$m
pitch XV patterned strips with a stereo
angle of about 40 degrees.
The full system consists of about 30,000 FEE channels.
\par
This innovative technology provides us with an ultra-light and
fully sensitive detector, fulfilling the stringent requirement on the
material budget needed to minimize the multiple scattering effect for low-momentum tracks. The detector is composed by thin (50~$\mu$m) Copper clad
 polyimide foils, acting as cathode, readout anode and multiplication
 stages.

\par
Moreover the high rate capability of the GEM (up to 1~MHz/mm$^2$
 measured \cite{bib:elba09})
makes this detectors suitable to be placed near the interaction
 point of a high-luminosity collider machine.
%%%%%%%%%%%%%%%%%%%%%%%%%%%%%%%%%%%%%%%%%%%%%%%%%%%%%%%%%%%%%%%%%%%%%%%
\subsection{Operating principles of a triple-GEM detector.}
%%%%%%%%%%%%%%%%%%%%%%%%%%%%%%%%%%%%%%%%%%%%%%%%%%%%%%%%%%%%%%%%%%%%%%%
A GEM (Gas Electron Multiplier) is made by a thin (50 $\mu$m) kapton foil, copper clad on each side, with
a high surface density of holes~\cite{bib:sauli}. In the standard technique
each hole has a bi-conical structure with external (internal) diameter
of 70 $\mu$m (50 $\mu$m); the hole pitch is 140 $\mu$m. The bi-conical
shape of the hole is a consequence of the double mask process used in
standard photolitographic technologies.
The GEM foils are manufactured by the
% The Design and Manufacture of Electronic Modules (DEM) Group of the
% Engineering Support & Technology Division (EST)
 CERN EST-DEM \cite{bib:cern-est-dem} workshop.
% following our global geometrical design.
A typical voltage difference of 350 to 500 V is applied between the two copper sides, giving fields as
high as 100 kV/cm into the holes, resulting in an electron multiplication up to a few thousand.
Multiple structures realized by assembling two or more GEMs at close distance allow high gains to be
reached while minimizing the discharge probability~\cite{bib:nima479}.

The triple GEM detector can effectively be used as tracking detector, with good time and
position resolution performance.
A cross-section of the detector is shown in fig.~\ref{fig:gemscheme}
indicating all of the GEM stages: conversion and drift, transfer and induction.
\subsection{Simulation results}
Code to simulate the IT detector has been inserted in the
official KLOE Monte Carlo program, GEANFI \cite{geanfi}.
 We used a reference geometry where the IT
is made of five cylindrical triple-GEM layers, 70 cm long and
with the innermost plane of the triple-GEM positioned at the following
radii from the IP (in cm): 12.7, 14.9, 17.1, 19.3 and 21.5.
\begin{figure}[!h]
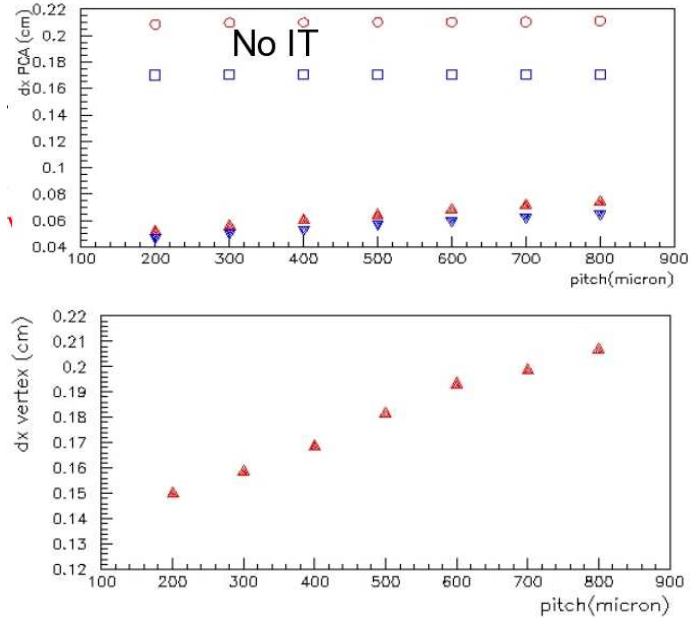

\centering
\figb PHYSpics/res_pca_ks;9; \kern0.1cm \figb PHYSpics/res_vertex_ks;9;
\caption{(Top)
Resolution on the point of closest approach to
the decay vertex in the transverse plane obtained for reconstructed tracks
in $K_S\rightarrow\pi^+\pi^-$ (BLUE triangle) and
$K_S\rightarrow\pi^+e^-\nu$ events (RED triangle).
%Minimum distance in the transverse plane of the reconstructed
%trackes in $K_S\rightarrow\pi^+\pi^-$  and
%$K_S\rightarrow\pi^+e^-\nu$ (RED triangle).
Empty point indicates the same quantities without the IT.
(Bottom)
 Vertex resolution obtained in the
transverse plane for $K_S\rightarrow\pi^+\pi^-$ decays using
 several values of the readout pitch. }
%Vertex resolution
%with respect to the strip pitch.}
\label{fig:pca_vert}
\end{figure}
\par
The simulation accounts for the following materials for each GEM layer: $300\ \mu m$ of kapton,
$2\ \mu m$ of copper, and 8 mm of gas mixture(Ar-CO$_2$). The grand total of the
material budget for the five layers of the IT is then $1.08\ \%\ $ of $X_0$.
% The step lenght and the cosine of the polar angle of the pion tracks in the
% GEM planes are shown in fig.\ref{fig:steps} and in fig.\ref{fig:costeta}
% for the golden $\phi$ decay channel for the IT design :$K_S \rightarrow \pi^+\pi^- ; K_L \rightarrow 3\pi^0$.
% \begin{figure}
% \vspace{6 cm}
% \special{psfile=PHYSpics/Step.eps hscale=50 vscale=50}
% \caption{\it Step of the pion tracks throgh the GEM active volume}
% \label{fig:steps}
% \end{figure}
% \begin{figure}
% \vspace{6 cm}
% \special{psfile=PHYSpics/Costeta.eps hscale=50 vscale=50}
% \caption{\it Cosine of the polar angle
% of the pion tracks throgh the GEM active volume}
% \label{fig:costeta}
% \end{figure}
Each GEM is simulated as a 2 mm gas stage, in which
ionization takes place, followed by the three multiplication stages
simulated as a whole. The simulated hits from the particles crossing the IT
are then digitized according to a two dimensional X-V strip readout for each GEM layer.
The X strips are parallel to the z beam axis
\footnote{In the following the coordinate system is defined with the
$z$-axis along the bisector of the \epm\ beams, positive along the positron
motion, the $y$-axis vertical
and the $x$-axis toward the center of the collider rings and origin at the
collision point.}, while the V strips
have a 40 degree angle with respect to the beam axis. Both X and V strips
have a $650\ \mu $m pitch. The Z coordinate is then obtained from the
crossing of both X and V strip readout as $Y=tan(50^\circ)\times X +
V/cos(50^\circ)$. The corresponding single hit resolution obtained with
digital readout is $\simeq 190\ \mu$m in the $r-\phi$ plane and 
$\simeq 400\ \mu$m in the z direction.

The digitization procedure takes properly into account the
presence of ghost hits due to the X-V projection.
To extract the helix parameters of the incoming particle,
 the digitized hits are grouped first into clusters, then the clusters
information is processed by a dedicated tracking code.
% exploiting the kalman filter technique.

\par
%For the purpose of the IT design
We start with the tracks already reconstructed
in the KLOE Drift Chamber and extrapolated to the IT.
The helix parameters obtained from the DC are updated with the Kalman filter
by using the information of the IT clusters. Finally the tracks are extrapolated to
the IP and given as input to a simple vertex finder based on minimum helixes distance.
%************************************************
%************************************************
\par
The simulation shows that pions coming from $K_S\rightarrow\pi^+\pi^-$ decays
cross nearly orthogonal the GEM layers,
due to the typical P-wave angular distribution of kaons
produced from $\phi$ decays.
%************************************************
%************************************************
Fig \ref{fig:pca_vert} top shows the resolution on the point of closest approach to
the decay vertex in the transverse plane obtained for reconstructed tracks
in $K_S\rightarrow\pi^+\pi^-$ and  $K_S\rightarrow\pi^+e^-\nu$ events.
\begin{figure}[!hb]
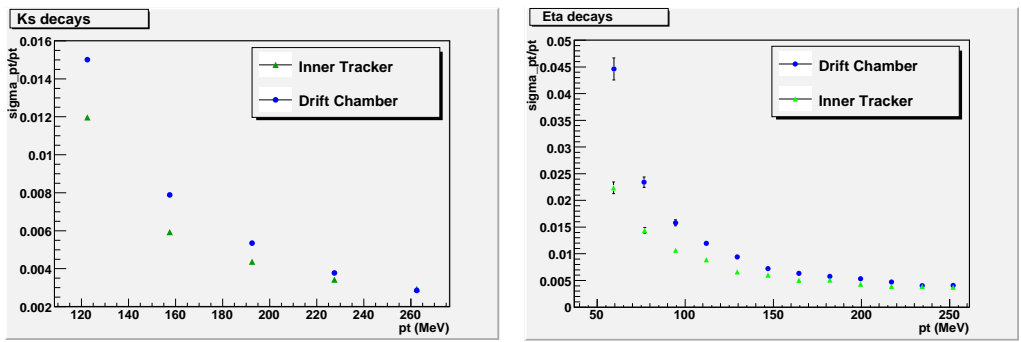

\centering
\figb PHYSpics/sigmaptoverpt_pt_ITvsDC;6.5; \kern0.2cm \figb PHYSpics/sigmaptoverpt_pt_ITvsDC_etappee;6.5;
\caption{
Resolution $\sigma_{p_T}/p_T$ on the transverse momentum measurement as a
function of $p_T$ for $K_S \to \pi^+\pi^-$ (Left) and  $\eta \to \pi^+\pi^-
e^+e^-$ decays (Right) with DC only (circles) and
adding the IT information (triangles). }
\label{fig:sigmapt_overpt}
\end{figure}
The comparison with the same quantity obtained without the IT shows an
improvement of a factor 3 with a $600\ \mu m$ pitch,  perfectly
compatible with the requirements of KLOE-2, discussed in the
previuos section. Fig \ref{fig:pca_vert} bottom shows the vertex resolution obtained in the
transverse plane for $K_S\rightarrow\pi^+\pi^-$ decays using
 several values of the readout pitch.
\par
The insertion of the Inner Tracker improves also the resolution on the
measurement of the transverse momentum,  $p_T$.
 The effect has been studied with $K_S \to \pi^+\pi^-$ and  $\eta \to \pi^+\pi^-
e^+e^-$ decays, assuming a pitch of 650 $\mu$m. Fig.\ref{fig:sigmapt_overpt} shows
the behaviour of $\sigma_{p_T}/p_T$ as a function of $p_T$,  obtained using
the DC only and adding the IT information. As expected, usage of the IT
improves the transverse momentum resolution for low momentum tracks,
as much as 20\% at $p_T\simeq$120 MeV for $K_S \to \pi^+\pi^-$ decays and
50\% at $p_T\simeq$60 MeV for $\eta \to \pi^+\pi^-e^+e^-$ decays.
%
%
% ATTENZIONE AGGIUNGERE UNA FRASE SUL PATTERN INDIPENDENTE!
%
\newpage
%@@@@@@@@@@@@@@@@@@@@@@@@@@@@@@@@@@@@@@@@@@@@@@@@@@@@@@@@@@@@@@@@@@@@@@@@@@@@@@@@@@@@@@@
\section{Prototype studies}
The idea of a cylindrical GEM detector was tested for the first time
with a small
prototype (7 cm radius and 24 cm length) \cite{bib:vienna}.
The very positive results obtained with this prototype paved the road
for the construction of a full-scale prototype for the first layer of the
IT.
Since then, the R\&D activity for the final detector has been focused on three main
items: the realization and test of a full-scale Cylindrical GEM prototype,
the detailed study of the XV readout, performed on dedicated small planar
chambers for simplicity and economical reasons, and the realization of
very large GEM
foils based on the new single-mask technology. In the following three
subsections, these three items will be discussed in detail.

\subsection{Full-scale CGEM prototype}
In 2007 we built a cylindrical GEM prototype with dimensions similar to
those of the first layer of the final IT: it has the same diameter of 300~mm but a reduced active length, 352~mm instead of 700~mm, due to a limitation in the availability of large area GEM foils. For the same reason a single cylindrical electrode (352x960~mm) was obtained as a joint of three identical GEM foils (352x320~mm).

% The foils have been glued
%with an overlap region 3~mm wide, where the copper has been completely etched.
%The cathode is realized as a unique polyimide foil copper clad on the internal face.
%The anode is realized as a join of 3 foils, made of polyimide, with a
%read-out segmentation composed by a 1-dimensional set of strips along the
%axis of the cylinder, to measure the $r\phi$ coordinate. The 1538 strips have a 650~$\rm \mu m$ pitch.

The detector has a geometrical configuration of the
gaps of 3/2/2/2~mm, respectively for drift/transfer1/transfer2/induction
(fig.~\ref{fig:gemscheme}), with the cathode being the innermost electrode.

\subsubsection{Construction}
\label{sec:construction}
%
% \begin{figure*}[htb]
% \centerline{\subfigure{\includegraphics[width=3.5cm]{CGEMpics/buildingCgem1}}
% \hfil
% \subfigure{\includegraphics[width=5cm]{CGEMpics/buildingCgem2}}
% \hfil
% \subfigure{\includegraphics[width=5cm]{CGEMpics/buildingCgem3}}
% \hfil
% \subfigure{\includegraphics[width=3.5cm]{CGEMpics/buildingCgem4}}}

\begin{figure}[htb]
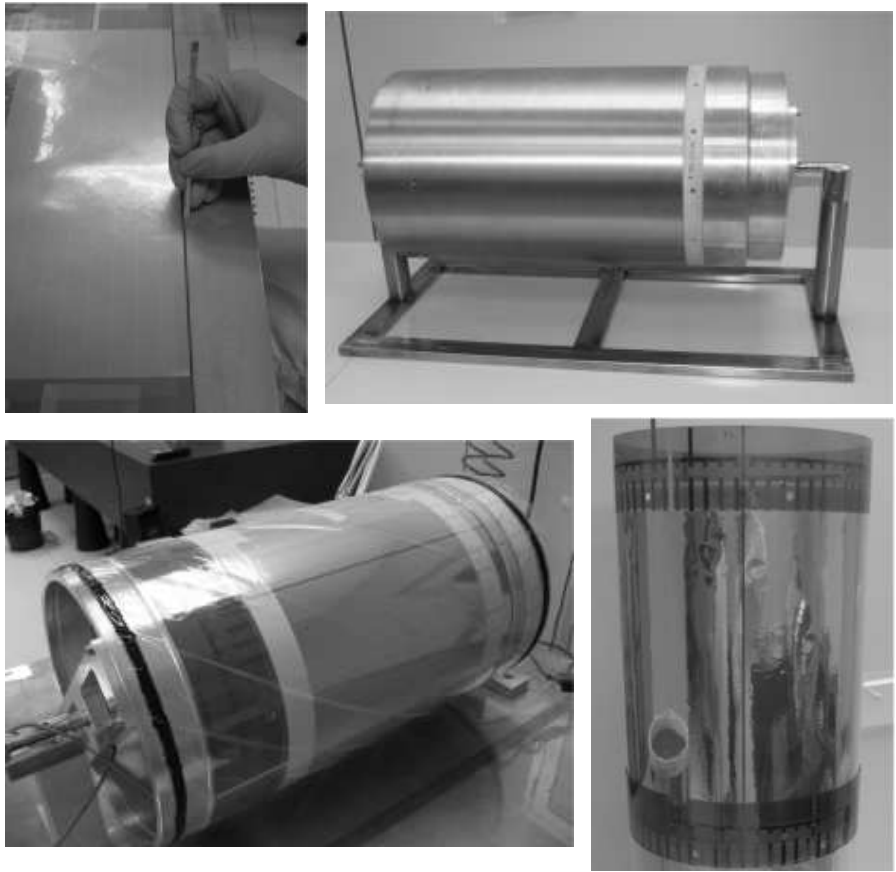

\centering
\figb CGEMpics/buildingCgem1;4.0;\kern0.2cm \figb
CGEMpics/buildingCgem2;7.5; \\
\figb CGEMpics/buildingCgem3;7.5;\kern0.2cm \figb CGEMpics/buildingCgem4;4.0;
\caption{The various steps of the construction of a cylindrical
  GEM. See the text for the description.}
\label{building}
\end{figure}
A special manufacturing technique has been developed to obtain cylindrical
GEM electrodes, as shown in fig.~\ref{building} %and described below.
At first three GEM foils are glued together to obtain the single large foil
needed to make a cylindrical electrode. We used an epoxy (Araldite),
applied on one of the short sides of the GEM foil, on a 3~mm wide region
(fig.~\ref{building}~top-left).
Then the foil is rolled on an aluminum mould coated with a very precisely
machined 400~$\rm \mu m$ thick Teflon film which provides a non-stick, low
friction surface (fig.~\ref{building}~top-right). Finally the mould is inserted in a
vacuum bag where vacuum is made with a Venturi system, resulting in a
uniform pressure of 0.8~Kg/cm$^2$ over the whole surface of the
cylinder (fig.~\ref{building}~bottom-left). At this stage, two fiberglass annular rings
are glued on the edges of the electrode, acting as spacers for the
gaps and providing all of the mechanical frames needed to support the
detector. After the curing cycle of the glue, the cylindrical electrode is
easily extracted from the mould thanks to the Teflon surface
(fig.~\ref{building}~bottom-right). Cathode and anode are obtained with the same
procedure as well.
%At the end the five electrodes are inserted one into the other and the detector is sealed with epoxy on
%both sides.

% \begin{enumerate}
% \item An epoxy adhesive (typically Araldite) is distributed along one
%   edge (3~mm wide) of the foil.
% \item The foil is rolled on an Aluminum mould coated with a very
%   precise 400~$\rm \mu m$
%   thick machined Teflon film, to have a non-stick, low-friction
%   surface.
% \item The cylinder is enveloped in a vacuum bag and the vacuum is
%   obtained with a Venturi system, providing a high
%   ($\simeq$ 1~kg/cm$^2$) and uniform pressure throughout the surface
%   of the cylinder.
% \item The foil is easily extracted from the mould, thanks to its
%   Teflon surface, and a cylindrical GEM is obtained.
% \item With the same technique the cathode and anode foils are obtained
%   as well.
% \end{enumerate}

\paragraph{The cylindrical cathode}
The cathode is realized as a unique polyimide foil, 100~$\rm \mu m$ thick, with a copper cladding of 18~$\rm \mu m$ on the internal surface.
All the support mechanics of the chamber is composed by annular flanges
made of Permaglass (G11) placed on the edges of the cylinder. These flanges
house the gas inlets and outlets and their thickness defines the distance
between the various electrodes.

\paragraph{The GEM foils}
% The largest GEM foil has an area of 352x960~mm$^2$, being one of the
% largest ever built.
Three 352x320~mm$^2$ foils are spliced together in order to realize one
single prototype electrode.
For a safe detector operation the foil has independent high voltage sectors, in order to limit the capacitance and hence the energy released through
the GEM hole in case of discharges. Each foil has 20 sectors, with an area of about 56~cm$^2$,
corresponding to a width of 1.6~cm for the single sector.
\begin{figure}[!h]
\centering
\includegraphics[width=0.7\columnwidth]{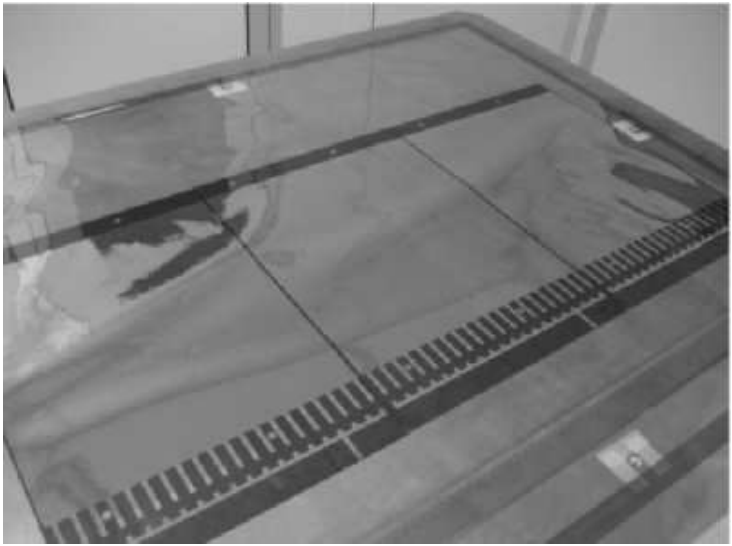}
\caption{The GEM obtained as a join of three foils. The dimensions are
960x352~mm$^2$.}
\label{fig:largefoil}
\end{figure}

\begin{figure}[!t]
\centering
\includegraphics[width=0.6\columnwidth]{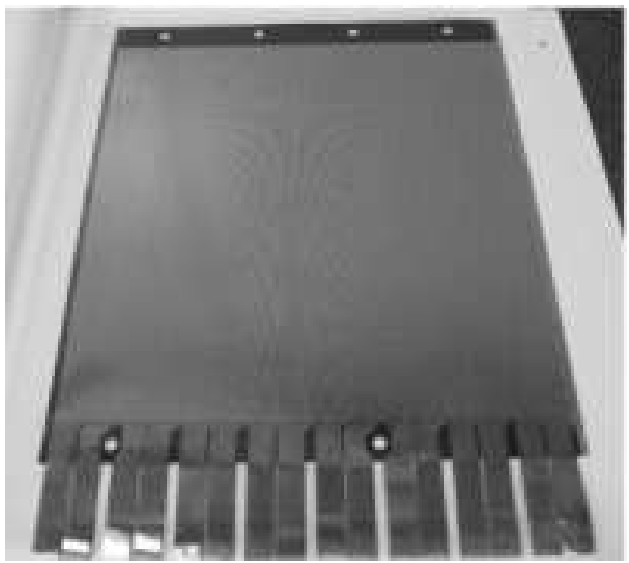}
\caption{The anode with the strips for the readout and the flaps for
  the bonding of the FEE.}
\label{anode}
\end{figure}

\paragraph{The anode readout}
Also the anode is realized as a splicing of three foils (fig.~\ref{fig:largefoil}), each with the readout copper strips and the ground.
The copper strips surface and the ground surface are placed on the opposite
faces of two different polyimide foils, staggered and glued together.
This avoids discontinuities in the readout strips and preserves the pitch value
across the overlap region. In order to bring out the signals, the strips end-up in polyimide flaps (see fig.~\ref{anode}),
each grouping 32 strips, where the FEE is plugged with ZIF connectors. The pitch of the
strips is 650~$\rm \mu m$ in the readout zone
and provides a $\sigma\simeq$~200~$\rm \mu m$ spatial resolution,when equipped with a digital readout,
fulfilling the KLOE-2 requirements.
In the flaps the pitch is reduced to 500~$\rm \mu m$ in order to match the pitch of the connector.

\begin{figure}[!t]
\centering
\includegraphics[width=4.5cm]{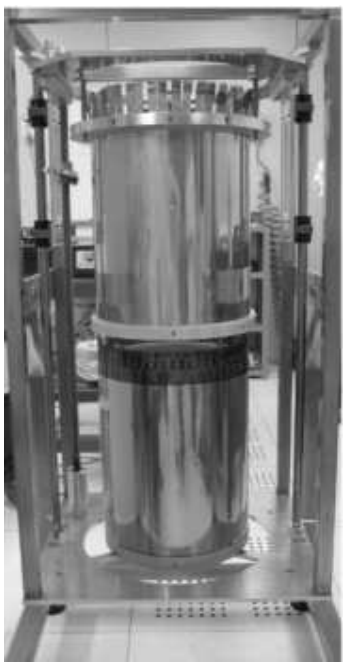}
\caption{Two electrodes fixed on the vertical insertion system used to
  assembly the detector.}
\label{vis}
\end{figure}

\paragraph{Assembly of the detector}
The five electrodes are extracted from the moulds by using a PVC ring, bound with pins to one of the annular flanges of the cylinder, and then are inserted one into the other.
To accomplish the insertion of the electrodes without damaging the GEMs, a
dedicated tool has been realized: the Vertical Insertion System (fig.~\ref{vis}).
The electrodes are fixed on two aluminum plates aligned on a vertical axis, and one
is pulled down with a very precise linear bearing equipment.

\begin{figure}[!t]
\centering
\includegraphics[width=0.6\columnwidth]{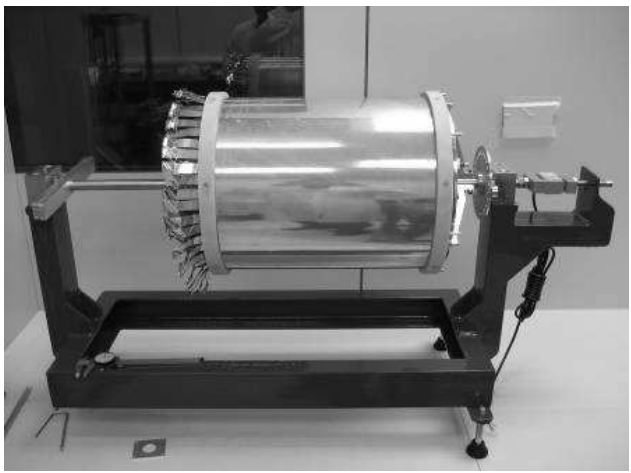}
\caption{The chamber mounted on the support system and stretched. The
  tension is measured by the load cell on the right end.}
\label{cell}
\end{figure}

After the assembly of all the electrodes, the detector has been sealed on
both sides, mounted on a support system and longitudinally
stretched with a tension of 200~g/cm ($\sim$100~kg of overall
tension) measured by a load cell (see fig.~\ref{cell}).
%%%%%%%%%%%%%%%%%%%%%%%%%%%%%%%%%%%%%%%%%%%%%%%%
\subsubsection{X-ray test}
The very first test of the CGEM prototype was performed using a X-rays source.
The prototype was flushed with a gas mixture of $Ar/iC_4H_{10}/CF_4 =
65/7/28$ and tested in current mode with a 6~keV X-ray gun.
A 10x10~cm$^2$ planar GEM was placed in the same gas line and
irradiated from a side opening of the gun. It was used as a reference
to account for possible gain variations due to changes in
atmospheric variables.
We estimate that we operated the
CGEM prototype at gains larger than 10$^4$.
\begin{figure*}[htb]
\centerline{
\subfigure{\includegraphics[width=0.4\linewidth]{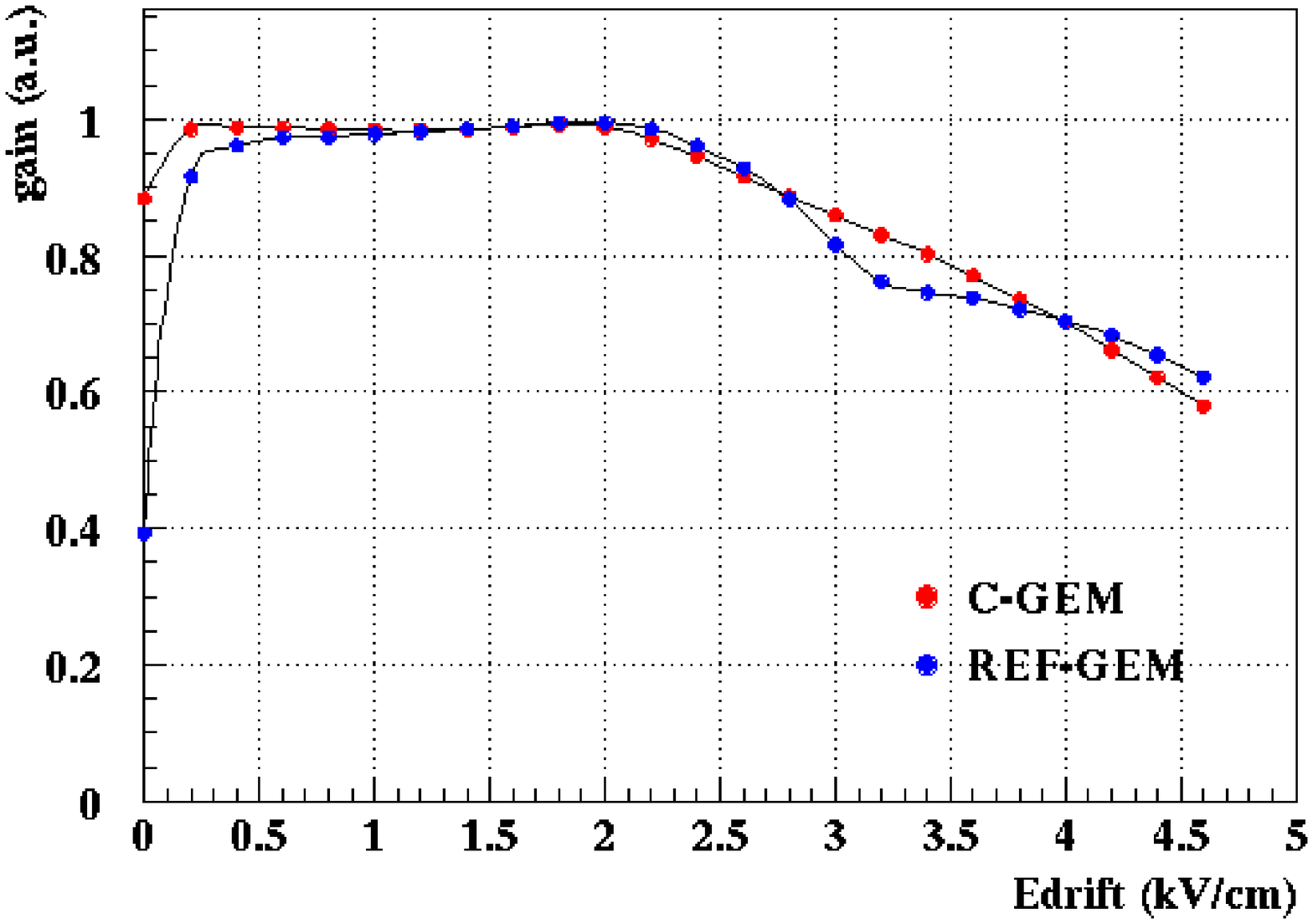}}
\hfil
\subfigure{\includegraphics[width=0.4\linewidth]{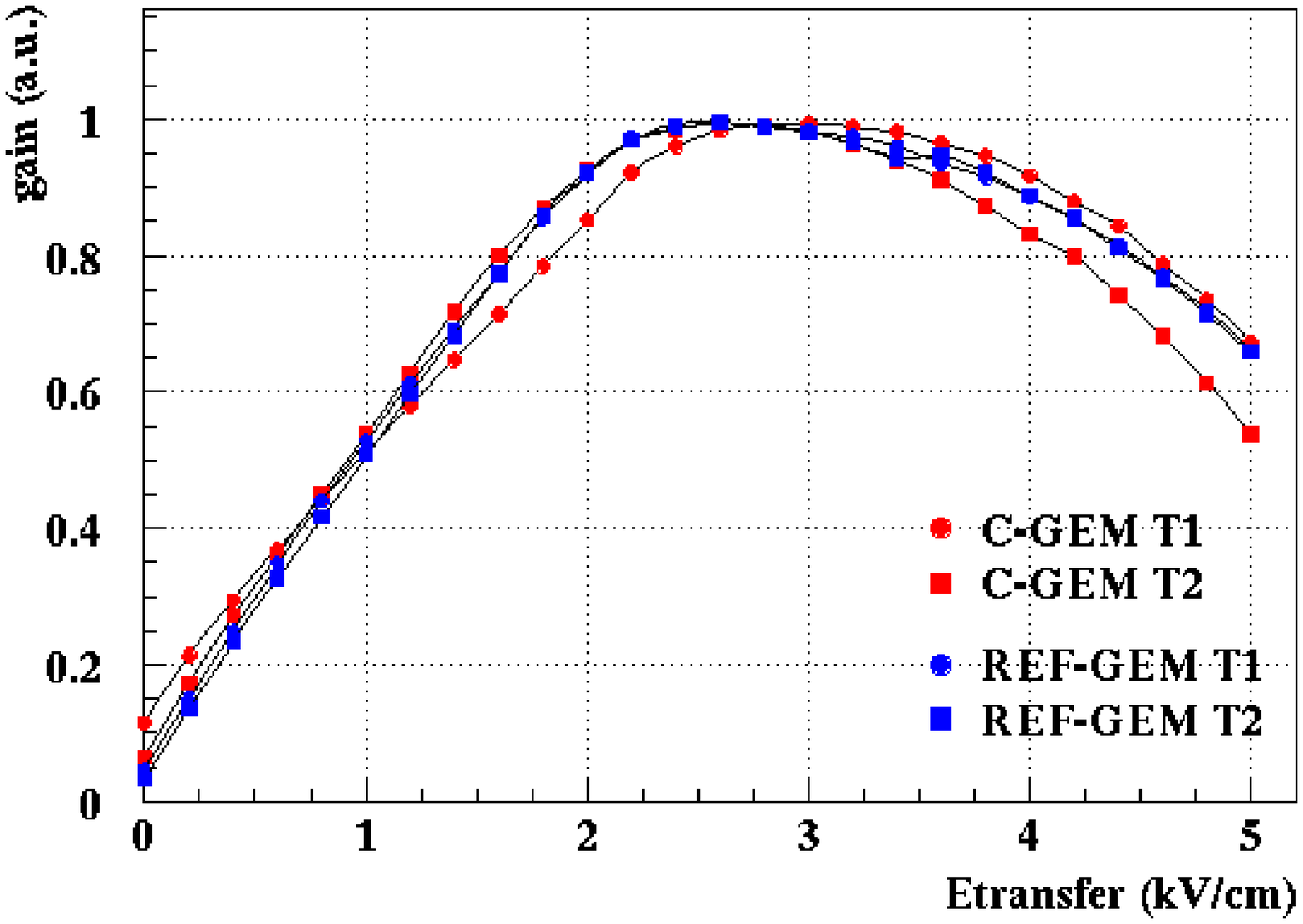}}
\hfil
\subfigure{\includegraphics[width=0.4\linewidth]{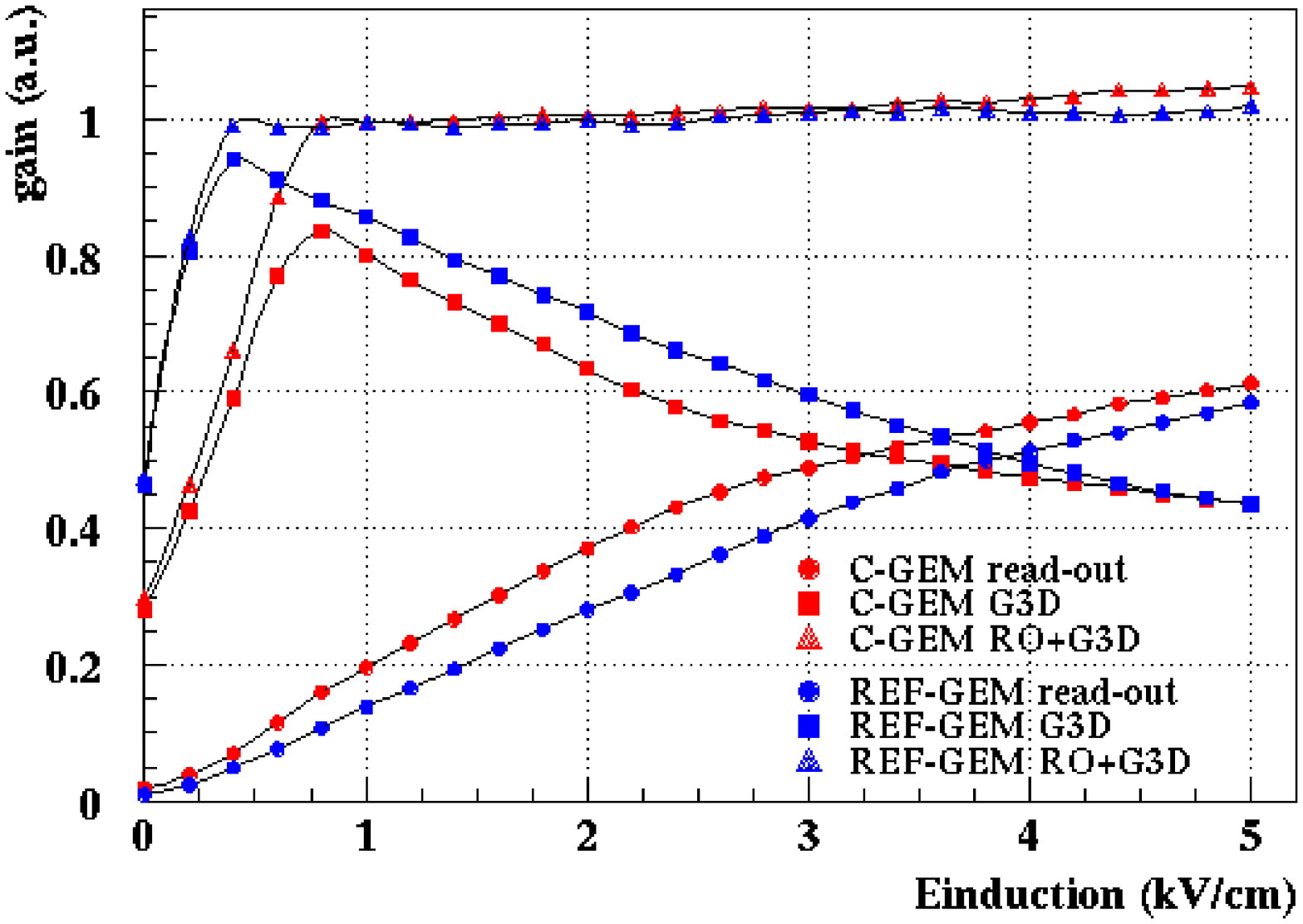}}
}
\vspace{-1.0cm}
\caption{Relative gain as a function of drift field (left), transfers fields T1 and T2 (center) and induction field (right) of the CGEM with respect to the planar reference chamber. All the electric fields not involved in each measurement are kept at constant values.}
\label{fig:cgemxray}
\end{figure*}
In order to characterize the CGEM detector and to find the optimal operating parameters,
the relative gain with respect to the reference was measured as a
function of the different electric fields in the GEM.
The first two plots of fig.~\ref{fig:cgemxray} show the electron
transparencies (defined as effective gain normalized on its maximum value)
as a function of the drift and the transfer fields: this is
a measurement of the focusing efficiency of the electrons into the three
GEM foils. The third plot of fig.~\ref{fig:cgemxray} shows the charge
sharing between the bottom surface of the third GEM and the anode readout, as a
function of the induction field: this is a measurement of the extraction
efficiency from the last GEM foil. The sum of the two charges is also
shown (red and blue triangle on the third plot); no dependence
from the induction field is observed. All the measurements have been performed with a CGEM
polarization of 375/365/355~V and are in good agreement with previous
results found in literature.
\begin{figure}[hbt]
\centering
\includegraphics[width=0.4\linewidth]{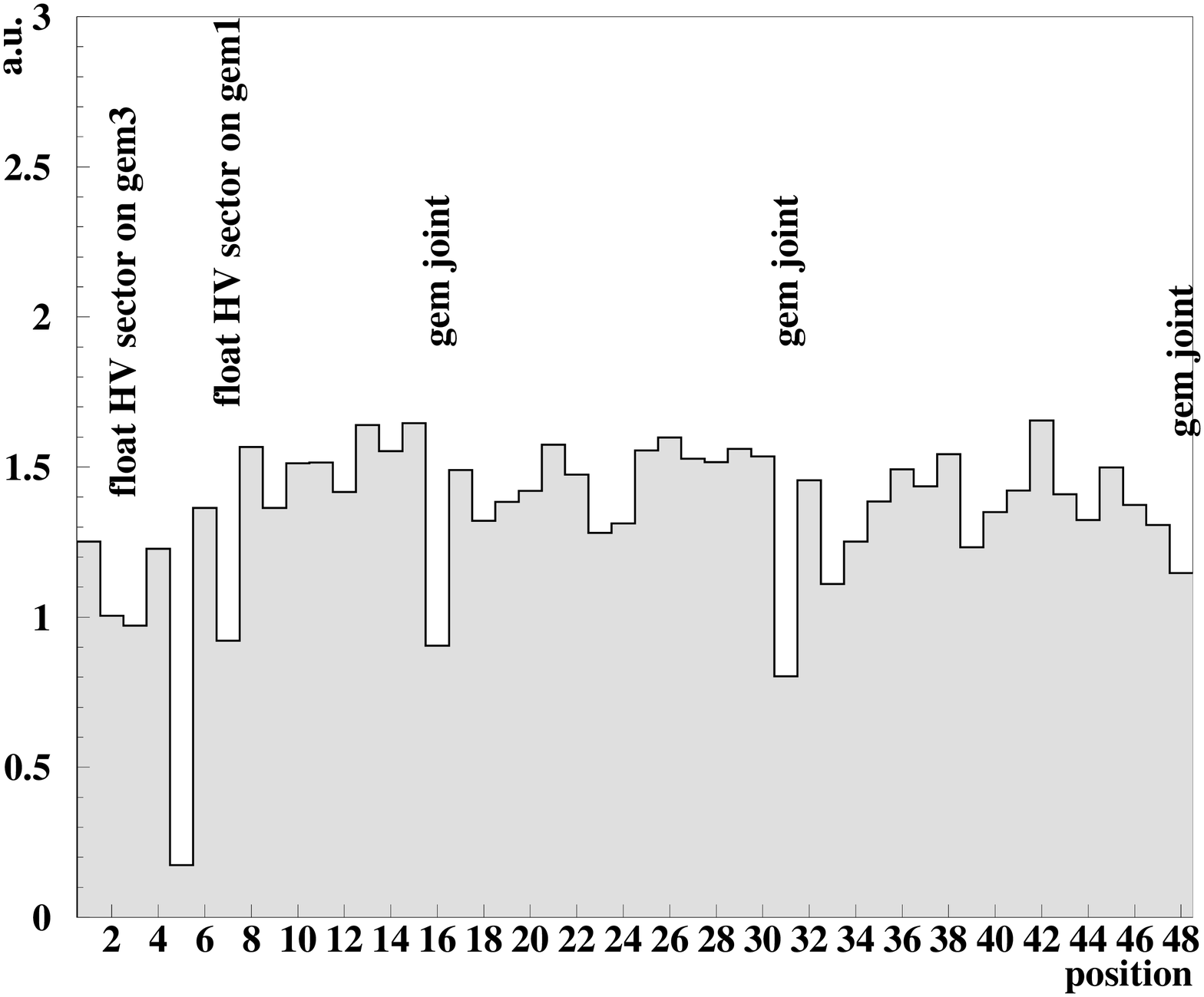}
\includegraphics[width=0.4\linewidth]{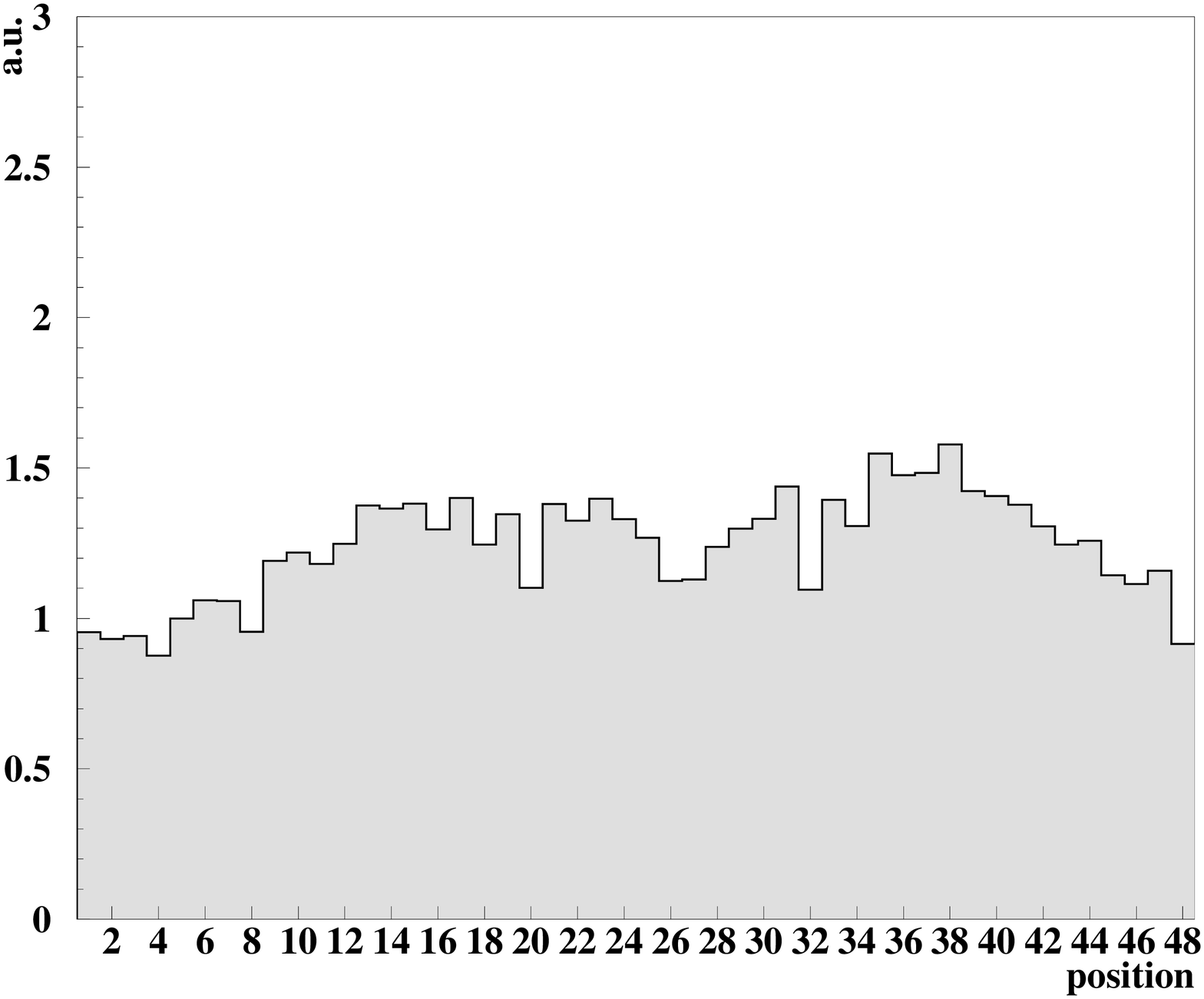}
\caption{CGEM gain uniformity (left) and induction gap thickness uniformity (right). Excluding sectors with intrincically weaker response (e.g. gem joints), both measured values have R.M.S. within 9\%.}
\label{fig:uniformity}
\end{figure}
\par
The uniformity of the CGEM has also been measured over the 940~mm of its
circumference (fig.~\ref{fig:uniformity}).
The fluctuations (R.M.S.) of the measured currents were within 9\%, showing a good uniformity on such a large surface.
Since the ratio between the current on the bottom side of the third GEM and the read-out electrode depends only on the applied induction field, we were also able to measure the uniformity of the induction gap thickness. The result indicates that the detector construction was realized with good mechanical precision.
%%%%%%%%%%%%%%%%%%%%%%%%%%%%%%%%%%%%%%%%%%%%%%%%%%%%%%%%%%%%%%%%%%%%%%%%%%%%%%%%%%%%%%%%%%%
\subsubsection{CGEM test beam at CERN}
\label{sec:CGEM_testbeam}
The CGEM prototype was also extensively tested with the 10 GeV
pion beam at the T9 area of CERN PS~\cite{CGEM-dresda08}.
Two different read-out devices were used: 128 channels
were equipped with the GASTONE ASIC, while
96 channels with the CARIOCA-GEM electronics.

GASTONE is a custom chip, developed to fulfill
the low-power consumption and high integration requirements of the KLOE-2
experiment.  Its
first release was tested during the test beam~\cite{CGEM-Gastone}. Results of this
test are reported in
sec.~\ref{sec:gastone}.

The CARIOCA-GEM chip, instead, has been developed for the LHCb Muon System GEM
detectors. CARIOCA is a very fast chip (10~ns peaking time) with  digital
readout and has been used to measure the timing performance of the
detector, coupled to a 100~ps resolution TDC.

The CGEM was flushed with a Ar/CO2 ( 70/30) gas mixture and operated with the following voltages:
 V$_{fields}$ = 1.5/2.5/2.5/4 kV/cm and
 V$_{GEM}$ = 390/380/370 V ($\sum$V$_G$ = 1140V, corresponding to a gas gain of 2$\times$10$^4$).

\begin{figure}[!h]
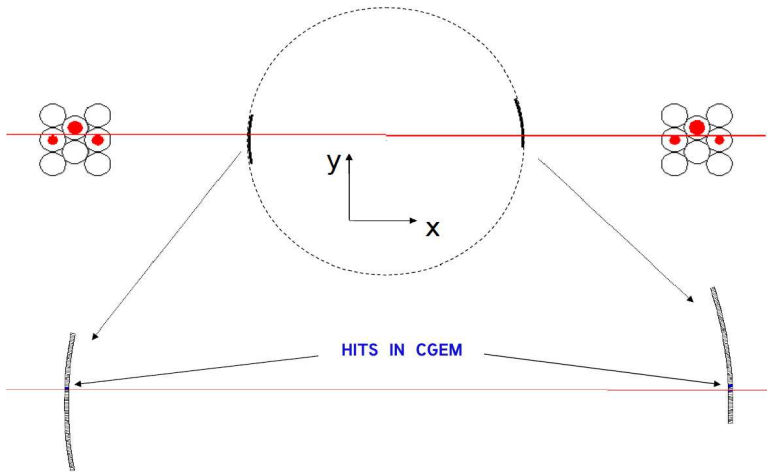

\centering
\figb TBpics/event_good;10.;
\caption{An event display example illustrating the tracker setup. Two sets with 3 planes of drift tubes (20 channels in total) are placed in front and behind the CGEM to measure the Y coordinate.}
\label{fig:event}
\end{figure}
\par
The track position was measured with an external tracker based on ATLAS
drift tubes with 30~mm diameter and 0.4~mm wall thickness.
These tubes were operated with a 100~$\mu m$ wire diameter, in streamer mode, using a gas mixture
Ar/C$_4$H$_{10} : 40/60$ at STP.

The tracker consisted of two sets of longitudinal stations, each with 8
channels arranged on 3 planes, placed in
front and behind the CGEM prototype and providing a
measurement of the Y coordinate (see fig.~\ref{fig:event}).
Only tracks with on-time hits in each of the 6 tracker planes have been used for the analysis.

The contribution to the spatial resolution due to the external tracker was measured to be
$\sigma_{tracker}$= 140$\mu$m.
The setup also included a trigger built with a coincidence of three scintillators placed
before and after the CGEM.

\begin{figure}[!h]
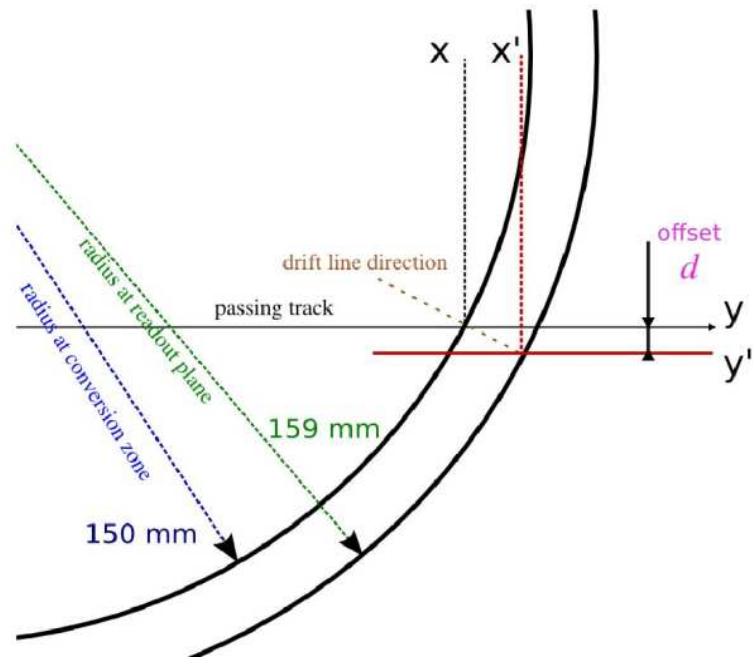

\centering
\figb TBpics/gem_sector;10.;
\caption{Radial tracks follow drift lines of electrons inside the GEM and
  pass the readout plane at the same position where the charge collection
  takes place. Non-radial tracks do not follow drift lines and the measured
  position on the readout plane is shifted with respect to the
  impinging position of the track (offset {\it d}). Moreover cluster created in
  the conversion zone are not distributed along the drift lines and are
  collected on several different readout strips thus increasing the cluster size.}
\label{fig:gemsect}
\end{figure}

\begin{figure}[!h]
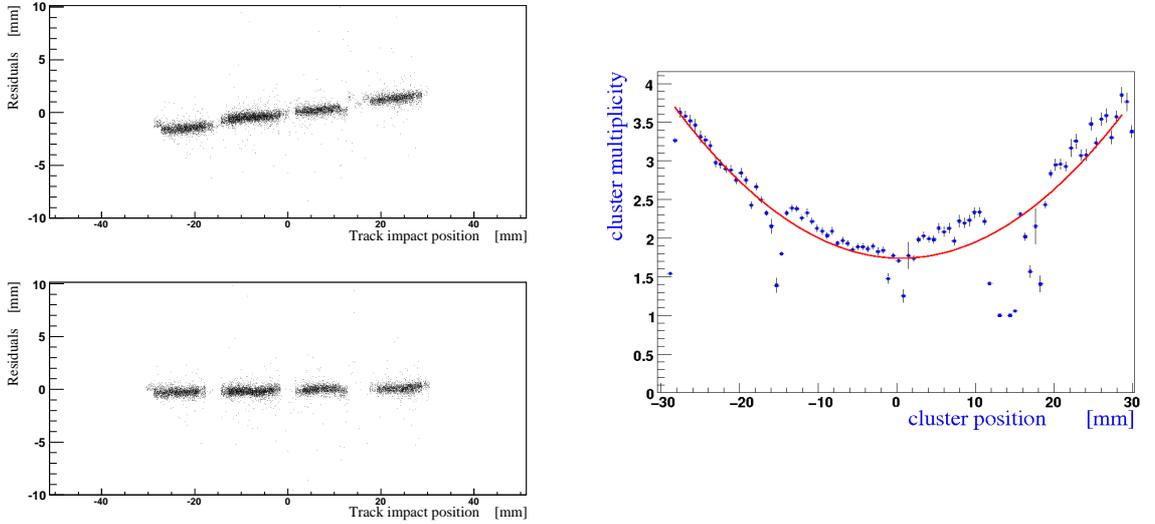

\centering
\figb TBpics/res_comparizon;8.;\kern0.2cm \figb TBpics/multipl_vs_pos1;7.;
\caption{GEM residuals as a function of the track impact point before (top-left) and after
  (bottom-left) the correction factor for the non-radial tracks. Cluster multiplicity (right) as a function of the track position shows the effect of the offset.}
\label{fig:gemcorr}
\end{figure}

The reconstruction procedure had to take into account the
fact that the measured particles do not generally cross the detector along
the radius of the cylinder i.e. along the drift lines inside the gems. This is shown in fig.~\ref{fig:gemsect}.
The effect increases moving from the center of the cylinder outwards,
and affects both the position resolution and the strip cluster size.
For each reconstructed track an offset is calculated, based on the
impact position of the track, thus allowing to correct the reconstructed position (see fig.~\ref{fig:gemcorr} left).
%

%The cluster behavior of the CGEM was studied.
Ionization clusters created in the conversion zone by non-radial tracks are not distributed along the drift lines and are
collected on several different readout strips. This effect leads to an
increase of the cluster size as a function of the track impact point in the
Y coordinate (see fig..~\ref{fig:gemcorr} right). For radial tracks
(i.e. Y=0) an average cluster multiplicity of 1.8 has been measured.
%
% \begin{figure}[!h]
% \centering
% \figb TBpics/gast_nr_clusters;5.;\kern0.2cm \figb TBpics/gast_nr_hits_in_cluster;5.;
% \caption{Cluster multiplicity (left) and cluster size (right) in the zone equipped with GASTONE.}
% \label{fig:gast_clu}
% \end{figure}
% %

\begin{figure}[!h]
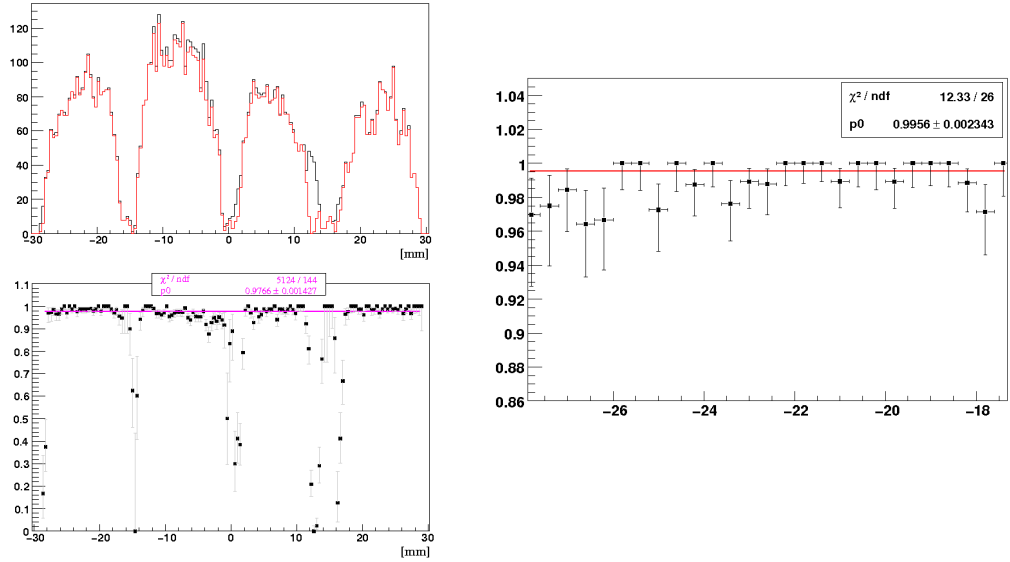

\centering
\figb TBpics/efficiency_gastone_official;7.;\kern-0.6cm \figb TBpics/glob_eff_zoom;9.;
\caption{Left top: Distribution of reconstructed (black) and measured (red) impact track position; left bottom:
efficiency as a function of the impact track position; right: detail of the efficiency in a high statistics region.}
\label{fig:gast_eff}
\end{figure}

Fig.~\ref{fig:gast_eff} shows the efficiency of the chamber measured for
different positions of the impact track, with the GASTONE chip threshold set
at 3.5 fC. The efficiency averaged over the
whole equipped region is 97.7\%. The
low statistics points are due to a lack of reconstructed tracks
in proximity of the walls of the drift tubes (30 mm diameter).
An efficiency of 99.6\% is obtained when rejecting these points.

\begin{figure}[!h]
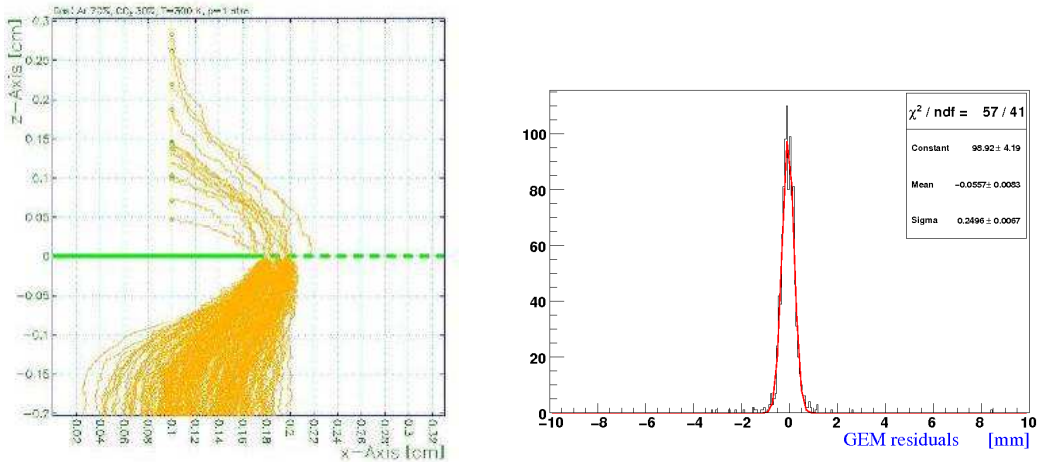

\centering
\figb TBpics/driftlines_sim;6.;\kern0.1cm \figb TBpics/gem_res_Rgt5;9.;
% \figb TBpics/ArCO2track_2;6.;\kern0.1cm \figb TBpics/gem_res_Rgt5;9.;
\caption{Left: Distortion of the field lines in the gluing region (GARFIELD
  simulation). Right: CGEM resolution.}
\label{fig:gast_res}
\end{figure}

In order to produce large area GEM foils, one needs to glue together
different smaller area GEMs (see sec.~\ref{sec:construction});
this technique may lead to a possible loss of efficency of the detector in
the overlapping region, which needs to be well studied and understood.
In the final detector design, this region accounts for less than 0.4\% of the total CGEM
foil. In order to investigate this effect, we studied the time distributions
obtained using the CARIOCA-GEM chip, readout by a  100ps resolution
 TDC and with gas mixture Ar/CO$_2$ (70/30) with drift velocities V$_d$ = 7
 cm/$\mu$s at 2 kV/cm (approximately 10 clusters in 3 mm).
\begin{figure}[!h]
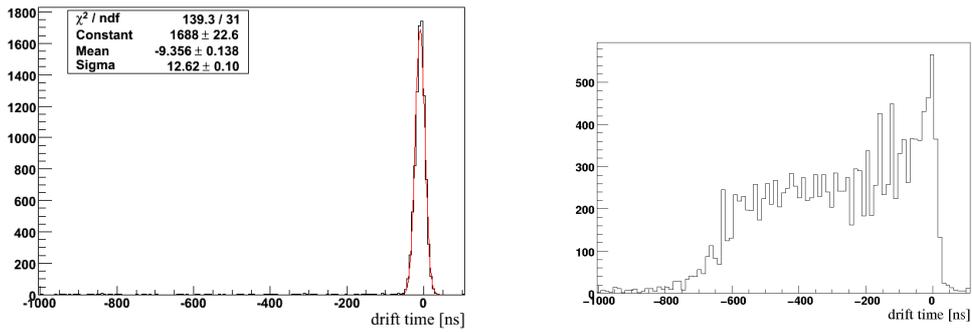

\centering
\figb TBpics/DriftShort1;7.;\kern0.2cm \figb TBpics/DriftTlong2;7.;\\
\caption{Time distributions in regular regions (left) and in the joint zone
(right).}
\label{fig:drifttimes}
\end{figure}
Time distributions in non overlapping (regular) regions and in the
joint zone
are shown, respectively in fig.~\ref{fig:drifttimes} left and right.
In the regular region a 13 ns RMS is obtained, in agreement with the performance expected for
the gas mixture in use. In the joint zone the spectrum is much broader, with
a 200 ns RMS. In particular the signals are delayed up to
800 ns, suggesting that a longer drift path had to be followed to reach the
anode. Ionization electrons originating from a track passing through an overlap region,
drift along the distorted field lines, are then efficiently driven
and focused in the multiplication holes of the GEM and finally are
picked up by the anode with longer collection time.
Such hypothesis has been confirmed by simulations studies using
ANSYS and GARFIELD, showing a distortion of the
field lines in the glueing regions, due to a space charge effect on the
dielectric (fig.~\ref{fig:gast_res} left).

Fig.~\ref{fig:gast_res} right shows the residuals of the clusters position with respect to the reconstructed
position of the track. If the
contribution of the tracker resolution ( $\sigma_{tracker}$= 140$\mu$m) is subtracted, the GEM spatial resolution is found to be
$ \sigma_{GEM}  = \sqrt{\sigma^2_{residuals} - \sigma^2_{tracker}} \simeq
200 \mu$m.
This is in good agreement with what expected from a digital readout of
650 $\mu$m pitch strips.
In conclusion, the full scale prototype was safely operated at CERN PS with 10 GeV
pion beam.
This test successfully validated the innovative idea of
the fully sensitive cylindrical GEM detector, constructed with no support
frames inside the active area.

The material budget used in the construction of the CGEM prototype and the
the measured spatial resolution are fulfilling the inner
tracker requirements.

\pagebreak
\subsection{Planar GEM for readout studies}
\begin{figure}[!h]
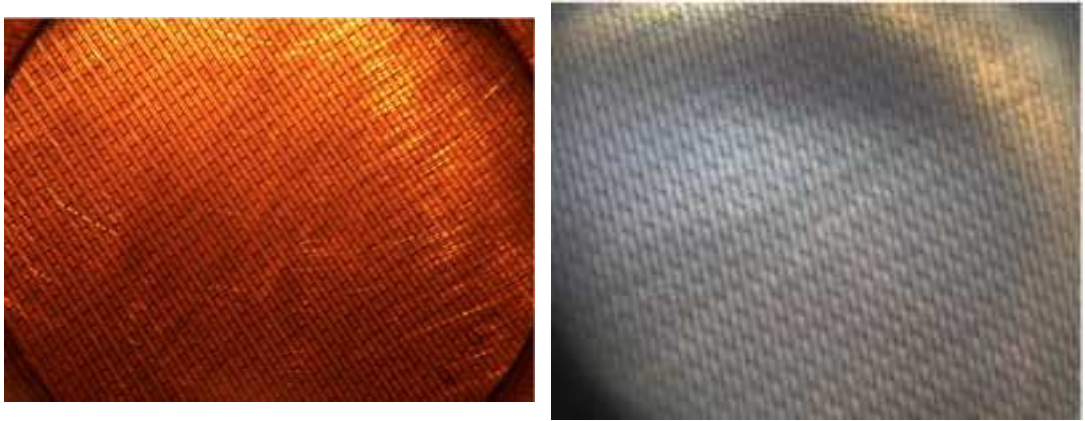

\centering
\figb PGEMpics/XY_readout;7.;\kern0.2cm \figb PGEMpics/XV_readout;7.;\\
%\figb PGEMpics/ ;7.;\kern0.2cm \figb PGEMpics/ ;5.;
\caption{Pictures of the 2D-read-out: (Left) X-Y strips and (Right) X-V strips}
\label{fig:2D_readout1}
\end{figure}

\begin{figure}[!h]
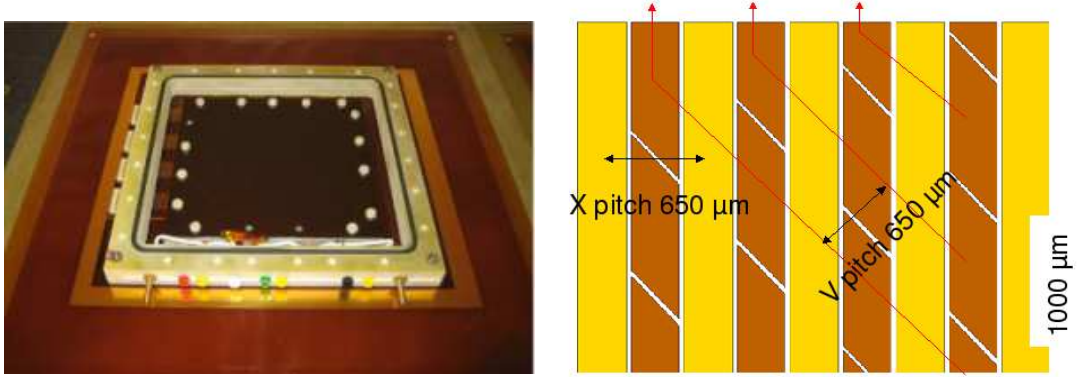

\centering
\figb PGEMpics/XY_chamber;7.;\kern0.2cm \figb PGEMpics/XV_readout_draw;7.;\\
%\figb PGEMpics/ ;7.;\kern0.2cm \figb PGEMpics/ ;5.;
\caption{Picture of assembled triple GEM plane with X-V readout (left), schematic details of X-V readout design (right).}
\label{fig:2D_readout2}
\end{figure}

A typical orthogonal X-Y readout can not be used for the IT,
due to its cylindrical geometry. However, a 2-D readout can still be obtained
making use of 650 $\mu$m pitch XV patterned strips at an angle of 40$^\circ$ (see sec.\ref{sec:readout_anode}).
The X-Y and X-V readouts are shown in figs.~\ref{fig:2D_readout1} and \ref{fig:2D_readout2}.
 The investigation of possible problems correlated with this specific
 arrangement (e.g. charge sharing, grounding, cross-talk) required a
 dedicated test. Moreover, the effects of the KLOE magnetic field on the cluster
 formation and electronics readout have
 to be studied. During the previous KLOE data taking, the magnetic field was set at 0.52
 T. For the future, in order to improve the acceptance for low momentum tracks,
 the option to run at
 a reduced magnetic field value, e.g. 0.3 T, is also under consideration.

\subsubsection{Operation in magnetic field}
\label{sec:mfield}
The presence of a magnetic field affects the drift motion of electrons produced by ionization in the gas, thus influencing the position reconstruction of primary particles.
 All the charged particles involved in the ionization process experience the Lorentz
 force and consequently their drift path increases.
 Since the Lorentz force depends on the velocity, the effect is larger for electrons rather than for ions.
The Lorentz angle is given by
$\tan\alpha_L=|v\left(\vec{E}\right)|\cdot|\vec{B}|/|\vec{E}|$; its
knowledge for each GEM gap is used to evaluate the
systematic shift of the collection area of the electrons from the track of
the incoming particle.
 Because of the non-linear dependence of drift velocity from electric field
 in many gas mixtures, a comparison between simulation results and direct
 measurements is needed.
The simulation has to take into account all the features of the gas
mixture, such as the diffusion and Townsend coefficients.
\begin{figure*}[htb]
\begin{minipage}[b]{6.5cm}
\centering
\includegraphics[scale=0.30]{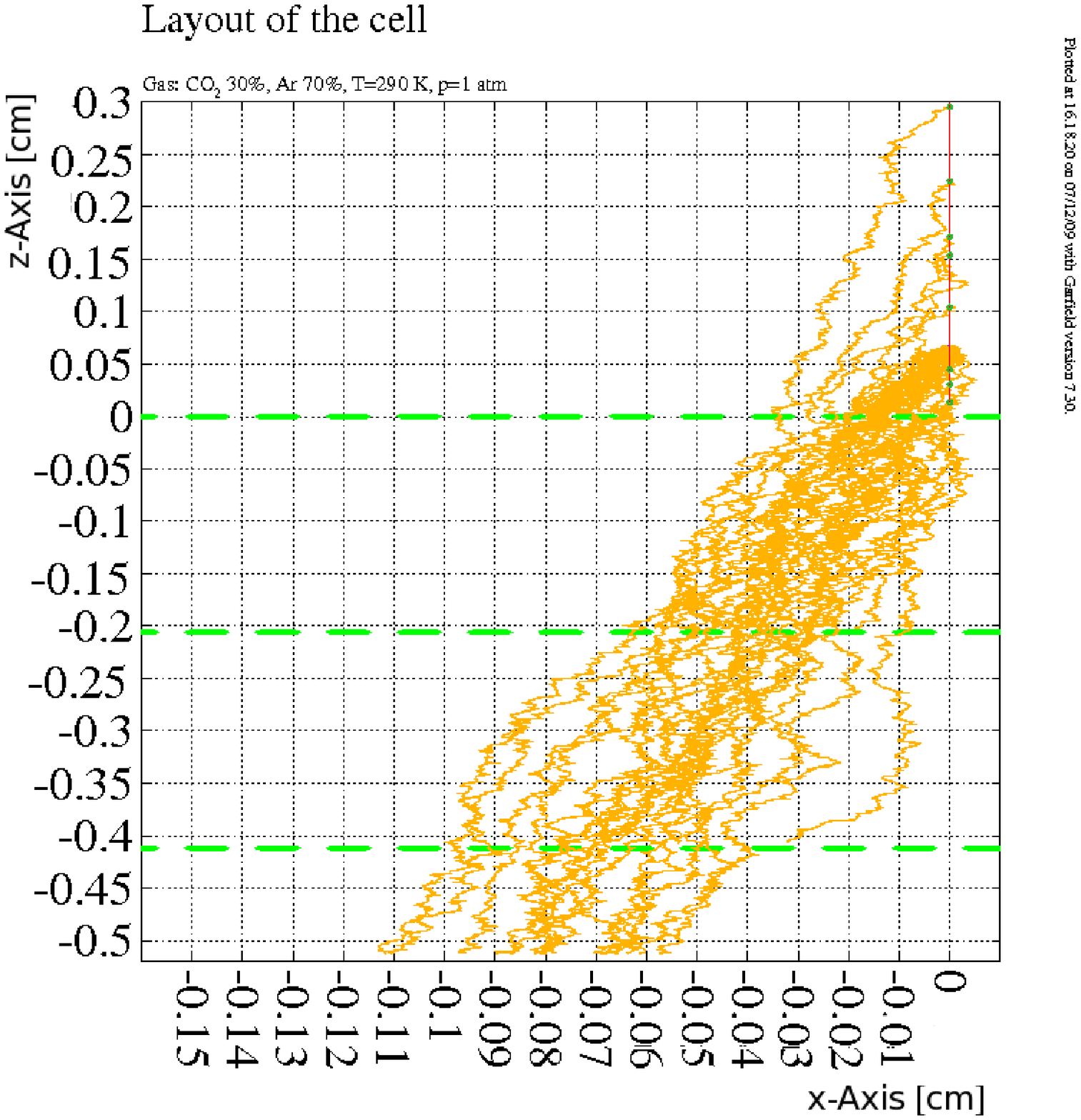}
\end{minipage}
\hspace{0.1cm}
\begin{minipage}[b]{6.5cm}
\centering
\includegraphics[scale=0.33]{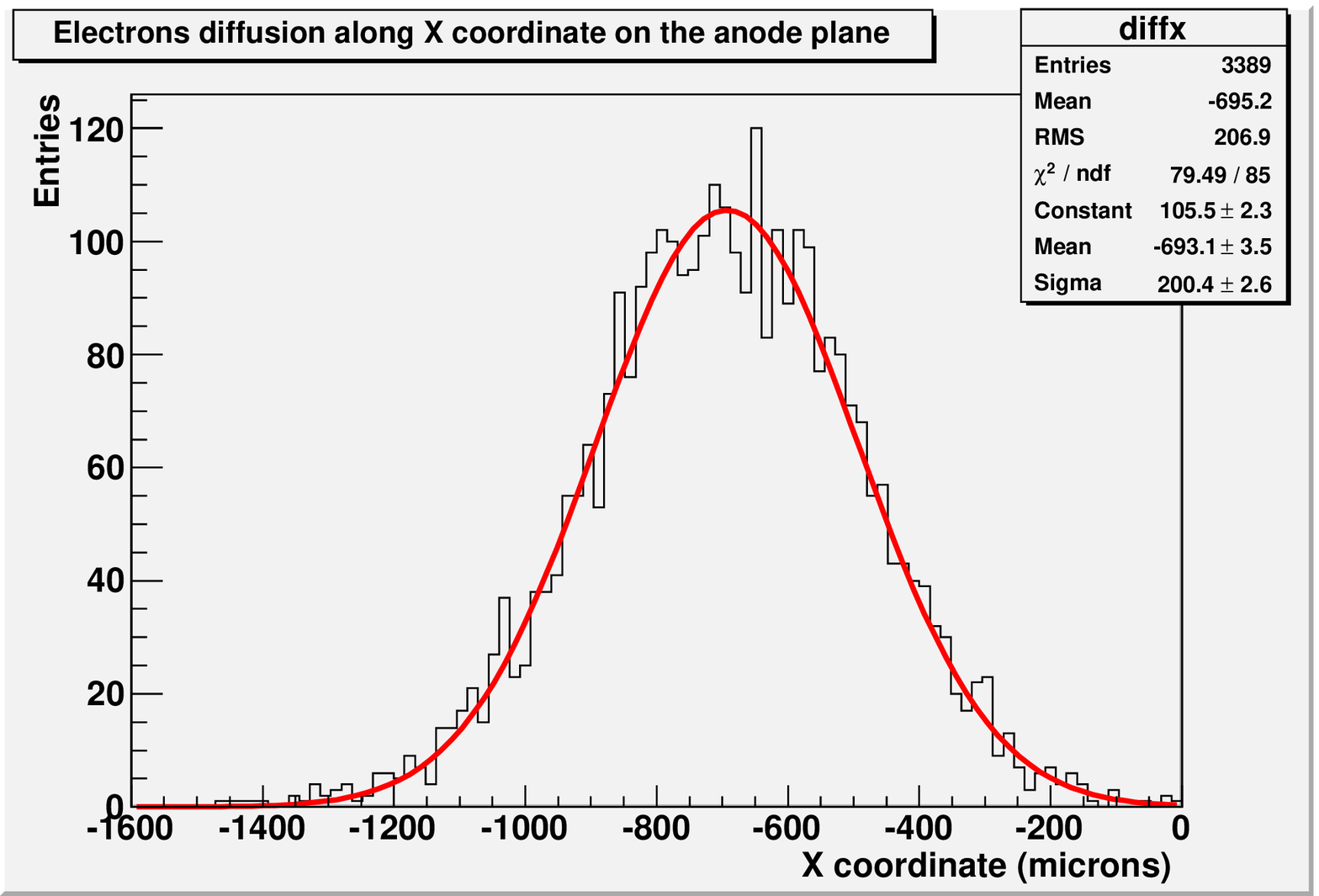}
\end{minipage}
\caption{The magnetic field effect in ArCO$_2$ $70:30$ on a single track simulated with GARFIELD(left). Position distribution of the collected charge on the X coordinate with the Lorentz shift simulated with GARFIELD (right).}
\label{arco2}
\end{figure*}
\par
We use a finite elements method in ANSYS program to simulate the chamber, set all the desired voltages and choose a gas mixture.
\begin{figure*}[htb]
\begin{minipage}[b]{6.5cm}
\centering
\includegraphics[scale=0.32]{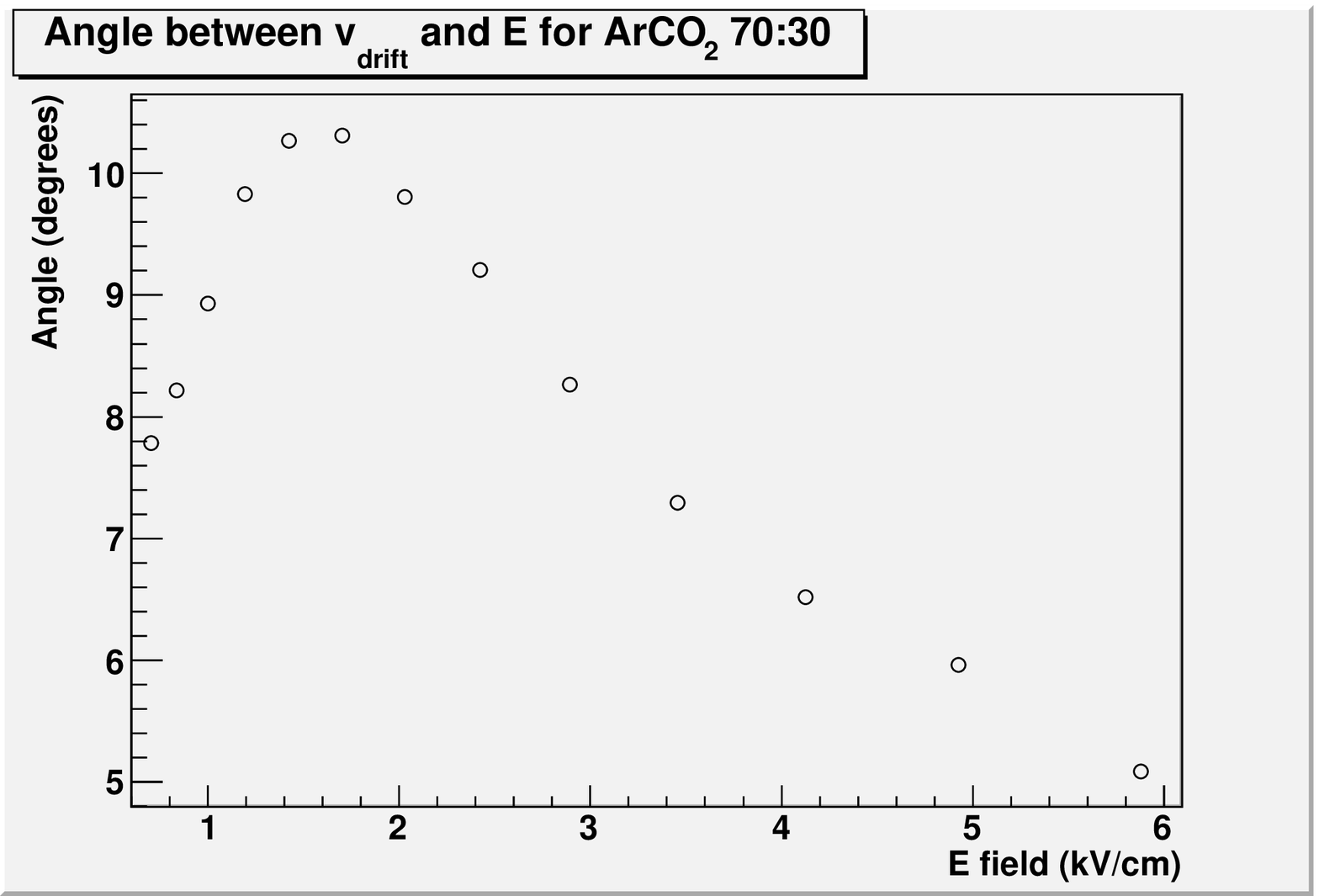}
\end{minipage}
\hspace{0.1cm}
\begin{minipage}[b]{6.5cm}
\centering
\includegraphics[scale=0.32]{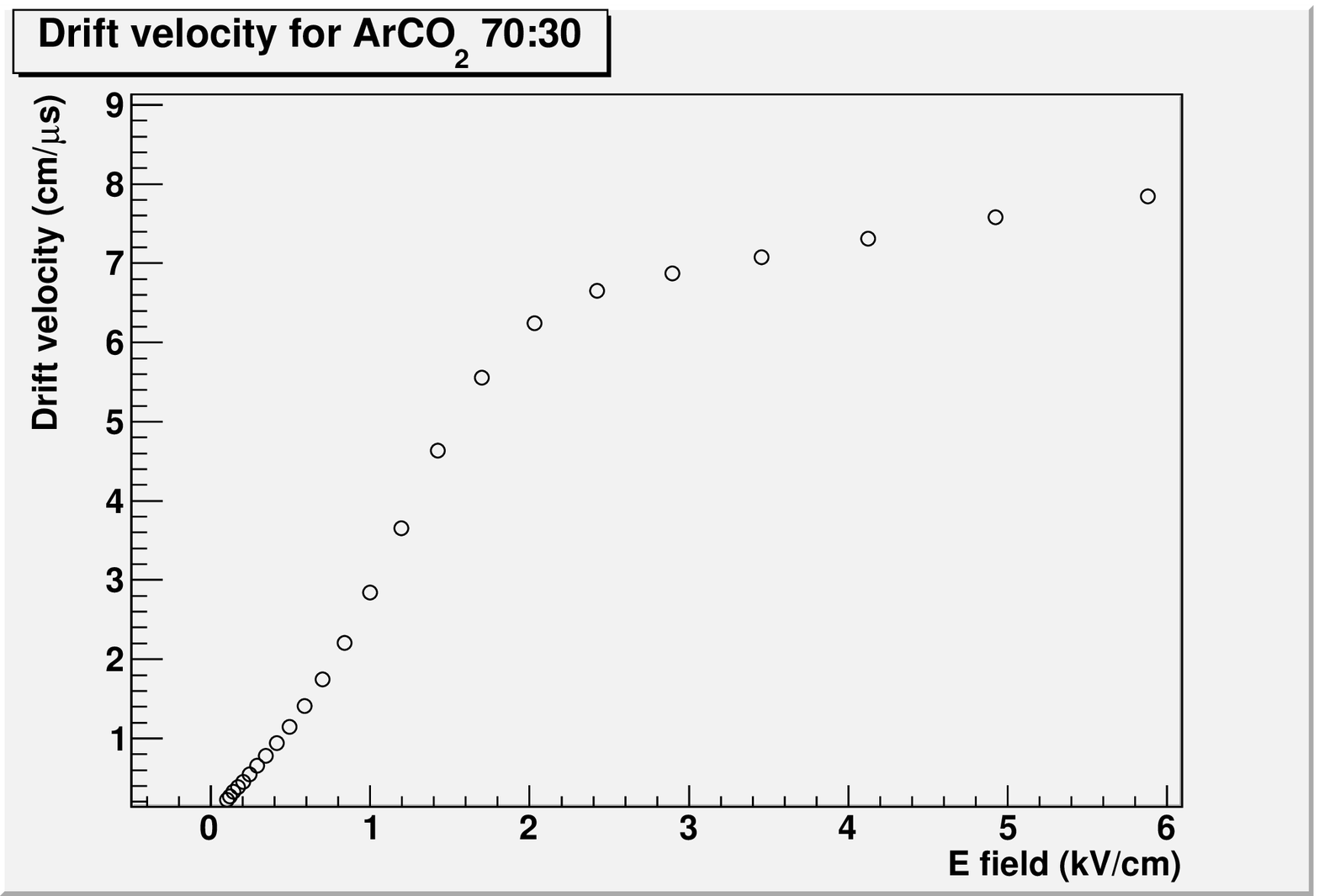}
\end{minipage}
\caption{Lorentz angle $\alpha_L$ (left) and drift velocity (right) for the ArCO$_2$ $70:30$ mixture as a function of the electric field.}
\label{garfield1}
\end{figure*}
\begin{figure*}[htb]
\begin{minipage}{6.5cm}
\centering
\includegraphics[scale=0.32]{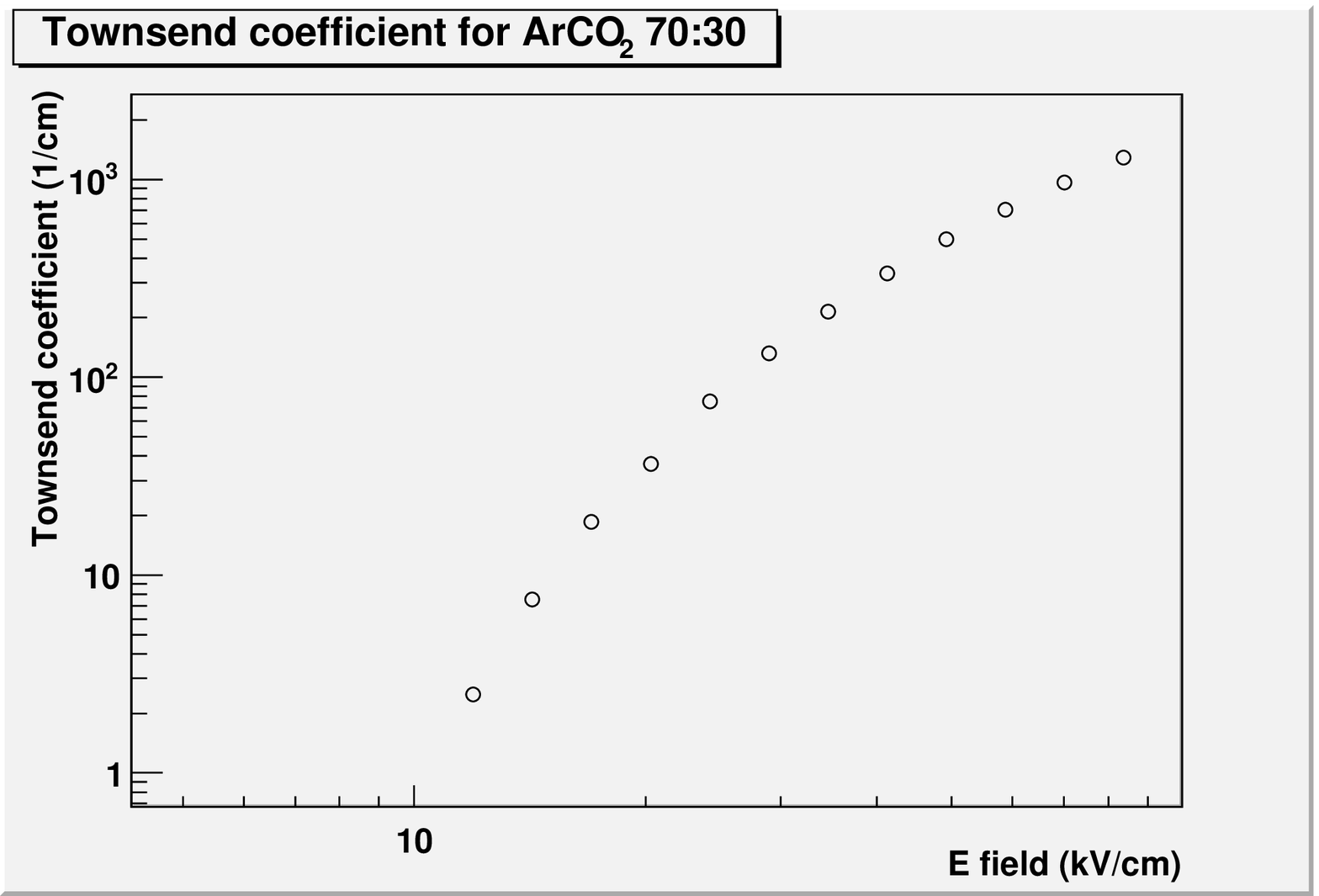}
\end{minipage}
\hspace{0.1cm}
\begin{minipage}{6.5cm}
\centering
\includegraphics[scale=0.32]{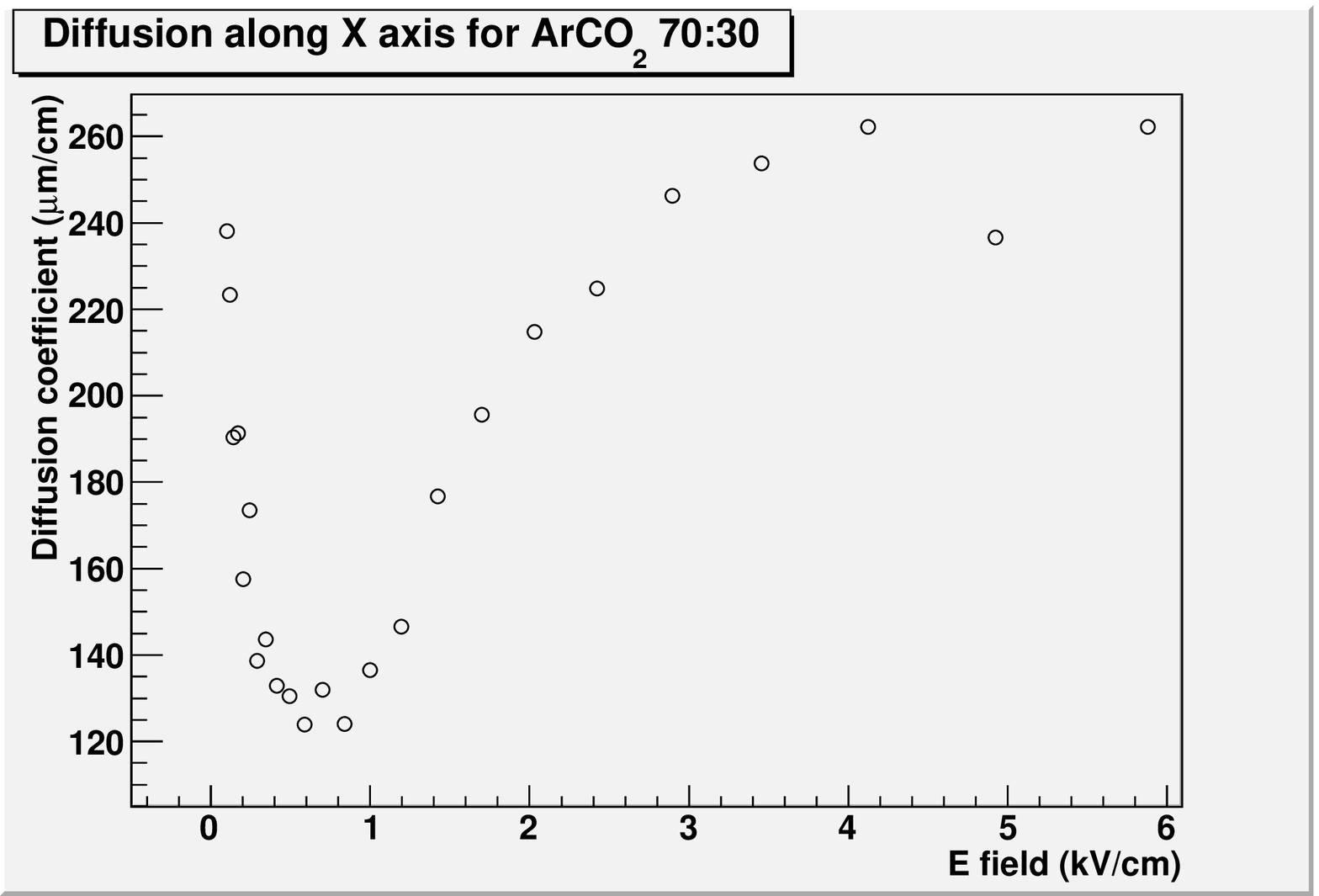}
\end{minipage}
\caption{First Townsend coefficient (left) and the diffusion coefficient along the electric field for the ArCO$_2$ $70:30$ mixture as a function of the electric field.}
\label{garfield2}
\end{figure*}
\begin{figure*}[!htb]
\begin{minipage}{6.5cm}
\centering
\includegraphics[scale=0.32]{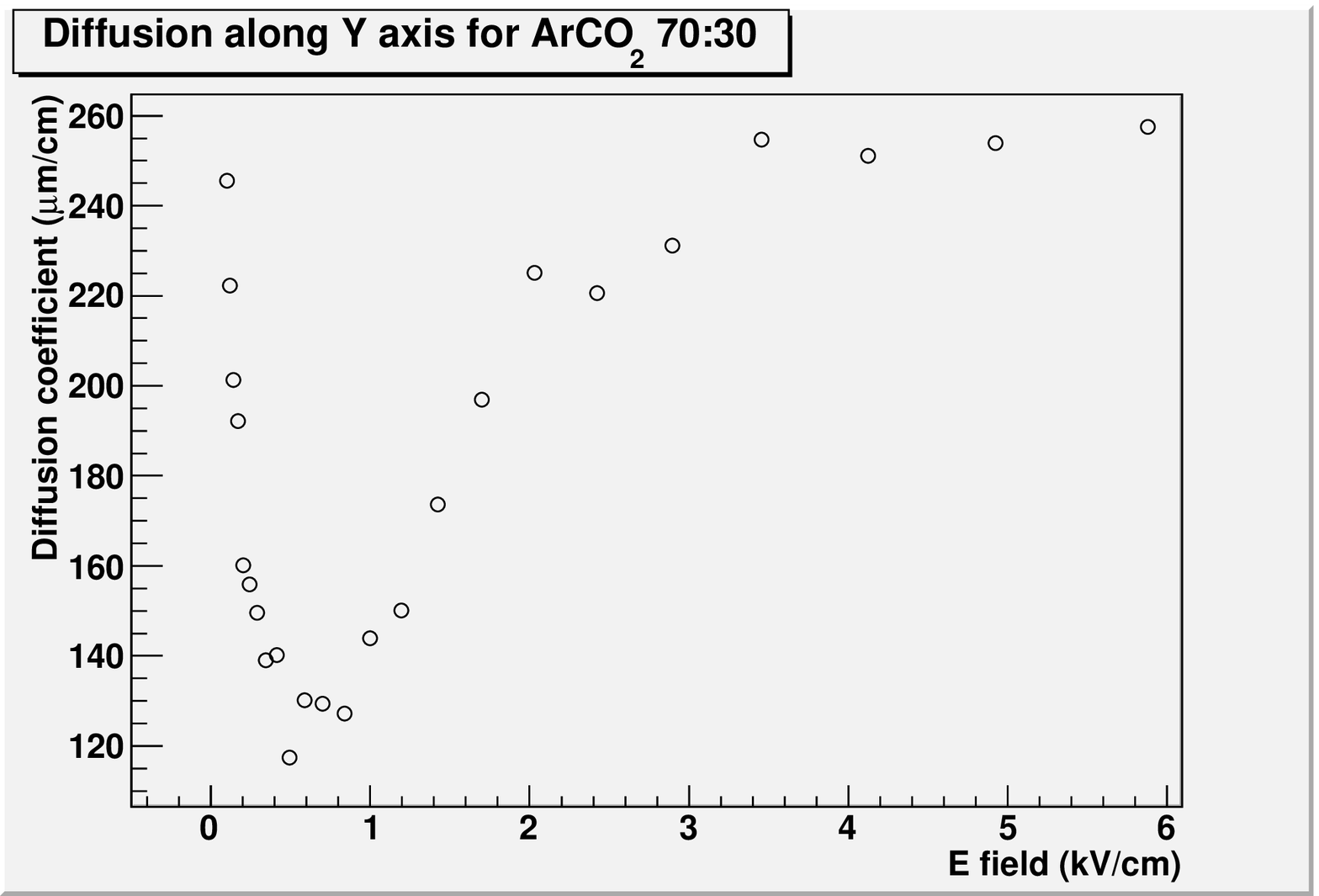}
\end{minipage}
\hspace{0.1cm}
\begin{minipage}{6.5cm}
\centering
\includegraphics[scale=0.32]{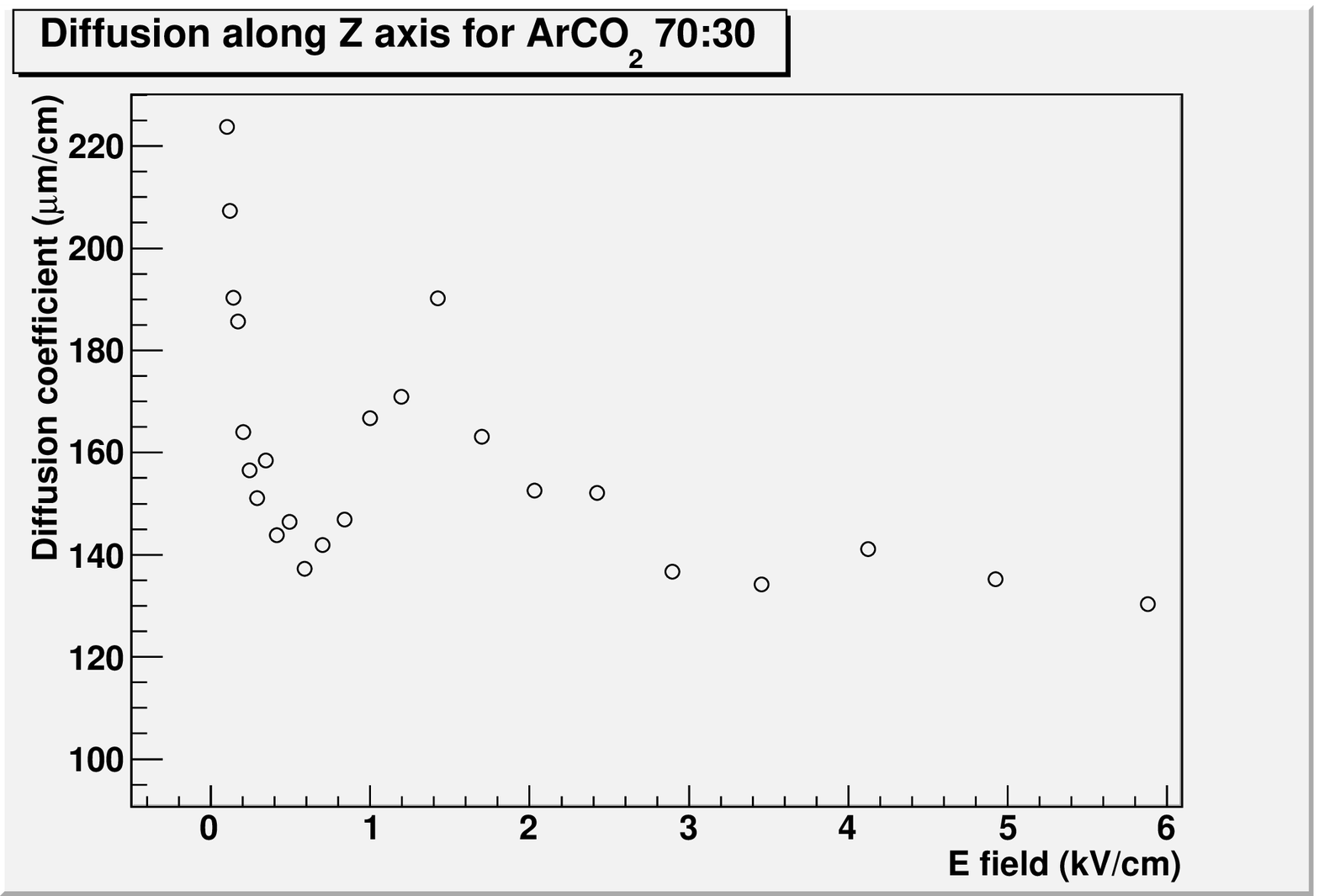}
\end{minipage}
\caption{Diffusion coefficient orthogonal (left) and parallel (right) to the magnetic field for the ArCO$_2$ $70:30$ mixture as a function of the electric field.}
\label{garfield3}
\end{figure*}
The GARFIELD program simulates the electron drift through the chamber
in presence of magnetic field, according to the transport
properties of the gas mixture, as a function of the electric field computed by
MAGBOLTZ. Fig.~\ref{arco2}-left shows the deflection of electrons produced by a minimum ionizing particle in the
different GEM gaps (three dashed lines), obtained with a 0.5 T magnetic field
orthogonal to the electric fields.
The displacement of the collected charge with respect to the entry position
 of the track can be evaluated using this
 simulation.
For $B$=0.5 T and for the standard working point of the chamber,
we obtained an offset $|\Delta x|$ = 0.693 mm, as shown in fig.~\ref{arco2}-right.

To study the properties of the gas mixture in the magnetic field we performed detailed calculation of its transport properties in the electric field range
 $10^2\div10^5~V/cm$ and the most relevant parameters are shown in figs.~\ref{garfield1}, \ref{garfield2},
 \ref{garfield3}.
%%%%%%%%%%%%%%%%%%%%%%%%%%%%%%%%%%%%%%%%%%%%%%%%%%%%%%%%%%%%%%%%
In the following section the results of the simulation have been compared with our measurements obtained with planar GEMs in a  magnetic field.

%%%%%%%%%%%%%%%%%%%%%%%%%%%%%%%%%%%%%%%%%%%%%%%%%%%%%%%%%%%%%%%
\subsubsection{Planar GEM test beam at CERN}
\label{sec:PGEM_testbeam}
%%%%%%%%%%%%%%%%%%%%%%%%%%%%%%%%%%%%%%%%%%%%%%%%%%%%%%%%%%%%%%%
%To address these issues of the XV readout and operation in magnetic field,
A dedicated test was done at SPS-H4 North Area beam line at CERN~\cite{PGEM-orlando09}.
The H4 area is equipped with the GOLIATH magnet providing a field
adjustable from 0 to 1.5 T perpendicular  to the horizontal beam-plane
(X-Z).  We used 150 GeV/c $\pi^+$ beam.

\begin{figure}[!h]
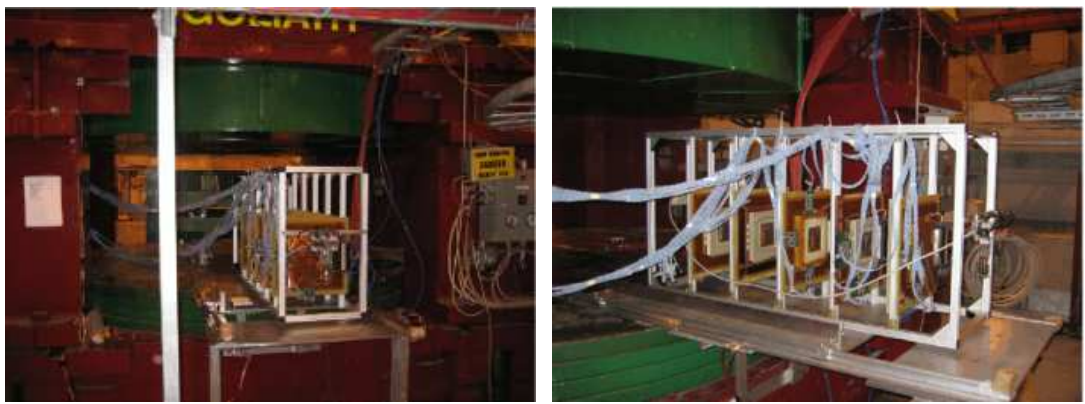

\centering
\figb PGEMpics/setup_xv_testbeam2;7.;\kern0.2cm \figb PGEMpics/setup_xv_testbeam;7.;\\
\caption{Setup of the test beam at CERN with planar chambers with 2D-read-out.}
\label{fig:setup_xv_testbeam}
\end{figure}

The X-V readout (fig.~\ref{fig:2D_readout2} right) was tested
with a tracking telescope realized with five 10$\times$10 cm$^2$ planar triple-GEMs
detectors with 650 $\mu$m pitch: four chambers were equipped with
standard X-Y readout and the fifth with the X-V readout under
investigation. In the following the X-V readout chamber will be
referred to as the XV chamber while the X-Y readout chamber as the XY chambers.
The five detectors were equally spaced between each other,
 with the XV chamber placed at the center (see fig.~ \ref{fig:setup_xv_testbeam}). The
entire setup was 1 meter long.

The GEMs were partially equipped with 22 digital readout GASTONE boards,
32 channels each, four on each XY chamber and six on the XV chamber.
This was enough to fully cover the area illuminated by the SPS beam.
The coincidence of 6 scintillators (3x3  cm$^2$) read-out by silicon
multipliers provided the trigger signal for the acquisition.

We have used the same working point as for the CGEM prototype (sec.\ref{sec:CGEM_testbeam}):
Ar/CO2 ( 70/30) gas mixture and operating voltages V$_{fields}$ = 1.5/2.5/2.5/4 kV/cm and
 V$_{GEM}$ = 390/380/370 V ($\sum$V$_G$ = 1140V). The GASTONE threshold was
 set at 3.5 fC.

\begin{figure}[!h]
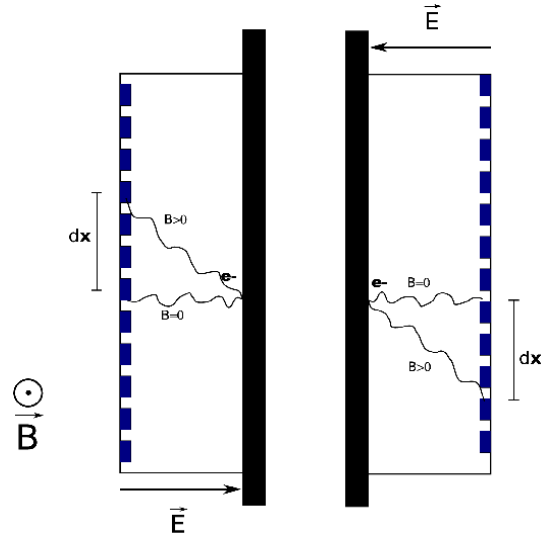

\centering
\figb PGEMpics/lorentz_angle;7.;\\
\caption{Test beam setup and definition of the measured quantity.}
\label{fig:lorentz1}
\end{figure}

In the presence of a magnetic field a displacement {\bf dx} on
the readout plane must be obserevd, due to the effect of the Lorentz force.
% (see section \ref{sec:mfield}).
The displacement is related to the direction of the electric field and, if the
 detector is rotated, the displacement also changes direction. We arranged
 the test-beam setup to be able to measure this effect (fig.~\ref{fig:lorentz1}).

\begin{figure}[!h]
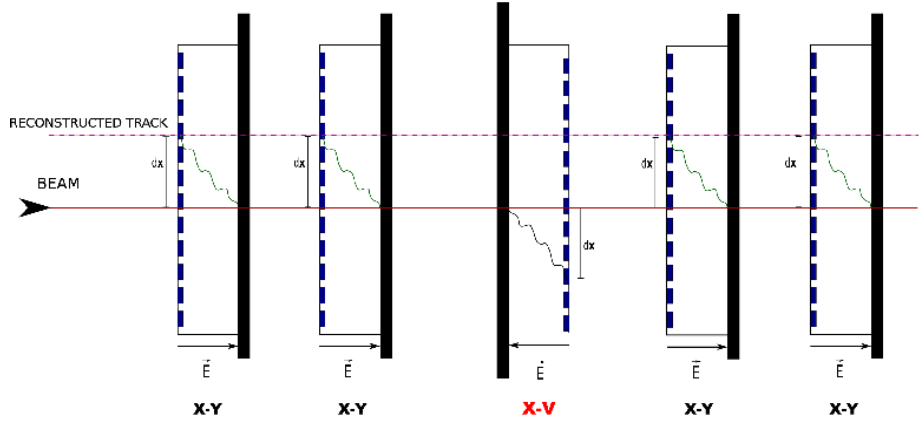

\centering
\figb PGEMpics/setup_lorentz_angle;12.;\\
\caption{Test beam setup and definition of the measured quantity.}
\label{fig:lorentz2}
\end{figure}

All XY chambers in the tracking telescope were placed with the
same anode-cathode configuration while we reversed the XV chamber to
cathode-anode arrangement. We aligned the setup with zero magnetic field
(B=0T) to a few micrometer precision. In the presence of a magnetic field we
reconstructed the track using only the four XY chambers. Since all four of them are
subject to the same Lorentz force in the same direction, the reconstructed
track will be shifted by the same offset  {\bf dx} with respect to the true
track trajectory.
The XV chamber with the reversed cathode-anode arrangement is
affected by the Lorentz force; since the electric field has opposite
direction, the displacement inside this chamber is of the same
magnitude ({\bf dx}) but with opposite direction. Hence the measurement of
the total displacement between the track reconstructed by the XY telescope and the
value measured in the XV chamber is D = 2$\cdot$ {\bf dx}. (See
fig.~\ref{fig:lorentz2}).

\begin{figure}[!h]
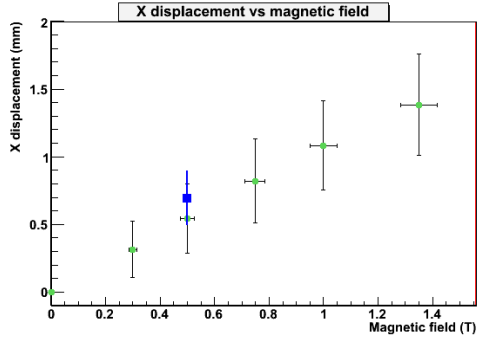

\centering
\figb PGEMpics/lorentz_angle_bfield;7.;\\
%\figb PGEMpics/ ;7.;\kern0.2cm \figb PGEMpics/ ;5.;
\caption{Displacement dx as a function of the magnetic field. The blue point comes from simulation results with GARFIELD presented in section \ref{sec:mfield}.}
\label{fig:lorentz_bfield}
\end{figure}

The displacement  {\bf dx} was measured for 5 values of the magnetic field
and found in good agreement with the value obtained from the GARFIELD simulation at B=0.5 T (sec.\ref{sec:mfield}).
See fig.~\ref{fig:lorentz_bfield}. Such effect will be properly taken into account in the reconstruction procedure.

\begin{figure}[!h]
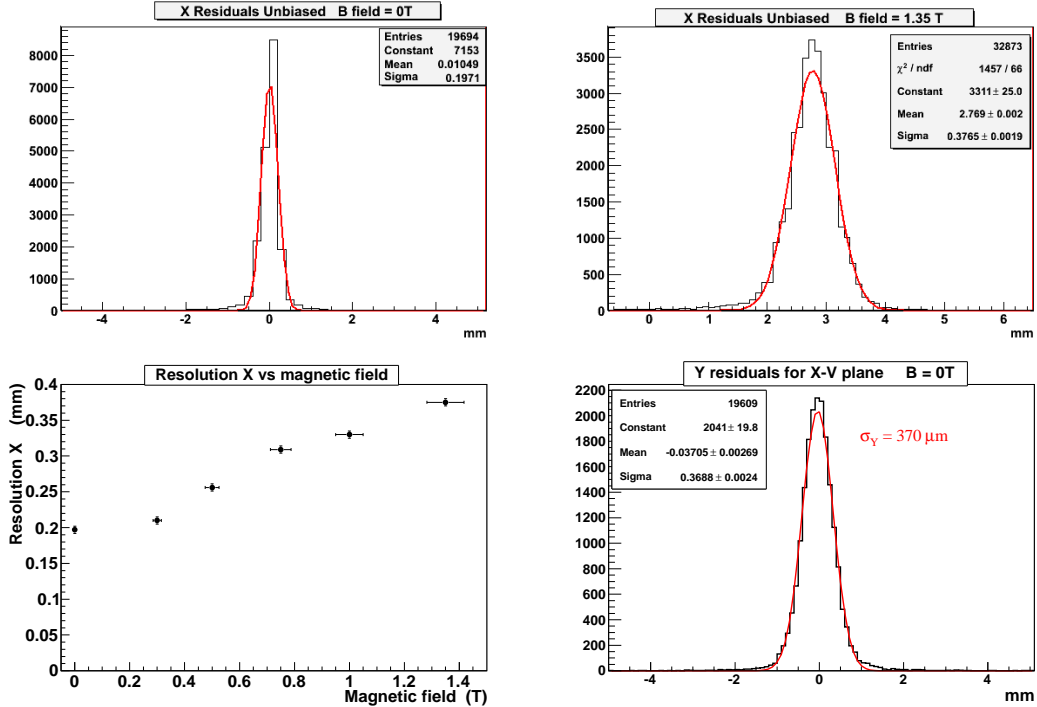

\centering
\figb PGEMpics/Xresolution_bfieldzero;7.;\kern0.2cm \figb PGEMpics/Xresolution_bfield1e35tesla;7.;\\
\figb PGEMpics/xres2;7.;\kern0.2cm \figb PGEMpics/vres2;7.;\\
\caption{(Top:) Residuals in the X view on the plane with X-V readout,
  without magnetic field (left) and with 1.35T field (right). (Bottom:)
  Resolution on the X coordinate as a function of the magnetic field
  (left). Resolution on the Y coordinate with indicated the corresponding value on V coordinate (right).}
\label{fig:resolu}
\end{figure}

To study the XV chamber performance, we measured the position resolution
defined as the sigma value from the gaussian fit to the residual plot of
the XV chamber.
The resolution obtained as a function of the magnetic field is shown
in fig.~\ref{fig:resolu} together with examples of residual plots at B=0 T and
B=1.35 T.
There is a clear effect on the resolution with increased
magnetic fields, due to the Lorentz force, which on our test-beam setup
affects the X coordinate only.

The resolution on the X coordinate ranges from 200 $\mu$m at B=0
T up to 380 $\mu$m at B=1.35 T. The Y coordinate is obtained from the crossing of both X and V strip readout as $Y=tan(50^\circ)\times X + V/cos(50^\circ)$ and shows a $\sim$370 $\mu$m resolution, in agreement with what expected from the digital readout of the two views.

\begin{figure}[!h]
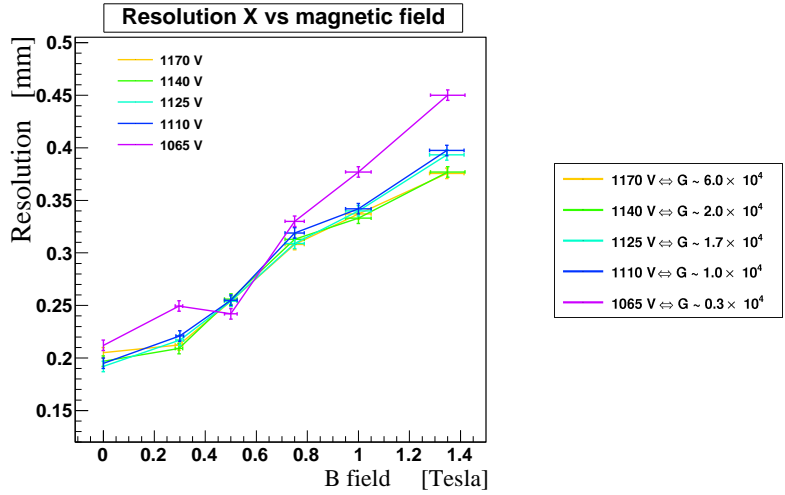

\centering
\figb PGEMpics/res_bfield_gain;7.;\kern0.2cm \figb PGEMpics/conversion_table_best;3.;\\
\caption{Measured resolution in function of the GEM voltage (gain) and magnetic field.}
\label{fig:pgemresgain}
\end{figure}

We have also measured the position resolution as a function of the GEM voltage (gain).
(See fig.~\ref{fig:pgemresgain})
It is stable in  a rather broad range fof the voltage settings around to the nominal value;
an appreciable decrease can actually be observed only for values much below the nominal,
 corresponding to gains G$\sim$0.3$\times10^{4} $.

\par
\begin{figure}[!h]
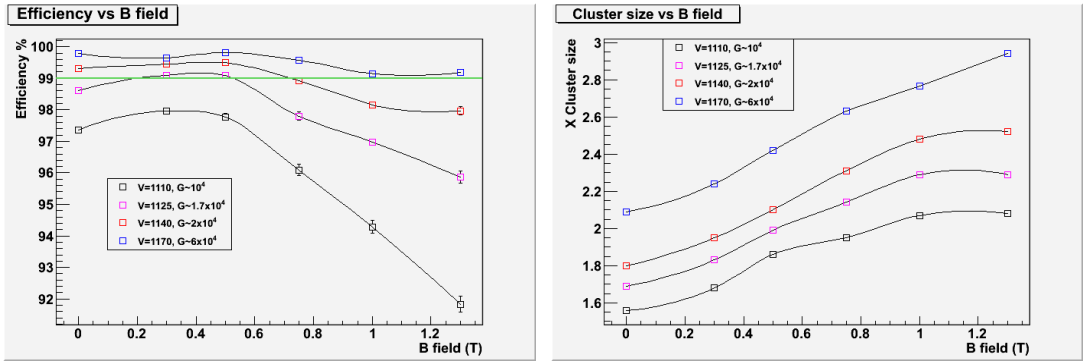

\centering
\figb PGEMpics/efficiency_vs_Bfield_Gains;7.;\kern0.2cm \figb PGEMpics/ClusterSize_vs_Bfield_Gains;7.;\\
\caption{ GEM efficiency {\bf left})  and cluster size {\bf right})  in function of the magnetic field given for four different gain values. }
\label{fig:pgemeff}
\end{figure}

The performance of the front-end chip GASTONE have been studied measuring
the cluster size and the reconstruction efficiency, defined as the presence
of a cluster in the XV chamber when a candidate track was reconstructed using four XY
chambers. GEM efficiency and cluster size were measured as a function of the magnetic
field and of the GEM voltage (fig.~\ref{fig:pgemeff}).
The efficiency for the nominal KLOE B field value and voltage settings  was measured to
exceed 99$\%$, slightly decreasing at higher B fields.
%The measurment proves high efficiency of the GEM (above 99\%) for the nominal magnetic field and the volta%ge values and slow decreese with higher B field.

\begin{figure}[!h]
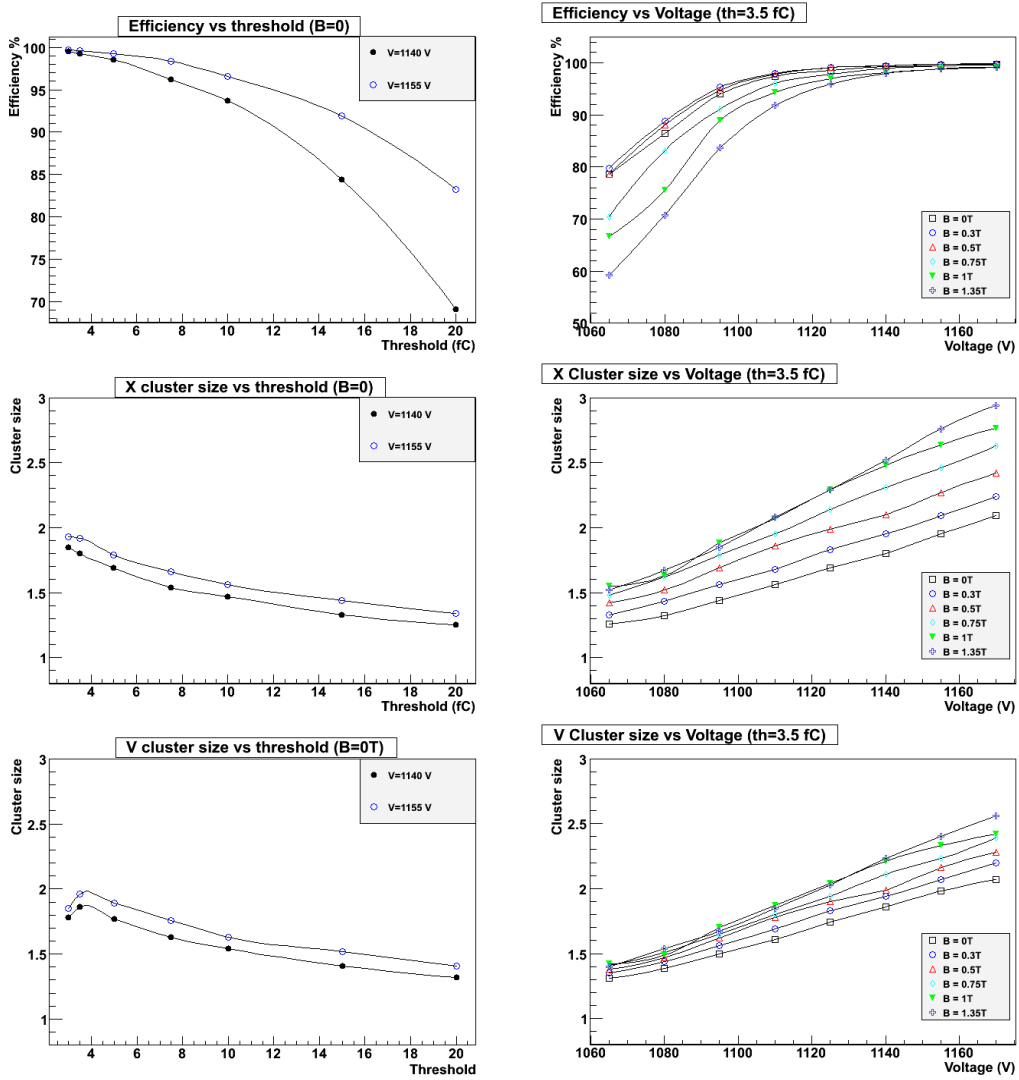

\centering
\figb PGEMpics/efficiency_thresholdscan;7.;\kern0.1cm\figb PGEMpics/efficiency_bfield-voltagescan;7.;\\
\figb PGEMpics/Xclustersize_thresholdscan;7.;\kern0.1cm \figb PGEMpics/Xclustersize_bfield-voltagescan;7.;\\
\figb PGEMpics/Vclustersize_thresholdscan;7.;\kern0.1cm \figb PGEMpics/Vclustersize_bfield-voltagescan;7.;\\
\caption{({\bf Left}) Threshold scan: efficiency (Top) X cluster size (Middle) V
  cluster size (Bottom); ({\bf Right}) Voltage and magnetic field scan:
  efficiency (Top) X cluster size (Middle) V  cluster size (Bottom).}
\label{fig:Scans}
\end{figure}

The size of the clusters formed in both X and V views were studied
as a function of the value of the GASTONE threshold, of the GEM voltage
and of the magnetic field (fig.~\ref{fig:Scans}).

In principle the charge sharing, grounding, cross-talk could be different
for X and V views due to the different readout geometry. However, our measurements
showed essentially identical response  of both X and V readout views.
 The variation of the magnetic fields within the
KLOE-2 planned values (0.3-0.5 T) has a negligible effect on the
reconstruction efficiency in the voltage range around our working point
($V_{ref}$= 1140 V).

% \begin{figure}[!h]
% \centering
% \figb PGEMpics/efficiency_thresholdscan;5.;\kern0.1cm \figb PGEMpics/Xclustersize_thresholdscan;5.;\kern0.1cm \figb PGEMpics/Vclustersize_thresholdscan;5.;\\
% \caption{Threshold scan: (Left) efficiency (Center) X cluster size (Right)
%   V cluster size.}
% \label{fig:ThresholdScan}
% \end{figure}

% \begin{figure}[!h]
% \centering
% \figb PGEMpics/efficiency_bfield-voltagescan;5.;\kern0.1cm \figb PGEMpics/Xclustersize_bfield-voltagescan;5.;\kern0.1cm \figb PGEMpics/Vclustersize_bfield-voltagescan;5.;\\
% \caption{Voltage and magnetic field scan: (Left) efficiency (Center) X cluster size (Right)
%   V cluster size.}
% \label{fig:VandBScan}
% \end{figure}
%%%%%%%%%%%%%%%%%%%%%%%%%%%%%%%%%%%%%%%%
\subsection{Large area GEM}
\label{sec:large_area_gem}
The GEM technology was born at CERN and CERN is still the main producer of
GEM foils. Presently the size of the foils manufactured at the CERN EST-DEM workshop
~\cite{bib:cern-est-dem} is limited to $\rm{450\times450~mm^2}$.
Such a limitation arises from two different reasons:
\begin{enumerate}
\item the raw material presently used for GEM foils is the Novaclad G2300
  from Sheldahl, a polyimide (kapton) foil with an adhesiveless copper
  cladding on both sides. This material is normally provided in 457~mm wide rolls.
    Considering the space needed for the handling of a foil, 450~mm is a
    limit for one dimension. Rolls as large as 514~mm are available from the same company, but are non-standard and
    should be ordered in bulk. Anyhow this option could be considered when large production of detectors is foreseen.
\item the manufacturing of GEM foils is based on the photolithographic
  process commonly employed in the printed circuit industry: the hole
  pattern is transferred by UV-exposure from transparent masks to
  photoresistive layers deposited on both sides of the raw material. The
  foil is then etched in an acid bath, which removes copper from the holes
  left in the photoresist. Since the hole has a diameter of 70~$\mu$m, the
  two masks have to be aligned with a precision of few $\mu$m, not a
  trivial goal to meet as the area increases. With the present plastic
  masks, 450~mm is considered a limit to preserve the homogeneity and the
  quality of the hole geometry.
\end{enumerate}

\subsubsection{Single-mask procedure}
\begin{figure}[hbt]
\centering
\includegraphics[width=0.4\linewidth]{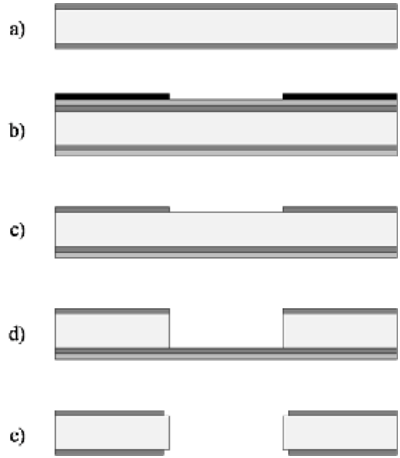}
\caption{Schematic procedure for fabrication of a single-mask GEM. See the text for detailed description.}
\label{fig:singleMask}
\end{figure}

In order to overcome the size limitation due to the troublesome alignment
of the two masks, and thus fulfilling the demand of large area foils from a
consistent part of the GEM community, a single-mask procedure has been
developed by the CERN EST-DEM workshop together
with the Gas Detector Development (GDD) group of CERN \cite{bib:GDD}. It is schematically highlighted in fig.~\ref{fig:singleMask}.
\begin{enumerate}[label=\alph*)]
  \item the raw material is a polyimide foil (kapton) 50~$\mu$m thick, with a double-sided 5~$\mu$m copper cladding;
  \item a photoresist layer is laminated on both surfaces; a lithographic mask with the pattern of the holes is placed on the top face of the foil and exposed to UV light;
  \item the top metal is etched while the bottom metal is protected where the photoresist layer is still present;
  \item the polyimide is etched all through from top;
  \item using the polyimide as a mask, the hole is opened by etching the bottom metal; the top metal is actively preserved from etching as it is biased to a more negative voltage with respect to the chemical bath, working as the cathode of the electrochemical cell and thus being protected \footnote{This technique, called ``Cathodic protection'', is commonly used for example to protect oil pipes from corrosion}.
\end{enumerate}

At the end of this process the shape of the hole is still slightly conical with a top diameter of 70~$\mu$m and a bottom diameter of 60~$\mu$m. This has to be compared with the usual double-conical shaped hole obtained with the double-mask procedure, with diameters of 70-50-70~$\mu$m, respectively for top, middle and bottom parts of the hole.

\subsubsection{Measurements}
To characterize the new foils, two identical $\rm{10\times10~cm^2}$
Single-GEM chambers have been assembled: one with single-mask foil and
one with the standard foil, used as a reference.
The two detectors have been put on the same gas line and flushed with $\rm{Ar:CO_2 = 70:30}$. They have been simultaneously irradiated with a 6~keV X-rays gun and tested in current mode.

At first the single-mask foil has been mounted in a bottom-open
configuration, i.e. with the larger section of the holes facing the
anode. Then the foil has been turned , in a top-open configuration,
i.e. with the larger section of the holes facing the cathode. In the
following these two
configurations will be referred to as 60-70 and 70-60,
respectively.
\begin{figure}[hbt]
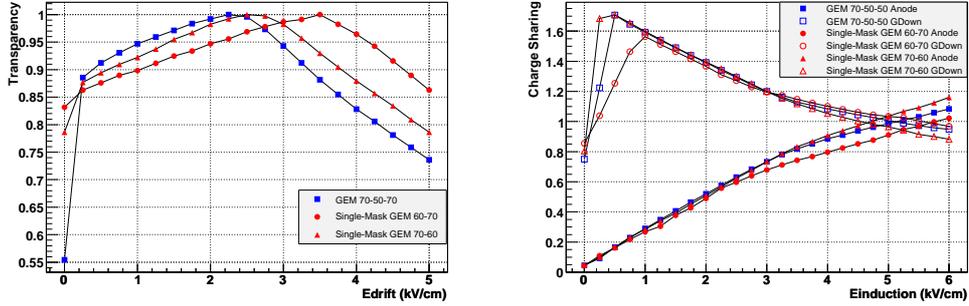

\centering
\figb CGEMpics/driftall;6.6; \kern0.1cm \figb CGEMpics/inductionall;6.6;
\caption{Left: Electron transparency as a function of the Drift
  field. Right: Charge sharing between bottom surface of the GEM and anode
  as a function of the Induction field. The equal sharing values are 4.6~kV/cm,
  5.2~kV/cm and 5.7~kV/cm respectively for 70-60, 70-50-70 and 60-70 hole
  configurations.}
\label{fig:drift}
\end{figure}
%
% \begin{figure}[hbt]
% \centering
% \includegraphics[width=0.5\linewidth]{CGEMpics/inductionall}
% \caption{Charge sharing between bottom surface of the GEM and anode as a
%   function of the Induction field. The equal sharing values are 4.6~kV/cm,
%   5.2~kV/cm and 5.7~kV/cm respectively for 70-60, 70-50-70 and 60-70 hole
%   configurations.}
% \label{fig:induction}
% \end{figure}
% %
\begin{figure}[hbt]
\centering
\includegraphics[width=0.5\linewidth]{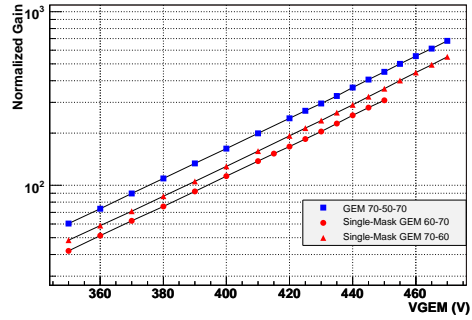}
\caption{Gas gain in $\rm{Ar:CO_2 = 70:30}$.}
\label{fig:gain}
\end{figure}

Fig.~\ref{fig:drift}-left shows the electron transparency as a function of the
drift field. The single-mask foil exhibits a different behavior depending
on the orientation: in the 60-70 configuration the full focusing efficiency
is reached for a higher value of the field with respect to the other two configurations.

In fig.~\ref{fig:drift}-right the charge sharing between the bottom surface of
the GEM and the anode is plot as a function of the induction field. It
is common practice to choose the equal sharing point as a reasonable operating value
for the induction field. The equal sharing values are 4.6~kV/cm, 5.2~kV/cm
and 5.7~kV/cm respectively for 70-60, 70-50-70 and 60-70 hole
configurations. A clear dependency of this parameter from the GEM orientation can be seen.

In fig.~\ref{fig:gain} the gas gain of the GEM is shown as a function of the
voltage drop between the two faces of the foil. The gain of the new GEM is 20$\div$30\%
smaller with respect to the standard one, meaning that additional 10$\div$20~V must be
applied in order to operate the new chamber at the same gain of the
70-50-70 configuration.

\subsubsection{Large planar prototype}
\begin{figure}[hbt]
\centering
\includegraphics[width=0.5\linewidth]{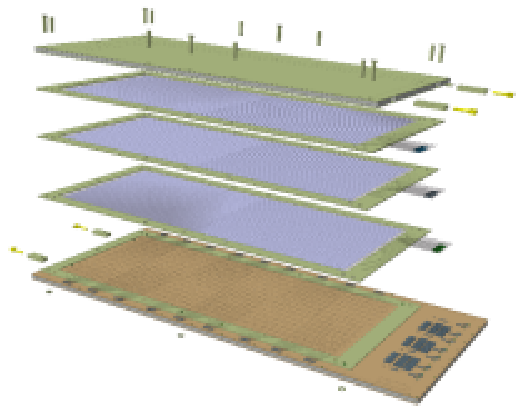}
\includegraphics[width=0.45\linewidth]{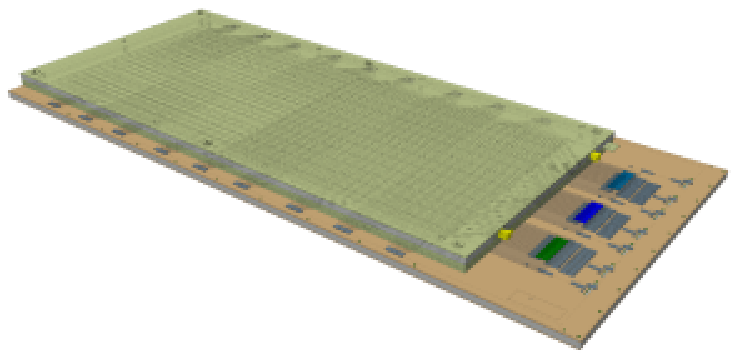}
\caption{The project for the large area planar GEM prototype.}
\label{fig:esploso}
\end{figure}

\begin{figure}[hbt]
\centering
\includegraphics[width=0.6\linewidth]{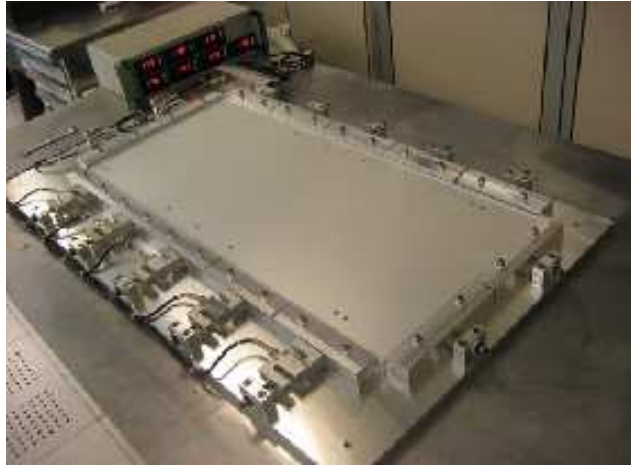}
\caption{Tensioning tool for large area GEM foils.}
\label{fig:tendigem}
\end{figure}

The CGEM prototype was limited in size since the single-mask technology for
the manufacturing of the large foils was not yet available at that time.
Now that the procedure has been finally established, we are building
a large area planar detector ($\rm{300\times700~mm^2}$) to test the
gain uniformity of a single-mask GEM over the same surface needed for the
construction of the Inner Tracker (fig.~\ref{fig:esploso}). For this purpose, a dedicated special tensioning tool has been realized (fig.~\ref{fig:tendigem}). ANSYS simulations indicate that even on such a large area, with a tension of 1~kg/cm the maximum sag due to combined gravitational and electrostatic effects is only 20~$\mu$m.
%
% Items to be covered:
% \begin{itemize}
% \item Short paragraph on the mechanical simulation
% \item Single mask (G. Bencivenni)
% \item Construction of the prototype
% \item Test and results: uniformity (1) and electric current (2) studies
%       \begin{enumerate}
%       \item {\bf will not be available for September}
%       \item {\bf available 10x10 cm$^2$ single-mask prototype only}
%       \end{enumerate}
% \end{itemize}
%%%%%%%%%%%%%%%%%%%%%%%%%%%%%%%%%%%%%%%%%%%%%%%%%%%%%%%%%
\newpage
\section{Mechanics and construction}
\begin{figure}[!h]
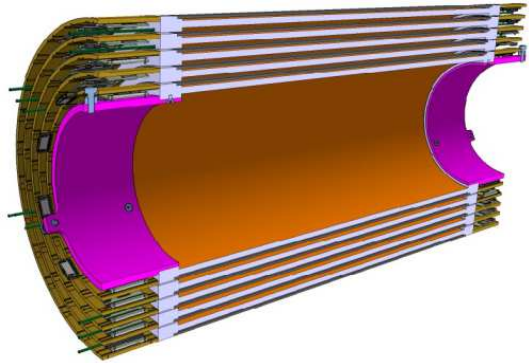

\centering
\figb MECpics/1b;7.;\kern0.2cm\\
%% \figb MECpics/cr;7.;\\
%\figb PGEMpics/ ;7.;\kern0.2cm \figb PGEMpics/ ;5.;
\caption{(Left) Inner Tracker Global view}
\label{fig:1b}
\end{figure}

%%%%%%%%%%%%%%%%%%%%%%%%%%%%%%%%%%%%%%%%%%%%%%%%%%%%%%%%%
%
The Inner Tracker  will be composed by five concentric layers of
CGEM detectors (fig.~\ref{fig:1b}), as described in section \ref{sec:detector_layout}.
% The CGEM is a triple-GEM detector composed by concentric
% cylindrical electrodes: from the cathode (the innermost electrode),
% through the three GEM foils, to the anode readout (the outermost electrode).
% The innermost CGEM layer has a radius of 127 mm, the outermost 230 mm, the
% active length for all layers is 700mm. The anode readout of each CGEM is
% segmented  with 650 $\mu$m pitch XV patterned strips with a stereo
% angle, slightly increasing with the radius, of the order of 40 degrees.
% The full system consists of about 35,000 FEE channels.
The technology used to build the cylindrical electrodes does not require
the presence of internal frames as a support and therefore
allows us to realize an intrinsically dead-zone-free detector.
Actually, since the shape of the detector is cylindrical, the typical sag of the
electrodes  is negligible: $<$ 5 $\mu$m for an overall
stretching mechanical tension of $\sim$ 50 kg applied at the ends of the
detector.\\

The detector construction will be entirely carried out in a class 1000
clean room: it is a clean procedure and protective clothings will be always worn.
The clean room will include:
\begin{itemize}
\item the storage areas for GEM and readout circuit foils and apart
  those dedicated to the storage of the fiber-glass components;
\item the GEM testing areas, for optical inspection and HV test;
\item the cleaning area for fiber-glass components;
\item the planar and cylindrical gluing stations;
\item the detector assembly area;
\item the area for the final detector test: gas leak and HV test.
\end{itemize}
The closed detector will be then moved to the laboratory where it will be tested first
with X-rays, in current mode, and then with cosmics, after the installation of
the front-end electronics.
%
%%%%%%%%%%%%%%%%%%%%%%%%%%%%%%%%%%%%%%%%%%%%%%%%%
\subsection{Detector components}
\begin{figure}[!h]
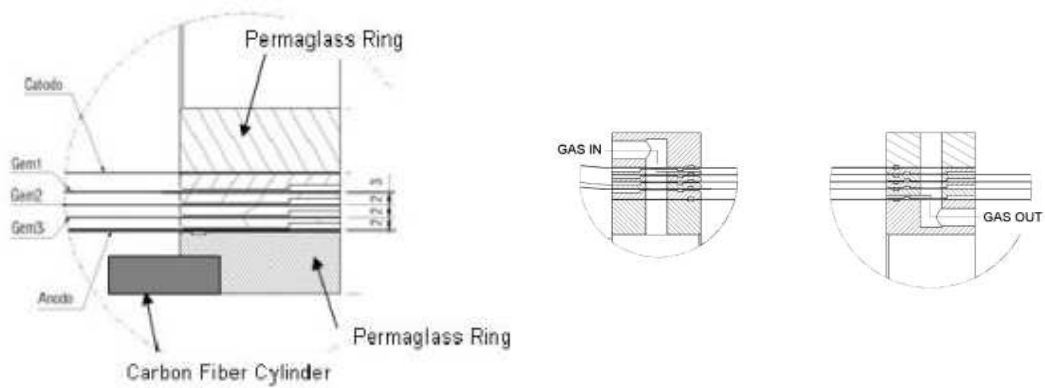

\centering
\figb MECpics/dettaglio_cgem;7.;\kern0.2cm \figb MECpics/gas_in_out;7.;\\
%\figb PGEMpics/ ;7.;\kern0.2cm \figb PGEMpics/ ;5.;
\caption{(Left) Detail of the CGEM structure. (Right) Detail of the gas inlets.}
\label{fig:dettaglio_cgem}
\end{figure}

%%%%%%%%%%%%%%%%%%%%%%%%%%%%%%%%%%%%%%%%%%%%%%%%%
%
The design of the detector components and construction toolings as well as
the choice of the materials originates both  from the experience of the construction
of the GEM detectors\cite{LHCb-GEM0} of
the LHCb Muon apparatus, and, more specifically, for the construction of the
full scale prototype of the CGEM \cite{LHCb-GEM1, LHCb-GEM2}.\\
In particular, the materials used for the CGEM were largely tested and
validated for high-rate environments \cite{LHCb-GEM3} and for
different gas mixtures (Ar, i-$C_4 H_{10}$, $CF_4$, $CO_2$ )
\cite{LHCb-GEM4}.
Such validations were performed with global large area irradiation
tests, with 1.25 MeV $\gamma$ from a 25 kCi $^{60}$Co source at the  ENEA
Casaccia, and with discharge tests performed at PSI
 with high intensity low momentum $\pi$-p particles fluxes on
reduced beam spot area ($\sim$ 15 $cm^2$).
The general design and the construction procedure for each CGEM of the
IT is substantially the same used for the realization of
the  full scale prototype.
\begin{figure}[!h]
\centering
\includegraphics[angle=90.,width=9.cm]{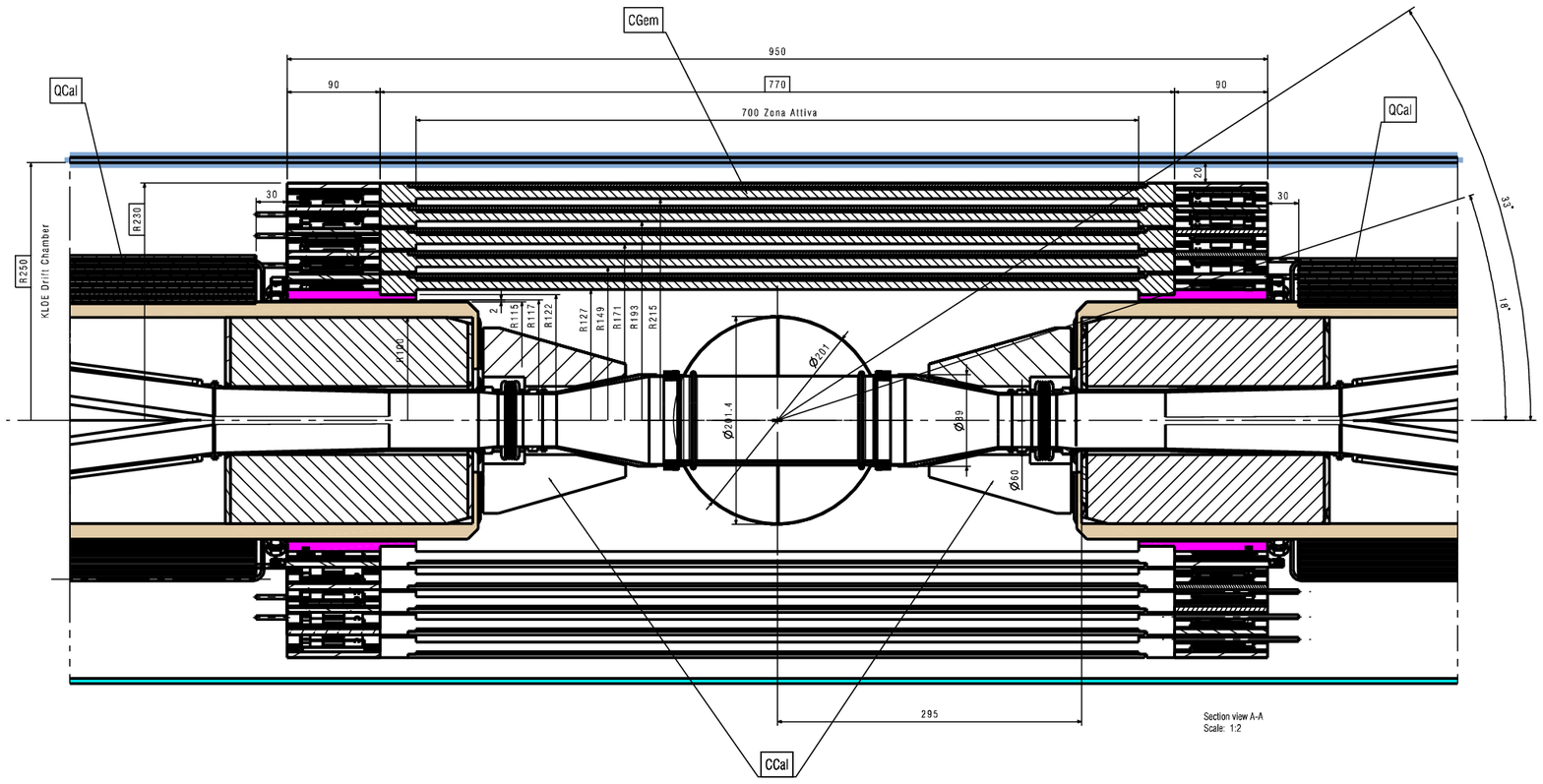}
%\figb MECpics/New-assy-3-4a;13.;\\
%\figb PGEMpics/ ;7.;\kern0.2cm \figb PGEMpics/ ;5.;
\caption{Inner Tracker Global cross section.}
\label{fig:new-assy-3-4a}
\end{figure}
In  fig.~\ref{fig:new-assy-3-4a} is reported the global cross-section of the IT.
As shown in fig.~\ref{fig:dettaglio_cgem} left the CGEM has the typical
triple-structure of such a kind of micro-pattern gas detector.
The gaps among the different electrodes of the detector (cathode-G1,G1-G2,G2-G3 and
G3-anode)  define the various regions of the detector itself:
drift (3 mm wide), transfers (both 2 mm wide) and induction (1-2mm).
The gas, supplied in open mode through the six gas inlets realized on the
Permaglass annular frame of the cathode, flows through the holes of the three GEM
and then exits from the detector by the six outlets of the anode frame
fig.~\ref{fig:dettaglio_cgem} right. Frames are realized in Permaglass, a not stratified
homogeneous composite material realized by the Resarm ltd (Belgium), that allows
very precise and almost fibers- and spikes-free machining: a crucial feature
against spurious discharges. The frames, glued at the edge of the
cylindrical electrodes outside their active area, define the various gaps
of the detector.
As successfully demonstrated by the full scale prototype, no support
frames inside the active area of the detector are required for a safe
operation of the CGEM.\\
The most relevant modification with respect to the prototype design is
represented by the {\it embedded-anode}, consisting of a very light
honeycombed carbon fiber cylinder (CFC) on which the anode readout circuit
is glued. The CFC acts as a rigid support for the whole detector layer.

\begin{figure}[!h]
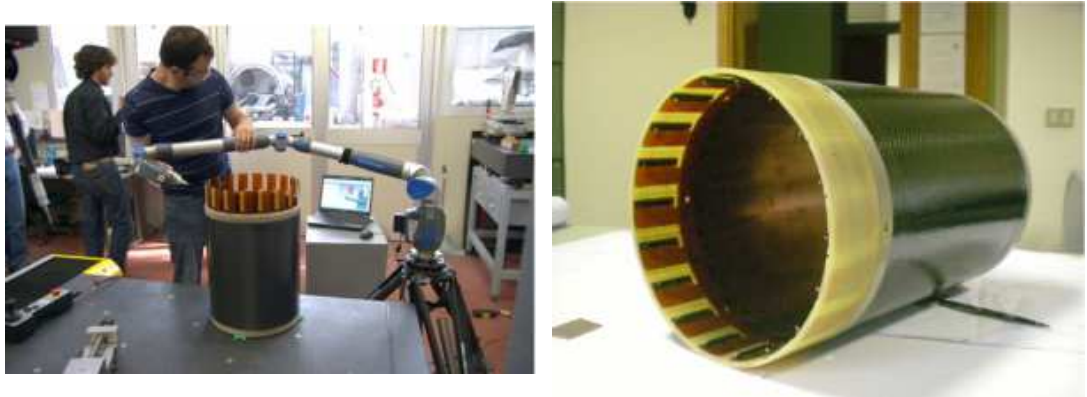

\centering
\figb MECpics/CRW_0330;7.;\kern0.2cm \figb MECpics/DSCN1288;7.;\\
%\figb MECpics/CRW_0321;7.;\\
\caption{(Left) Carbon fiber cylinder prototype with the
  "embedded-anode". (Right) The annular frame for FEE support coupled with the carbon fiber cylinder.}
\label{fig:CFC-Riba}
\end{figure}

A complete mechanical model of one of the IT layers has been realized and
includes: a prototype of the CFC with $\sim$300 mm radius and 360 mm length;
the annular flange (Service Flange-SF) foreseen as a support for the FEE
cards, HV connectors and gas piping.
The CFC has been realized at the RIBA srl (fig.~\ref{fig:CFC-Riba} top-left).
The SF prototype (fig.~\ref{fig:CFC-Riba} top-right) made by the Nuova Saltini
srl company, is described in sec.\ref{sec:fee_integration}.
The simulation studies of the CFC deformation
indicate that for an axial load of 600 kg applied to the CFC
the maximum radial displacement is about 70$\mu$m for the prototype dimension and 150
$\mu$m for a CGEM with the dimension of the first layer (up to 700 mm
length). The applied load is well above the 100 kg overall load
required for good detector stiffness and negligible sag ($<$5$\mu$m)~\cite{bib:lina}.

The buckling of the CFC is expected for an axial load
of about 9 tons.
The load tests have shown that the break-down of the CFC prototype occurs
at about 8 tons, well above the expected load on the final IT structure.\\
The 700 mm long GEMs will be realized with the new single mask procedure,
extensively discussed in sec.\ref{sec:large_area_gem}, that allows
large area foils to be build.
In order to overcome the limit on the width of the raw material used for
the GEM foils (453 mm maximum size), all the large electrodes of each
layers of the IT will be realized by splicing three smaller foils:
the gluing technique is the same used for the construction of the prototype
(planar gluing), exploiting the vacuum bag technique for epoxy
glue polymerization (sec.\ref{sec:construction}).
The epoxy used for the foils splicing is the two components Araldite
 AY103 + Hardener HD991, with a curing and polymerization time, at
ambient temperature, of about 1 hour and 12 hours respectively.
The epoxy, largely tested in harsh radiation environment, is the same used
for the assembly of the LHCb GEM chambers.
The GEM foils as well as the XV strip-pad patterned anode readout circuits
and the cathode foils will be realized on the typical 50$\mu$m polyimide
foils as substrates, on the basis of our design, by the CERN EST-DEM Printed
Circuit Board Workshop.\\
In table \ref{tab:assembly-materials} we summarize the materials used for the
construction of the detector.\\
\begin{table}[ht]
\caption{Materials used for the assembly of a CGEM}
\centering
\begin{tabular}{|l|l|l|}
\hline
{\bf Material} & {\bf Details} & {\bf Supplier/}  \\
               &               & {\bf Manufacturer}\\
\hline
Epoxy glue     & Araldite AY103 + HD991            & Ciba Geigy  \\
(2 comp.)&Ciba 2012 (for fast applications) & Ciba Geigy \\
\hline
Annular frames & 3/2/1mm thick Permaglass   & Resarm ltd (Be) \\
\hline
CFC support    & CF(250$\mu$m)-Nomex(3mm)-CF(250$\mu$m) & Riba srl (It) \\
\hline
SF support     & 8mm thick Permaglass annular-flange             & Nuova
Saltini (It)   \\
               & for FEE/HV/gas-piping location                  &           \\
\hline
GEM foils      & 50$\mu$m thick kapton, 3-5$\mu$m Cu;     & EST-DEM CERN \\
               & 70$\mu$m hole dia., 140$\mu$m hole pitch & Workshop     \\
\hline
Cathode        & 3-5$\mu$m Cu on a 50$\mu$m thick kapton  & EST-DEM CERN \\
\hline
Anode readout  & 3-5$\mu$m Cu on a 100$\mu$m thick kapton &  EST-DEM CERN \\
               & with XV strips-pad patterned             &               \\
\hline
Gas pipes      & 4 mm out.-dia. brass tube                    & LNF-workshop \\
\hline
Gas outlet     & 6 mm out.-dia. rilsan Pa11 tube & Tesfluid srl (Italy)  \\
               & (not hygroscopic) &  \\
\hline
\hline
\end{tabular}
\label{tab:assembly-materials}
\end{table}
Table \ref{tab:material-budget} shows the material budget of a CGEM
layer inside the active area.
\begin{table}[ht]
\caption{Material budget for a CGEM layer detector (active area)}
\centering
\begin{tabular}{|l|l|l|}
\hline
{\bf Component} & {\bf times x material ($X_0$) x quantity} & {\bf \% of $X_0$}  \\
\hline
3 GEMs & Copper: 6$\times$ 2$\mu$m Cu ($X_0$=14.3mm) $\times$0.8    & 0.067 \\
       & Kapton: 3$\times$ 50$\mu$m kapton ($X_0$=286mm) $\times$0.8 & 0.042 \\
       &                                                    & Total: 0.109  \\
\hline
1 Cathode & Copper: 1$\times$ 2$\mu$m Cu $\times$1     & 0.013 \\
          & Kapton: 1$\times$ 50$\mu$m kapton$\times$1 & 0.017 \\
       &                                               & Total: 0.030 \\
\hline
1 Readout & Copper: 1$\times$ 2$\mu$m Cu $\times$0.95  & 0.013 \\
anode     & Kapton: 2$\times$ 50$\mu$m kapton$\times$1 & 0.034 \\
       &                                              & Total: 0.047 \\
\hline
1 Shielding & Aluminum: 1$\times$ 10$\mu$m Al($X_0$=89mm)$\times$1  &
Total: 0.011 \\
\hline
1 Honeycomb & NOMEX: 1$\times$ 3mm Nomex($X_0$=13125mm)$\times$1 & Total: 0.023 \\
\hline
2 CF skins & CF: 2$\times$ 250$\mu$ CF($X_0$=250mm)$\times$1 & Total: 0.160\\
\hline
           &                                                 & Total: 0.380 \\
\hline
\end{tabular}
\label{tab:material-budget}
\end{table}
%
%%
%
%%%%%%%%%%%%%%%%%%%%%%%%%%%%%%%%%%%%%%%%%%%%%%%%%%%%%%%%%%%%%%%%
\subsection{Construction and tooling}
\begin{figure}[!h]
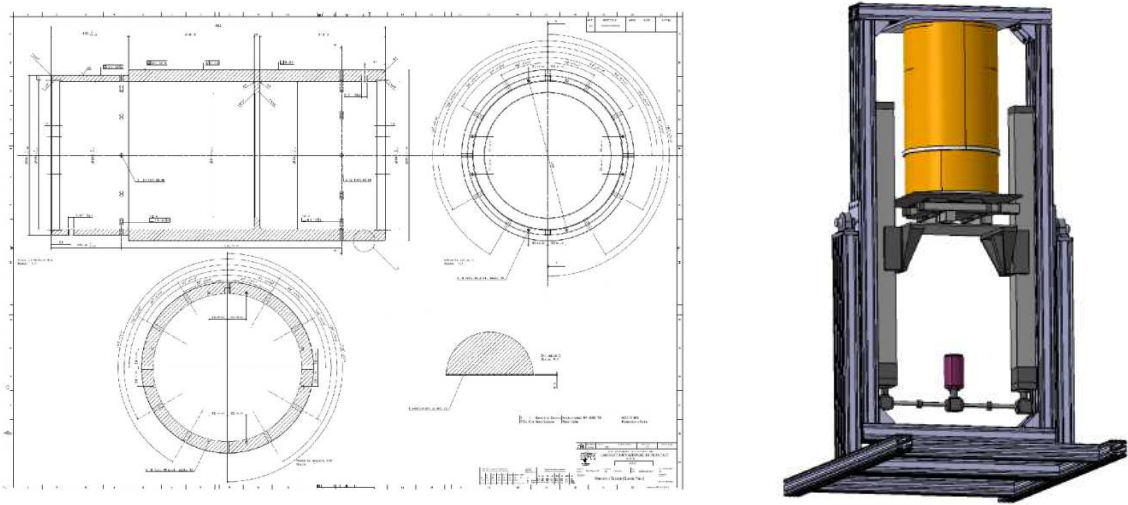

\centering
\figb MECpics/mandrino;9.;\kern0.2cm \figb MECpics/VIS-tool;7.;\\
%\figb PGEMpics/ ;7.;\kern0.2cm \figb PGEMpics/ ;5.;
\caption{(Left) Technical drawing of the cylindrical mould. (Right)
  Isometric view of the Vertical Insertion System.}
\label{fig:mandrino}
\end{figure}

%%%%%%%%%%%%%%%%%%%%%%%%%%%%%%%%%%%%%%%%%%%%%%%%%%%%%%%%%%%%%%%%
%
The construction strategy and the toolings of the CGEM layers composing
the IT are the same used for the prototype.\\
The main construction steps can be summarized as follows:
\begin{enumerate}
\item three GEM, as well as the anode and cathode, foils are preliminary glued
  together in order to obtain a single large foil needed to realize a
  cylindrical electrode. For this operation we exploit a precise Alcoa plane
  and the vacuum bag technique.
\item the large foil is then rolled on a very precise aluminum cylindrical
  mould covered with a 0.4 mm machined Teflon film for easy and safe extraction
  of the cylindrical electrode. The mould is then enveloped with the usual
  vacuum bag, and vacuum is applied for the glue curing time (about 12
  hours). In  fig.~\ref{fig:mandrino} left the technical drawing of one of the 25
  cylindrical moulds needed for the IT construction is shown.
\item the final assembling of a CGEM layer is performed by means of the
  Vertical Insertion System (VIS)  fig.~\ref{fig:mandrino} right, a tool that allows
  a smooth and safe insertion of the cylindrical electrodes one after the
  other.
  The system is designed to permit a very precise alignment of the
  cylindrical electrodes along their vertical axis. The bottom electrode is
  fixed, while the top one is slowly moved downwards by a manually
  controlled step-motor, coupled with a reduction gear system.
  The operation is performed with the help of three small web-cameras, placed
  at 120 degrees one to each other around the top cylindrical electrode, thus
  allowing the monitoring of the radial distance between the electrodes
  (2-3mm typically). The up-down rotation of the assembly tool allows an
  easy sealing of the detector on both sides.
\end{enumerate}
%%
%
%
%%%%%%%%%%%%%%%%%%%%%%%%%%%%%%%%%%%%%%%%%%%%%%%%%%%%%%%%%%%%%%%%%%%%%
\subsection{Material preparation \& Quality Controls}
%%%%%%%%%%%%%%%%%%%%%%%%%%%%%%%%%%%%%%%%%%%%%%%%%%%%%%%%%%%%%%%%%%%%%
%
Before the final assembly of the different parts of the detector, each
component follows a well defined preparation procedure that generally
includes a global optical inspection, a cleaning and an HV test.
In particular for GEMs the HV test
is repeated at each construction step, in order to avoid the assembly of
damaged GEM and to minimize the losses of precious components.
\subsubsection{GEMs}
\begin{figure}[!h]
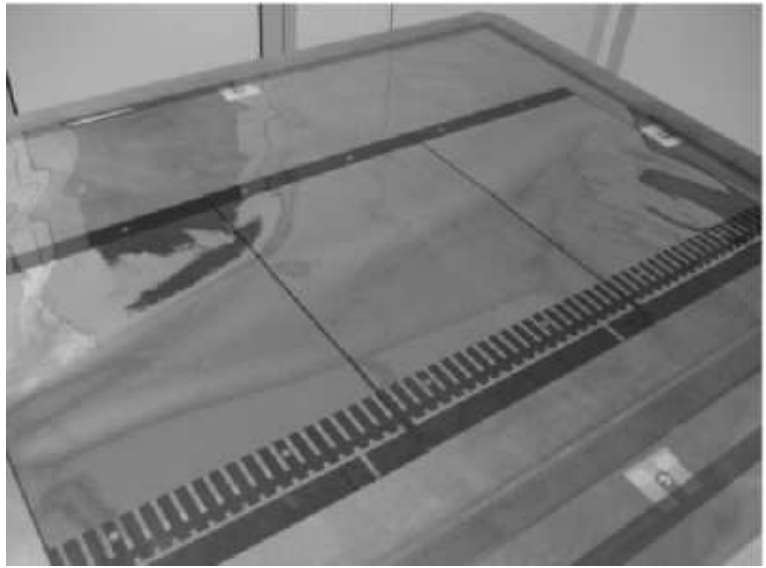

\centering
\figb MECpics/largeFoil;10.;
%\figb PGEMpics/ ;7.;\kern0.2cm \figb PGEMpics/ ;5.;
\caption{Detail of a large GEM foil.}
\label{fig:GEM-foil}
\end{figure}

GEMs are produced by the EST-DEM Workshop at CERN. A GEM foil realized for the construction of the full scale
prototype is shown in fig.~\ref{fig:GEM-foil}: clearly visible is the sectors structure realized on one side of
the foil with the aim of reducing the energy stored and then released in
case of discharge through the GEM hole.
The sectors in this case were 1.6-1.8 cm wide and about 36 cm long,
resulting in a sector area of about 60-65 $cm^2$, to be compared with those
of LHCb and Compass experiments respectively 80 and 90$cm^2$ large.\\
A first quality check is done at CERN by the producer: besides a global
optical survey each sector is supplied with a voltage up to 500 V and
checked for leak current that should be less than 5 nA.
This test is performed in clean room at ambient temperature and relative
humidity, that is without putting the GEM foil into a nitrogen flushed gas box.\\
GEMs are delivered inside a rigid clean plastic plates, each foil protected
between clean soft papers.
They are identified with a ID-number: "Z-type-XX", "Z" for the production
batch, the "type" identify the GEM (G1,G2 or G3) and "XX" indicates the
progressive number inside the batch.
In our laboratory they are stored in a dedicated cabinet inside the
clean room and always handled by operators suitable worn with clean clothings,
facial mask, gloves and hat.\\
GEMs do not require any special cleaning procedure, because in principle
they are clean: the only allowed cleaning procedure is done with a light
nitrogen flush that is generally used to gently remove possible dusts.
Any other procedure can instead damage the GEM and then is strongly
forbidden.\\
A first optical inspection is done by eye to look for areas with possible
defects, typically spots with unavoidable inhomogeneity in the material as
local absence of the metal and underlying kapton or lack of holes.
A more careful optical inspection of these spots is then performed under
microscope, equipped with a digital camera in order to evaluate the
dimension and type of the defect.
Such defects have generally no consequences in the operating behavior of
the GEM.
% {\bf unless the area or number of spots severely exceed XXX }.
For GEMs of the
LHCb detector not more than 1 defect with an area less than 1$mm^2$ per
each GEM sector were accepted.\\
The HV test of a GEM is performed inside a gas tight box flushed with
Nitrogen, in order to reduce the humidity down to 10\% Residual Humidity (time needed about
5 hours).
The HV is applied individually to each sector through a 500 M$\Omega$
limiting resistor, in order to avoid GEM damages in case of discharges,
while the non-sectored side is grounded.
The maximum current in the power supply is set to 100 nA. The HV is
slowly increased with eighteen steps up to a maximum of 600 V:
\begin{itemize}
\item from 0 to 400 V with 50 V steps of 20 seconds each;
\item from 400 V to 500 V with 25 V steps of 60 seconds each;
\item from 500 V to 600 V with 10 V steps of 2 minutes each.
\end{itemize}
The GEM sector pass the HV step if the current is less than 1 nA and no
more than 3 discharges occurred during the test time. In case these
acceptance requirements are not fulfilled the voltage ramp-up on the
sector is suspended and the test is repeated later on.
For each sector the test has a duration of about 30 minutes.
This figure, coming from our previous experience in LHCb, gives an
estimate of about 600 hours needed to test (only with HV) the about 1200
sectors foreseen for the IT, that become about 850 hours including the
time needed for nitrogen flushing (this must be done only about 50 times,
once per GEM foil), corresponding to about 5.5 man-months.
\subsubsection{Readout anode circuit}
\label{sec:readout_anode}
The readout anode circuit is manufactured by EST-DEM CERN Workshop, on the
basis of our design, starting from the 5$\mu$m copper clad, 50 $\mu$m thick
polyimide substrate, the same used for GEM foils. The 650 $\mu$m pitch,
300 $\mu$m wide X strips are parallel to CGEM axis, giving the r$\phi$
coordinates; while the V strips, realized connecting by vias a suitable
pattern of parallelogram shaped pads, and forming an
angle of about 40 degrees with respect the X-strips, gives the z
coordinate.\\
A first check to look for shorts between both coordinates is performed by the
producer. Connections test between pads and vias is also required.\\
A useful test for anode readout validation consists of measuring
the distance between first and last strip at different positions for both
coordinates, requiring a maximum deviation from the nominal not exceeding
100$\mu$m.\\
The readout anode circuits, as GEM foils, do not require any special
cleaning procedure: the only allowed cleaning procedure is done with a
light nitrogen flush to remove possible dusts.
Any other procedure can instead damage the circuit and then is strongly
forbidden.
They are stored in a dedicated cabinet in the clean room and always
handled by operators suitable worn with clean clothings,
facial mask, gloves and hat.\\
\subsubsection{Annular frames}
The annular frames, defining the gaps between electrodes, are 3 or 2mm (1mm)
thick and 35 mm wide. They are made of Permaglass and they are the only
supports for the cylindrical electrodes. The cathode and anode frames
contain the gas inlets and outlets respectively.\\
The frames delivered by the machine workshop before enter the clean room
are preliminary washed with isopropilic-alcohol and brush.
A carefully optical inspection to find and eliminate spikes and possible
broken fibers, is performed before and after the final cleaning done in a
ultra-sonic bath with demineralized water for 30 minutes.
After the cleaning the permaglass components are dried in a oven
at 40 $^o$C for 1 hours.\\
An HV test to verify its insulating is done placing the frame in between two
copper foils put at 4 kV within a clean box flushed with nitrogen.
If sparks occur, the frame is flushed for a couple of hours and then tested
again. If sparks still occur frames is inspected with a microscope.
Generally if the frame is perfectly clean neither sparks nor dark current
is observed.\\
\subsubsection{Final tests on closed chamber}
Once the chamber is closed we proceed with  gas leakage test.
The gas tightness is checked with a differential pressure device, a system
that allows us to measure the gas leak as the fall of pressure as function
of time.
The chamber to be tested and a reference volume (same volume of the chamber
with no leak, e.g. $\ll$ 1 mbar/day), flushed in parallel with an
overpressure of few mbars (typically 5 mbars), are connected to the two sensors
of the device. The reference volume allows us to take into account
for atmospheric and temperature variations during the measurement that can
lasts about 1 hour. The difference between the two pressure measurements
gives the gas leak rate of the chamber. All LHCb chambers \cite{LHCb-GEM5}
exhibit a gas leak of the order of less than 2 mbar/day,
corresponding to not more than 100 ppmV of residual humidity with a gas
flux of 80 cc/min.\\
The same system is also used to search for large gas leaks. In
this case the chamber connected to the sensor is inspected with a gas flow
all around the gluing regions: the possible sources of leak are easily
detected as a sudden increase of pressure, since the chamber is flushed
through the hole itself.
The holes are then sealed applying locally the same glue
used for the assembly of the chamber.
After glue polymerization the gas leak test is repeated and the chamber is
validated.\\
The chamber is then taken out from the clean room and moved to the
laboratory where it is flushed with the Ar/CO$_2$ (70/30) gas mixture
and tested in current mode with X-rays in order to check the gain
uniformity.
Gain variations of the order of 5-6\% \cite{LHCb-GEM5} are measured all
over the active surface of the detector.
%
%
%%%%%%%%%%%%%%%%%%%%%%%%%%%%%%%%%%%%%%%%%%%%%%%%%%%%%%
\subsection{Detector integration}
%%%%%%%%%%%%%%%%%%%%%%%%%%%%%%%%%%%%%%%%%%%%%%%%%%%%%%
%
%
%\newpage
% ERIKA mettere su unica pagina
\begin{figure}[!h]
\centering
\figb MECpics/New-assy-3-4;13.;
%\figb PGEMpics/ ;7.;\kern0.2cm \figb PGEMpics/ ;5.;
\caption{Technical drawings of the integration of the detectors at
  the interaction point.}
\label{fig:New-assy-3-4}
\end{figure}
\begin{figure}[!h]
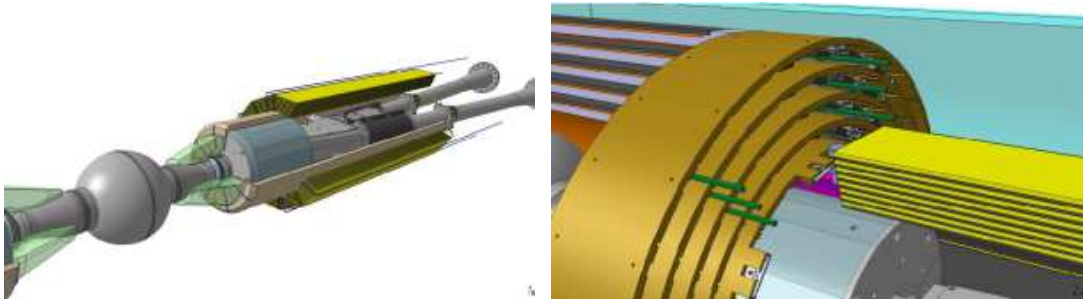

\centering
\figb MECpics/c7;7.;\kern0.2cm\figb MECpics/assy2;7.;\\
%\figb PGEMpics/ ;7.;\kern0.2cm \figb PGEMpics/ ;5.;
\caption{Isometric views of detectors integration (Left) IT and CCal integration  (Right)
  IT and QCal integration.}
\label{fig:c7}
\end{figure}
\begin{figure}[!h]
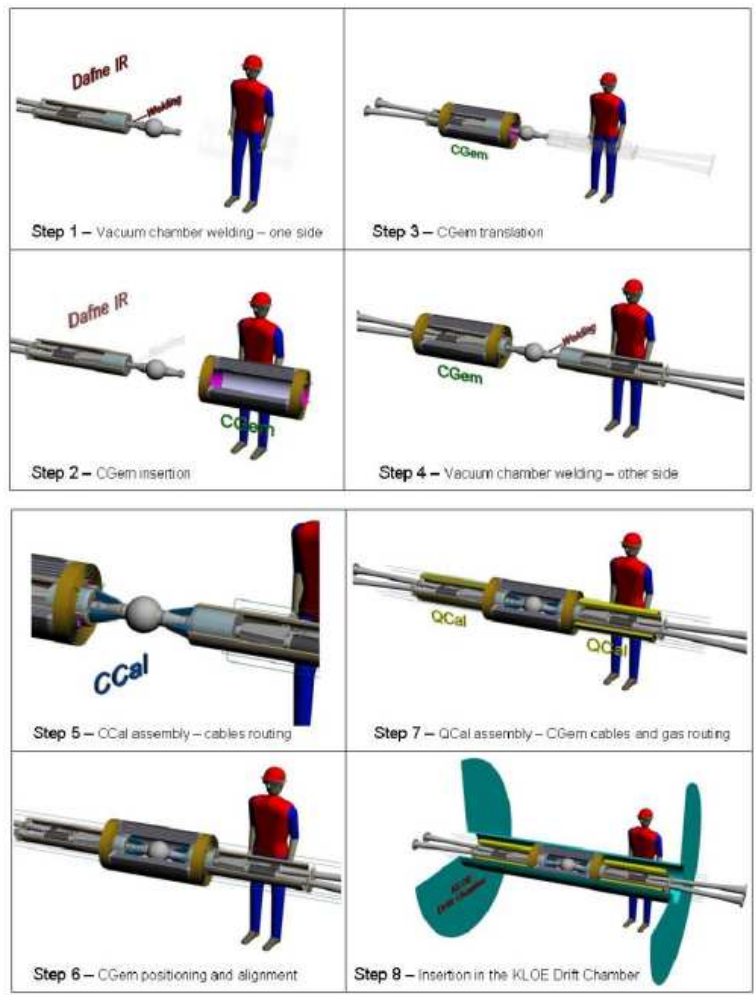

\centering
\figb MECpics/assembly_steps1;10.; \\ \figb MECpics/assembly_steps2;10.;\\
%\figb PGEMpics/ ;7.;\kern0.2cm \figb PGEMpics/ ;5.;
\caption{Preliminary studies of the detector integration sequence.}
\label{fig:assembly}
\end{figure}

The study of the IT integration around the DAFNE interaction point (IP) must
take into account the interferences between the IT and the internal tube of
the Drift Chamber and with the CCal and QCal calorimeters that are placed
around the beam pipe \cite{bib:cals}.\\
As shown in fig.~\ref{fig:New-assy-3-4}, a radial distance of 20mm between
the fifth layer of the IT and the internal surface of
the DC is foreseen in order to facilitate the critical insertion inside the
DC of the integrated system constituted by the beam pipe and detectors
complex.\\
The interference with CCal mainly concerns the routing of the calorimeter
cables that in the current design are foreseen to be embedded in the
aluminum support tubes of the interaction region (IR) fig.~\ref{fig:c7} left.\\
The isometric view reported in fig.~\ref{fig:c7} right shows the other critical
region between the IT ends and the two QCals surrounding the quadrupoles,
where gas piping, HV and signal cables of the IT (especially for the first
layer) must be routed out.\\
A very preliminary study of the integration sequence of the IT with the
beam pipe, CCal and QCal detectors is schematically shown in
fig.~\ref{fig:assembly}.
%{\bf Misc: Is this the final support with the final size?}
%
%%
%
%%%%%%%%%% insert HERE %%%%%%%%%%%%%%%%%%%%%%%%%%%%%%
\hyphenation{GA-STO-NE ele-ctro-nics wor-king ele-ctro-nics di-gi-tal te-chno-lo-gy si-gnal mo-no-sta-ble ge-ne-ra-tion ar-ri-ving ma-na-ged opti-cal inter-fe-ren-ce di-cte-te fun-ctio-na-li-ties}
\newpage
\section{Electronics}
\subsection{On-Detector Electronics}
\begin{figure}[h!]
\centering
\includegraphics[width= 5.5cm]{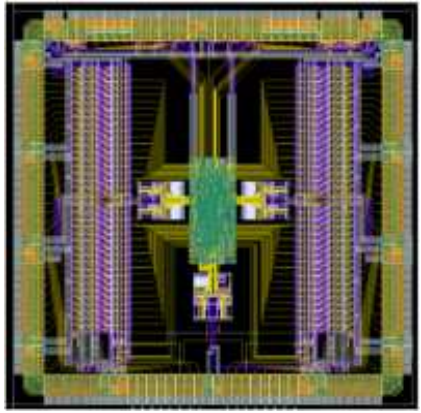}
\caption{GASTONE 64 channels layout}
\label{64chs}
\end{figure}

%
%\subsubsection{Introduction}
%
The IT will be readout by means of about 30000 XV
strips.  \\
Due to the different length of the V strips the parasitic capacitance will range between about 1 and 50~pF according to the position, making impossible the S/N optimization by capacitive matching. Moreover, because of the GEM moderate gas gain (e.g. with respect to Wire Chambers), the front-end electronics must be installed on the detector itself to maximize S/N ratio. Therefore both power consumption and I/O connections must be kept as low as possible.\\ \\
A further constraint concerns the KLOE DAQ timing: since the Level 1 trigger is delayed of about 200 ns with respect to the Bunch Crossing (BX), the IT discriminated signals must be properly stretched to be acquired. \\ \\
To fulfill the mentioned requirements a novel 16-channels front-end ASIC
prototype, named GASTONE ({\it Gem Amplifier Shaper Tracking ON Events}),
has been developed.  The sixteen channels ASICs have been used to readout
the  Cylindrical-GEM prototype in a test beam at CERN (sec~\ref{sec:CGEM_testbeam}). \\ \\
As the chip prototype fulfilled the IT readout requirements, a second
release, including a robust protection circuit to prevent damages caused by
discharge effects in the detector \cite{bib:nima479}, has been
implemented. This feature simplify significantly the layout of the front-end
electronic boards, avoiding the use of the classical, space-consuming,
external protection network made of resistors, diode and capacitors. The
new chips have been used to instrument the Planar GEM detectors for the
readout studies in a magnetic field performed in the testbeam at CERN (sec.~\ref{sec:PGEM_testbeam}).\\ \\
Finally a 64 channels prototype (fig.~\ref{64chs}) has been designed and will be used to instrument the 350x700 mm$^{2}$ large planar GEM detector (produced with the new single-mask technology) as final test of the GASTONE design before mass  production (about 700 chips).\\
\subsubsection{The GASTONE ASIC}
\label{sec:gastone}
\begin{figure}[h!]
\centering
\includegraphics[width= 8.5cm]{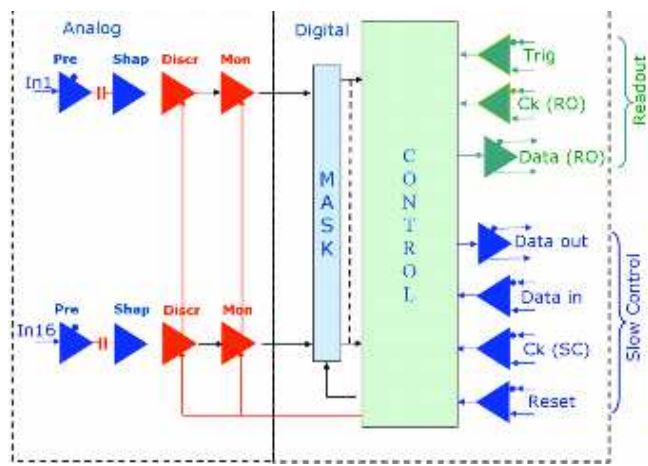}
\caption{Gastone Block Diagram}
\label{GastoneSch}
\end{figure}

\begin{figure}[h!]
\centering
\includegraphics[width= 12cm]{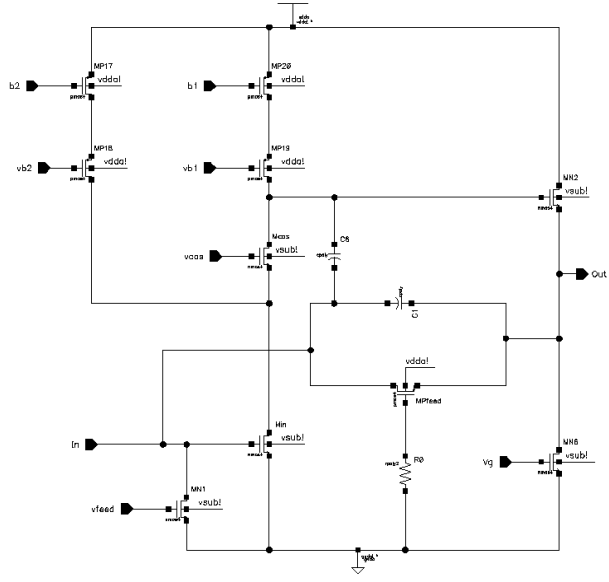}
\caption{GASTONE preamplifier}
\label{precircuit}
\end{figure}

\begin{figure}[h!]
\centering
\includegraphics[width= 7cm]{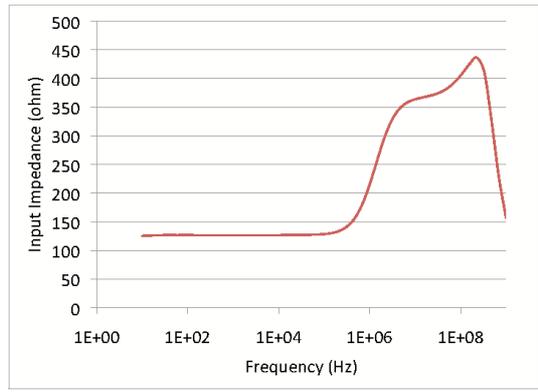}
\caption{Preamplifier input impedance vs frequency (I$_{feedback}$ = 300 nA, I$_{DC PRE} = 350 \mu$A ) }
\label{zinpre}
\end{figure}

\begin{figure}[h!]
\centering
\subfigure[]{
\includegraphics[scale=0.33]{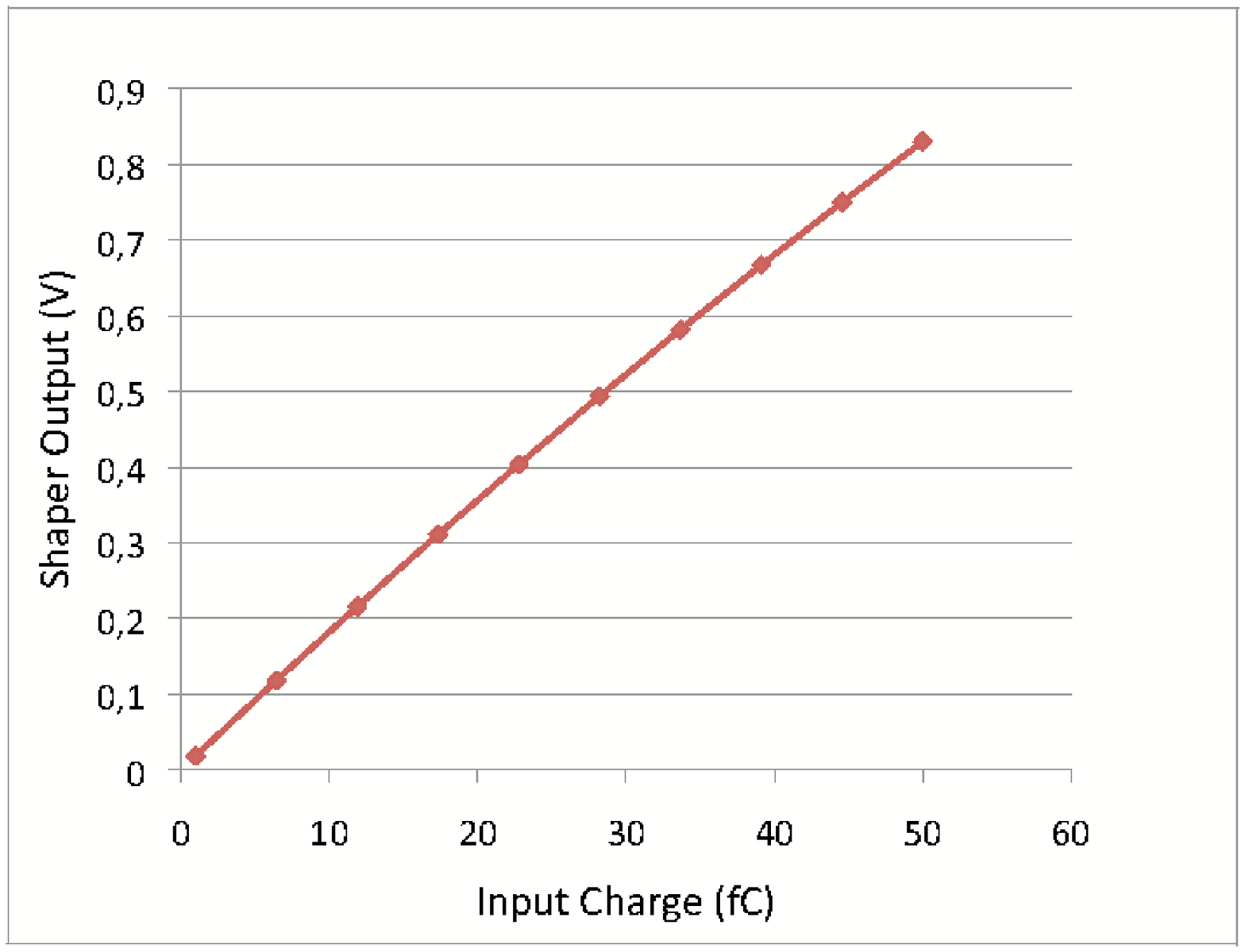}
\label{shaplin}
}
\subfigure[]{
\includegraphics[scale=0.335]{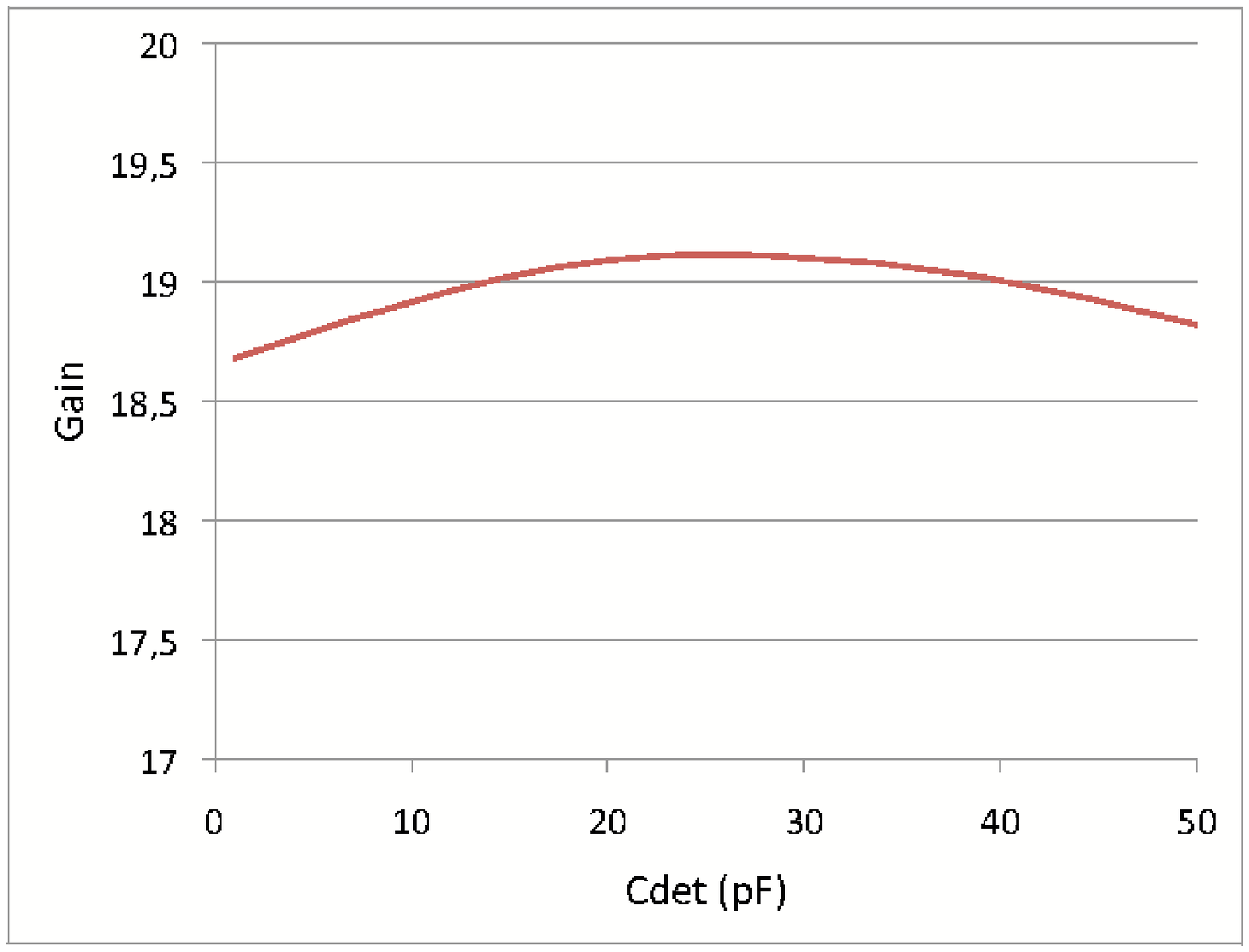}
\label{gainshapvscin}
}

\caption[Optional caption for list of figures]{Shaper output vs input charge  \subref{shaplin}, Overall gain vs C$_{in}$  \subref{gainshapvscin} }
\end{figure}

\begin{figure}[h!]
\centering
\includegraphics[width= 10cm]{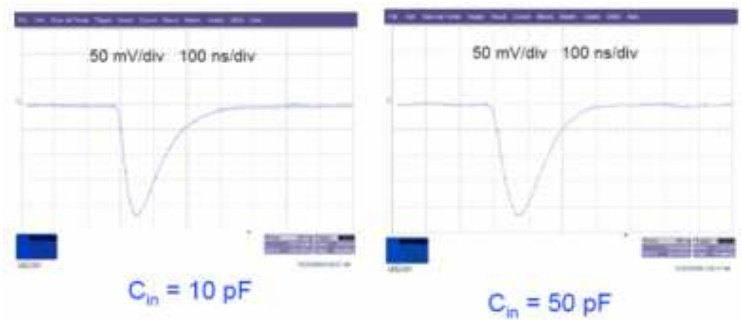}
\caption{Snapshot of the shaper output}
\label{snashap}
\end{figure}

The final version of the GASTONE chip is a mixed analog-digital circuit,
consisting of 64 analog channels %(16 channels in the first prototype)
followed by a digital section implementing the slow control and readout interface, as shown in  fig.~ \ref{GastoneSch}. \\
The main features of GASTONE are the low input equivalent noise (ENC) in
the detector capacitance range, the high modularity (to reduce the I/O
lines) and the low power consumption. The modularity has been set to 64
channels, as a compromise between the reduced number of I/O lines and the
overall system reliability (damaged devices can not be replaced as the
detector will be not accessible without extracting the beam-pipe). \\  \\
The amplified and shaped signals are digitized and serially readout using
both edges of a 50 MHz clock, achieving a 100 Mbps transfer rate. To avoid
interference between digital and analog sections, separate analog and
digital power supplies have been used and  I/O signals have been
implemented in the LVDS standard. The readout clock will be active only in
the download period (i.e. after the arrival of the Level1 trigger). \\
Since the expected integrated radiation dose is quite low, the ASIC has been designed in the mature and relatively cheap AMS 0.35 CMOS technology. \\
\paragraph{Analog section}
Each channel is made of four blocks: a charge sensitive preamplifier, an
amplifier-shaper, a threshold discriminator and a monostable. The charge
sensitive preamplifier converts the input current signal into a voltage
while the amplifier-shaper provides further amplification and noise
reduction. The discriminator generates the digital hit information which is
then stretched by the monostable so allowing the output signal to be sampled by the Level1 signal. The total power consumption of the analog channel is 1 mW.
\paragraph{Preamplifier features}
Due to the relatively low charge amplification of GEM devices
\cite{bib:cern98050}, the expected input charge will range from few fC to
some tenths of fC. The implemented amplifier should therefore combine a
high charge sensitivity with low noise level. \\
The input charge amplifier (fig.~\ref{precircuit})  consists of a common-source cascaded amplifier with an active feedback network made of a 150 fF capacitor and a PMOS transistor (W/L = 1.5/10).
The equivalent resistance is about 5 M$\Omega$ for a bias current of 300 nA.
The main preamplifier characteristics are: gain of $\sim$ 5 mV/fC @ C$_{IN}$
= 0 pF, non-linearity less than 1\% (0$\div$30 fC) and supply current of about 350 $\mu$A.  The input impedance is $\sim$  120 $\Omega$  over a 10$^5$ Hz frequency range as shown in fig.~\ref{zinpre}.
Table \ref{tab:prespec} summarizes the main features of the GASTONE preamplifier.
\begin{table}[h!]
\caption{Charge preamplifier main specifications}
\centering
\begin{tabular}{l l}
\hline\hline
Feedback capacitance & 150 fF  \\
Feedback resistance & 5 M$\Omega$ @ I$_{feedb}=300$ nA  \\
Trasconductance (g$_m$) & 6.4 mA/V \\
Gain & 5mV/fC @ C$_{in}$ = 0 pF \\
Z$_{IN}$ & 120 $\Omega$ up to 10$^5$ Hz \\
Non-linearity &  $\le$  1 $\%$ (0$\div$30 fC) \\
ENC (erms) & 800 e$_{rms}$ + 40 e$_{rms}$/pF \\
Supply Current & $\ \approx 350 \mu$A \\
\hline
\end{tabular}
\label{tab:prespec}
\end{table}
\paragraph{Shaper features}
The shaper provides semi-gaussian shaping for noise filtering and is
characterized by a voltage gain of 4. The non-linearity is less than 3\%
with a supply current of about 200 $\mu A$. The measured peaking time is
between 90 and 220 ns for a detector capacitance ranging between 1 and 50
pF. As shown in fig.~\ref{shaplin}, the overall preamplifier-shaper circuit has a global charge sensitivity of about 23 mV/fC,  while the behavior of the  gain as a function of the input capacitance is shown fig.~\ref{gainshapvscin}. The measured crosstalk is about 3\%.
Some snapshots of the shaper output for a fixed input charge of 10 fC and two preamplifier input capacitance (10 pF and 50 pF) are shown in fig.~\ref{snashap}.
\paragraph{Discriminator features}
A leading-edge discriminator follows the shaper stage. The threshold ranges between 0 and $\sim$200 fC while the supply current  is 180 $\mu$A. It is AC-coupled to the shaper to be independent from DC level variations and is characterized by a threshold spread below 2\% and an offset of  $\sim$ 2.2 mV rms over the full threshold range.
\paragraph{Monostable features}
The discriminator is followed by a monostable circuit. It stretches the pulse duration to allow the proper signal sampling upon the arrival of the Lev1 trigger.\\
\begin{figure}[h!]
\centering
\includegraphics[width= 7 cm]{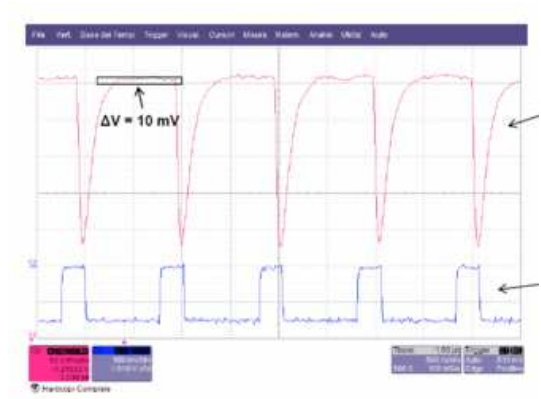}
\caption{Discriminator input and monostable output with a 10 fC, 1 MHz input signal rate}
\label{monlevshift}
\end{figure}
 The pulse width can be tuned between 200 ns and 1 $\mu$s. The range could be modified by choosing different  external bias resistors.

The fig.~\ref{monlevshift}, shows a snapshot of the monostable output. The maximum input rate is about 1 MHz, that is about 3 order of magnitude higher than the maximum expected rate. The fig.~\ref{monlevshift} also shows the effects on the discriminator input signal level for a 10 fC at 1 MHz input preamplifier signal. The shift, produced by the pile-up effect, is about 10 mV ($\sim$ 0.5 fC).
\subsubsection{The Digital section}
\begin{figure}[h]
\centering
\includegraphics[width= 10cm]{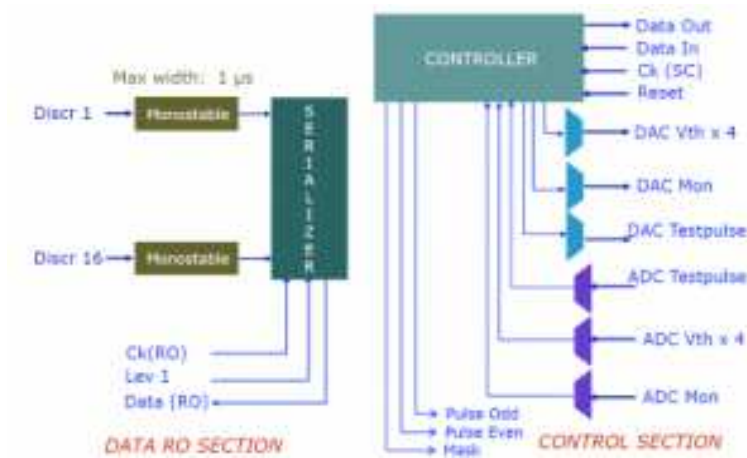}
\caption{Digital section block diagram}
\label{digblockdia}
\end{figure}

\paragraph{Readout Interface}
The digital section manages both  the control logic for threshold sensing/setting and the discriminated signal serialization for data readout (fig.~ \ref{digblockdia}). The 50 MHz readout clock is active only after the arriving of Lev 1 trigger signal, then avoiding possible crosstalk with the analog section. In  the 16 channels prototype the digital section has been already implemented to manage the 64 channels of the final chip. \\ \\
Upon the arrival of the Lev 1 trigger signal, the discriminated signals are stored into a 96-bits register according to the event frame described in tab. \ref{tab:frame}. The arrival of 50 MHz readout clock starts the data transmission procedure that will last 960 ns (the readout circuit uses both edges of the clock signal).  \\
\begin{table}[ht]
\caption{Gastone data transmission frame}
\centering
\begin{tabular}{l l l}
\hline \hline
Number of bits & Name & Description  \\  [0.5ex]
8 & Sync & 10101010 synchronization receiver pattern  \\
2 & START & 11 (consecutive data precursor bits)  \\
5 & TRIGGER\_NUM & Trigger number \\
9 & CHIP\_ID & chip identification \\
64 & CH\_DATA & Transmitted data \\
8 & ZERO & Ending transmission bits \\
\hline
\end{tabular}
\label{tab:frame}
\end{table}
The full digital section power consumption is about 40 mW and is dominated by the LVDS output drivers.
\paragraph{Slow Control-SPI interface}
The Slow Control section of the chip is managed by a SPI interface with a clock running at 1 MHz. It consists  of twenty-eight 8-bit registers (listed in tab \ref{tab:spireg}) for configuring the chip functionality, setting the four DACs threshold and reading back the four ADCs (16 channels modularity). A fifth DAC/ADC is devoted to set/read  the width of the monostable output pulse and, finally, a sixth DAC/ADC allows to set/read the amplitude of the internal test pulse. The threshold is set to  $(\sim$ 3 fC) at power-on.  \\
An internal pulsing procedure has been implemented to inject a selectable charge  for calibration and test purpose. The chip also generate a global OR signal that can be used in self-triggering applications.\\
\begin{table}[ht]
\caption{List of SPI registers}
\centering
\begin{tabular}{c l}
\hline \hline
\# BITS & Content  \\  [0.5ex]
8 & Mask  \\
4 & Threshold(1 threshold for 16 channels)  \\
1 & Pulse amplitude (1 per 64 channels) \\
5 & Read back from 5 ADCs \\
1 & Test pulse configuration \\
8 & Test pulse result \\
1 & Control register \\[1ex]
\hline
\end{tabular}
\label{tab:spireg}
\end{table}
\subsection{The 16 channels chip prototype}
\begin{figure}[h!]
\centering
\includegraphics[width= 6 cm]{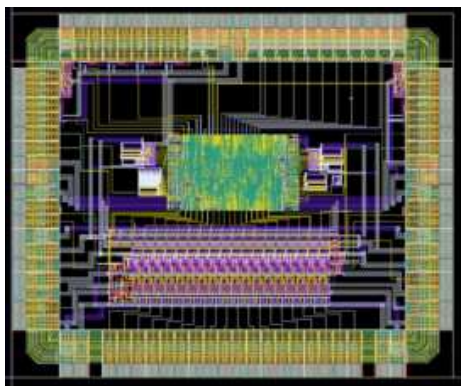}
\caption{Layout of the 16 channels GASTONE chip}
\label{layout16chs}
\end{figure}

\begin{figure}[h]
\centering
\includegraphics[width= 7cm]{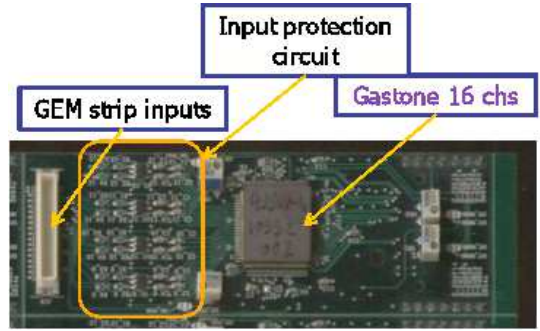}
\caption{Front-End Board developed to host the first release of the chip. It includes the protection circuit connected to the input pads of the chip.}
\label{feeboards}
\end{figure}

The 16 channels GASTONE prototype layout is shown in
fig.~\ref{layout16chs}. The chip occupies a silicon area of 3.2 x 2.2
mm$^2$ and has a power consumption of 0.6 mA/ch.
A custom Front-End Board (FEB) has been developed (fig.~\ref{feeboards})
to host the chips. The board contains  a passive network to protect
the chip against possible spark events due to gas discharges and hosts
two chips, for a total amount of 32 channels.\\
The main features of the FEB (whose dimension are 30x95 mm$^2$)  are:
\begin{itemize}
\item 2 chips/board (32 channels)
\item 2 independent LVDS serial data readout lines
\item LVDS I/O for Slow Control/Reset/Trigger
\item 16 chs OR output for self-triggering
\item HW ID
\end{itemize}
\subsubsection{Lab test result}
\begin{figure}[h]
\centering
\includegraphics[width= 10cm]{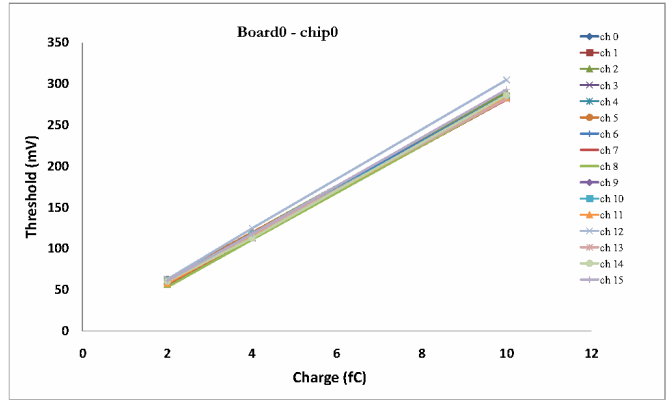}
\caption{Charge sensitivity (threshold vs input charge) }
\label{chargesensitivity}
\end{figure}

\begin{figure}[h!]
\centering
\includegraphics[width= 10cm]{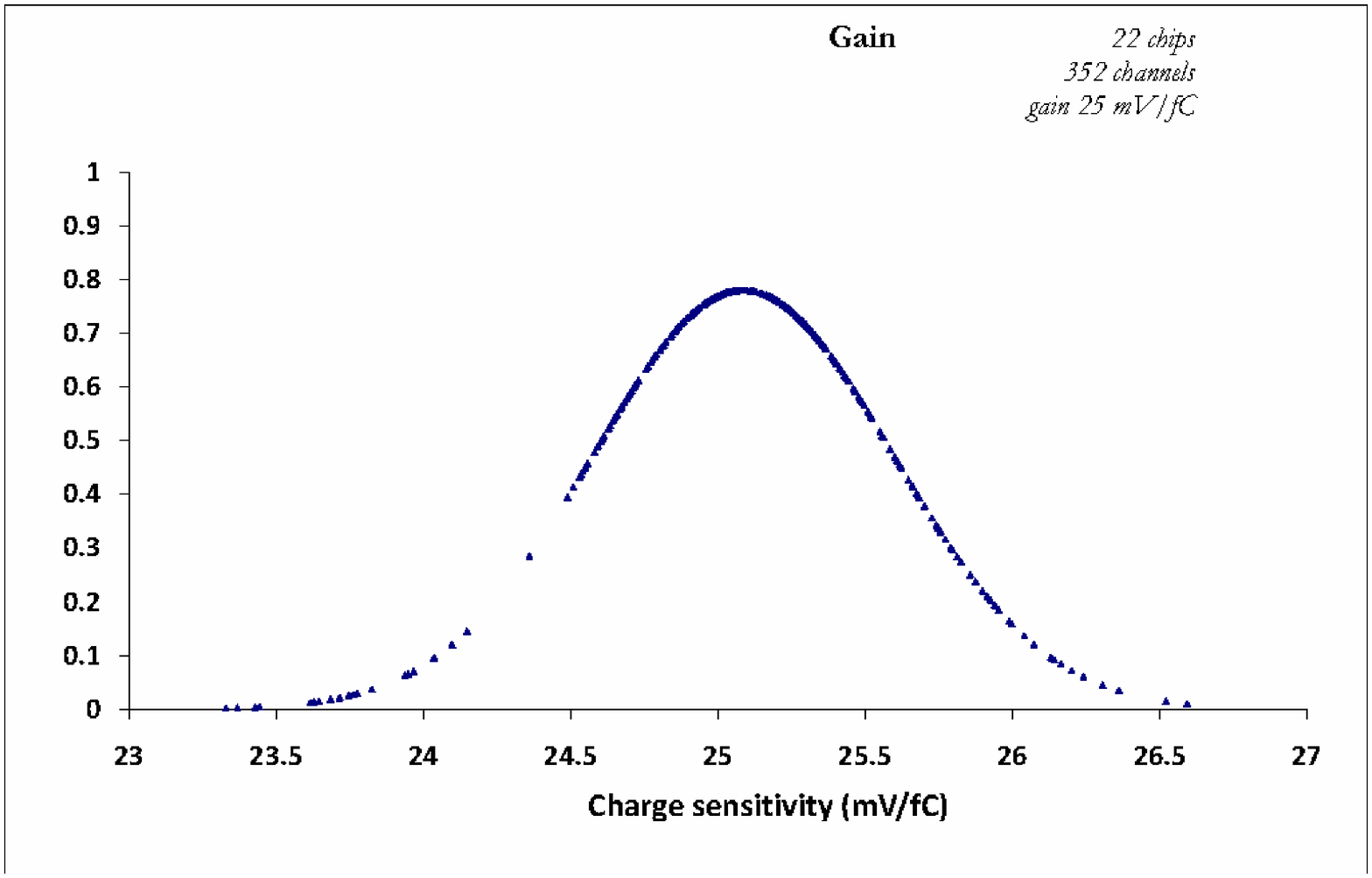}
\caption{Gain distribution}
\label{gaindistrib}
\end{figure}

\begin{figure}[h!]
\centering
\includegraphics[width= 10cm]{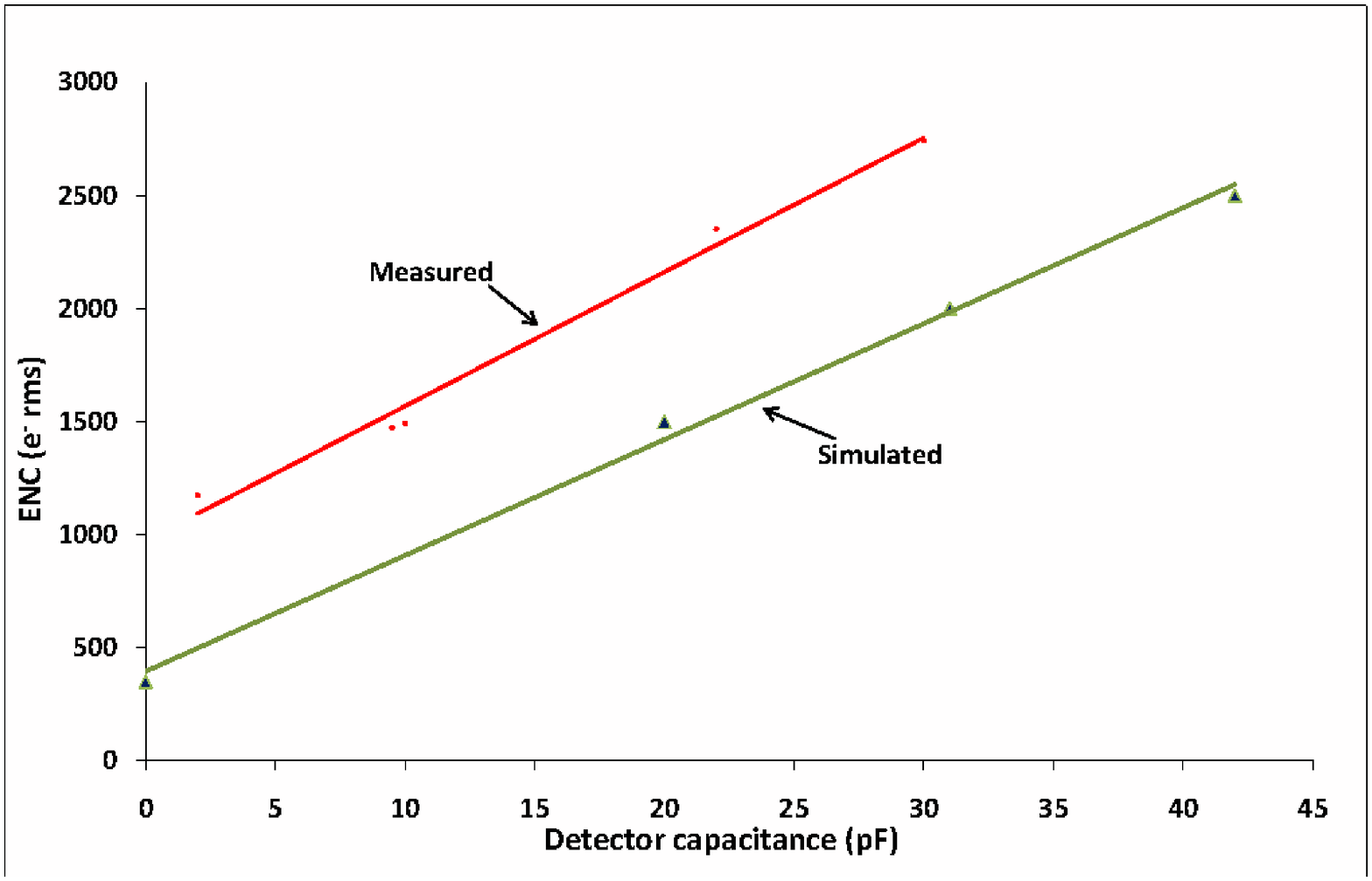}
\caption{Simulated and measured ENC distributions}
\label{erms}
\end{figure}

\begin{figure}[h]
\centering
\includegraphics[width= 12cm]{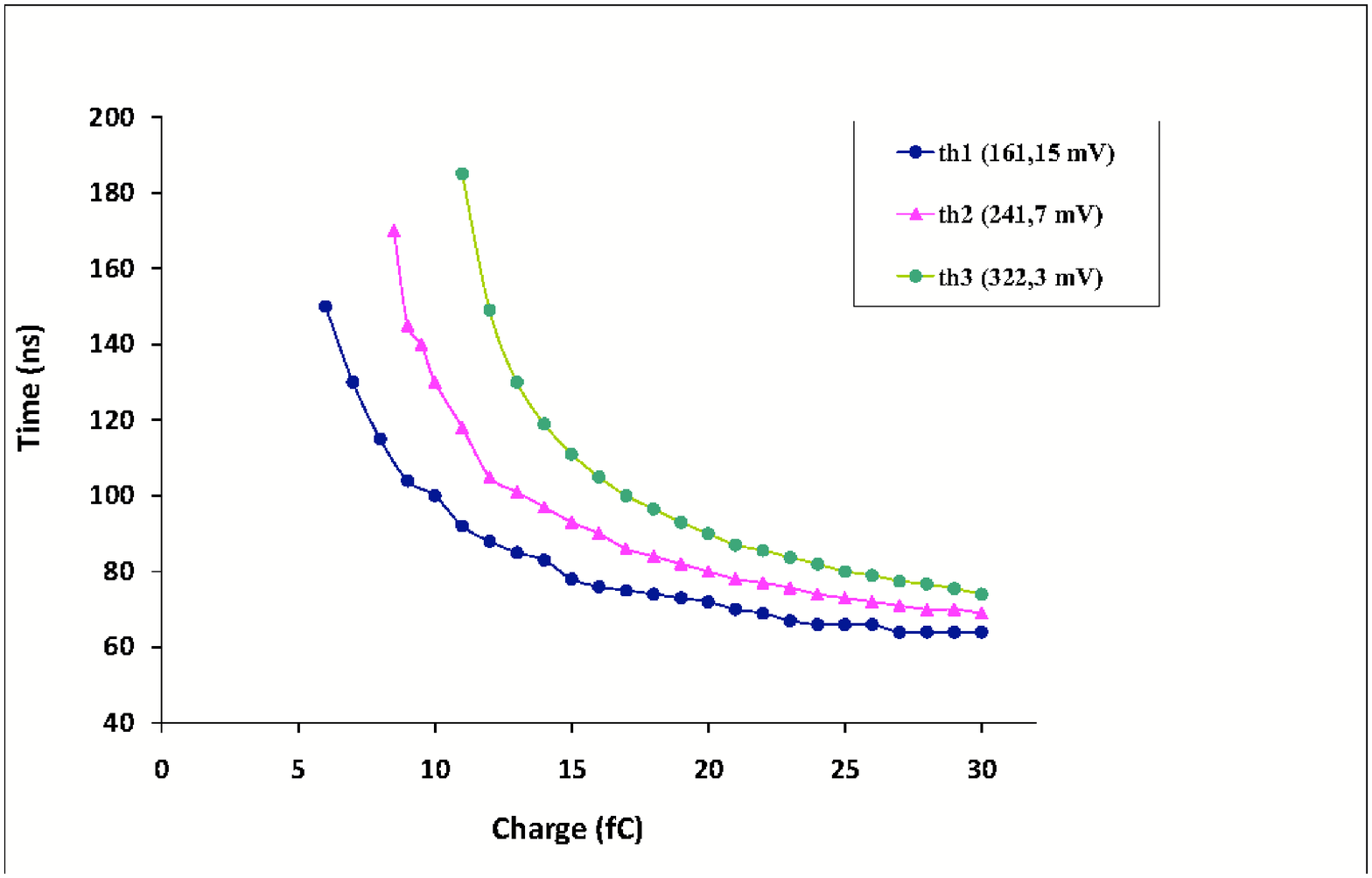}
\caption{Time walk vs Q$_{in}$ at C$_{det}$ = 10 pF }
\label{timewalk}
\end{figure}

Twenty-two chips have been produced in the run. All chips have been  characterized  by means of a dedicated  test bench based on a custom VME board implementing both the SPI control and the readout protocol. An Agilent Pulse Generator 81110A has been used to drive the charge injection used to test the devices.
\paragraph{Charge sensitivity and gain measurements}
The charge sensitivity over the 16 channels of one chip (chip 0 / Board 0) is shown in fig.~\ref{chargesensitivity}. A good linearity and uniformity over the various channels can be easily inferred. Doing the same measurements over the  22 chips (352 channels), an average gain of 25mV/fC has been obtained (fig.~\ref{gaindistrib}).
\paragraph{ENC measurements}
The ENC directly measured on a triple-GEM 10$\times$10 cm$^2$ prototype
(fig.~\ref{erms}) is about 974 e$_{rms}$+ 59 e$_{rms}$/pF. The measured
value corresponds to $\sim$ 0.63 fC for the longest strip and sets a
minimum level of $\sim  $1.5 fC to the discriminator threshold.
The difference between measured and simulated values is due to the PCB
lines and protection diodes parasitic capacitance (between 5 and 10 pF).
\paragraph{Time-Walk}
The time response of the chip as a function of the input charge has been
measured for different thresholds. As expected, the time response increases
as the charge approaches the discriminator threshold, because of the used leading
edge technique. In fig.~\ref{timewalk} the results obtained for C$_{det}$ = 10 pF are shown.
\subsection{Off-Detector electronics}
\begin{figure}[h!]
\centering
\includegraphics[width= 12cm]{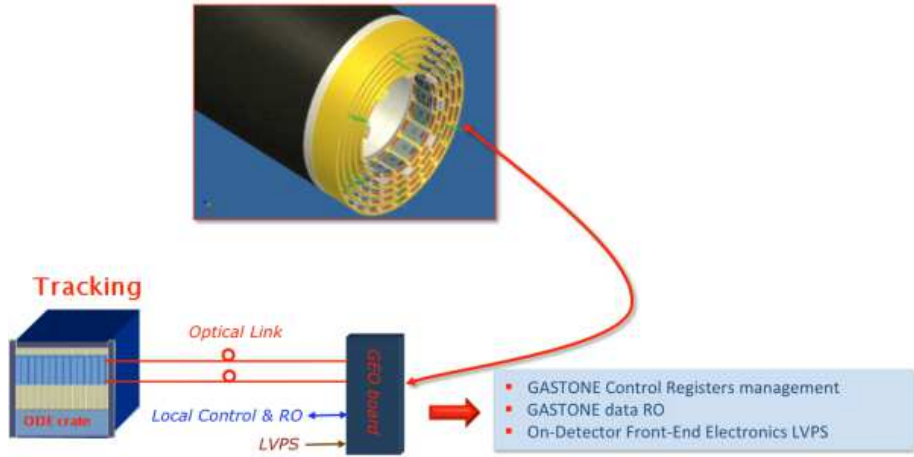}
\caption{Inner Tracker readout block diagram}
\label{ReadoutBlockDiagram}
\end{figure}

The IT off-detector electronics is made of two sections.
The first one (GASTONE Electronics Off-detector - GEO)  will be located
just outside  KLOE with the aim  of keeping the connections between the
boards and the on-detector electronics as short as possible
(about 3.5 m have been foreseen in the final design); the second one
(CONCENTRATOR) will be located on the platform near the DAQ electronics
(about 15 m apart). The connections between the two sections will be
implemented by means optical links, thus avoiding any trouble due to
ground-loops and/or interference with other readout sections of the KLOE
apparatus. A block diagram of the system is shown in fig.~\ref{ReadoutBlockDiagram}.
\subsubsection{GEO boards}
\begin{figure}[h!]
\centering
\includegraphics[width= 14cm]{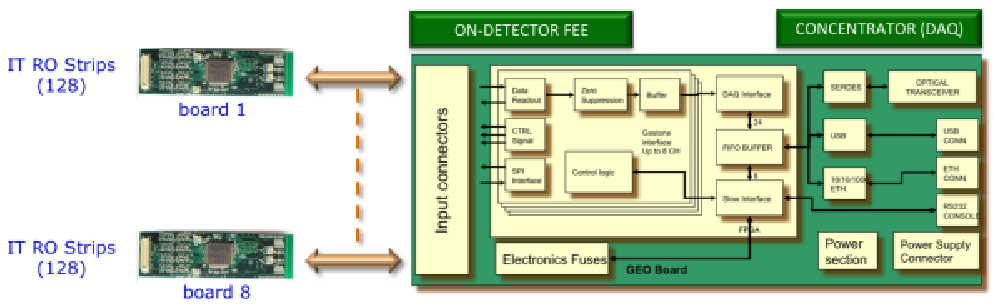}
\caption{GEO readout board Readout block diagram}
\label{GEO-BlockDiagram}
\end{figure}

\begin{figure}[h]
\centering
\includegraphics[width= 13 cm]{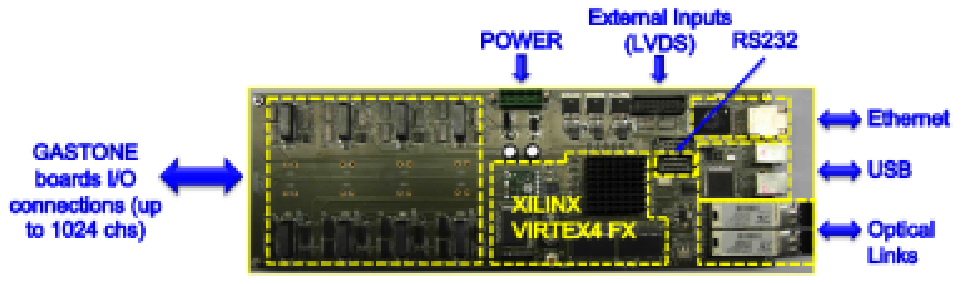}
\caption{GEO readout board prototype}
\label{GEO-Board}
\end{figure}

These boards,  besides supplying low voltage to the on-detector
electronics, provide full control over the GASTONE boards.
More in detail, with these boards it is possible to set/read the
GASTONE chips internal registers, to test different sections of the
chain (then implementing an efficient debug procedure) and, finally,
download the 64 bits data pattern if a Lev1 trigger  signal is issued
by the KLOE Trigger System.
The board is foreseen working together
with the CONCENTRATOR board; anyway some local control capability has
been foreseen as well (Ethernet and USB) both to simplify the detector
setup procedure and to use the board as stand-alone DAQ system.\\ \\
Fig.~\ref{GEO-BlockDiagram} shows the block diagram of the GEO board. The two main circuit sections can be easily identified: the on-detector interface  on the left side and the DAQ interface on the right side.\\ \\
The first prototype of the board is shown in fig.~\ref{GEO-Board}.
A Xilinx Virtex4 device has been used to manage both the Experiment
Control System (ECS) path (through the embedded Power PC processor)
and the DAQ path. Data transmission to/from GASTONE boards has been
implemented by means of LVDS connections (on the left-side of the board),
while communications with the experiment DAQ and ECS systems have been
implemented through two separate optical links (on the right side of the
board). LVDS inputs have been foreseen to manage (if required)
external signals. \\
The readout procedure is started as soon as the Lev1 trigger signal is issued. As a consequence of the Lev1 signal detection the GEO board internal state machine takes the following actions:
\begin {itemize}
\item propagates the Lev1 signal to the 468 sixty-four channels GASTONE chips to load the internal readout register with the 96 bits data frame;
\item start the GASTONE boards readout procedure by delivering a 50 MHz readout clock (effective data throughput is 100 Mbit/s as both sides of the readout clock are used).
\end {itemize}
Zero suppression on incoming data has been foreseen, as well, to reduce data transfer size in case of low detector occupancy.
\subsubsection{CONCENTRATOR board}
\begin{figure}[h!]
\centering
\includegraphics[width= 8 cm]{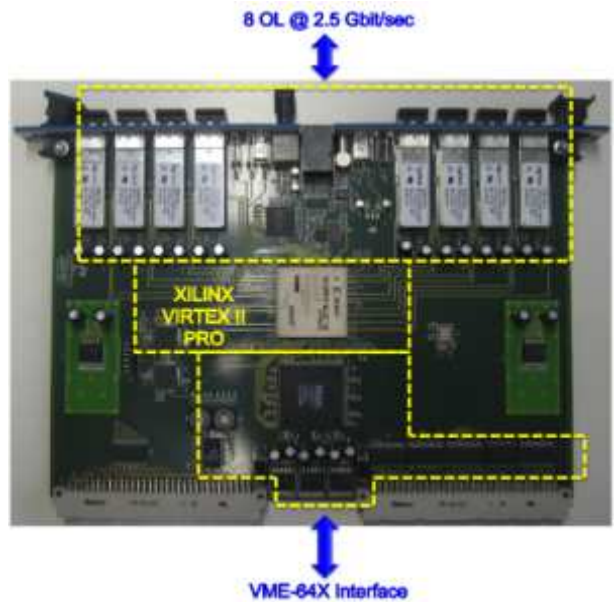}
\caption{CONCENTRATOR board prototype}
\label{ConcentratorProto}
\end{figure}

As already said, the CONCENTRATOR boards  (fig.~\ref{ConcentratorProto}) are located on the platform, about 15 m far from the detector. All communications (Slow Control, DAQ and Trigger) to/from the GEO boards are managed by these boards (through optical links).  \\ \\
\begin{figure}[h!]
\centering
\includegraphics[width= 7cm]{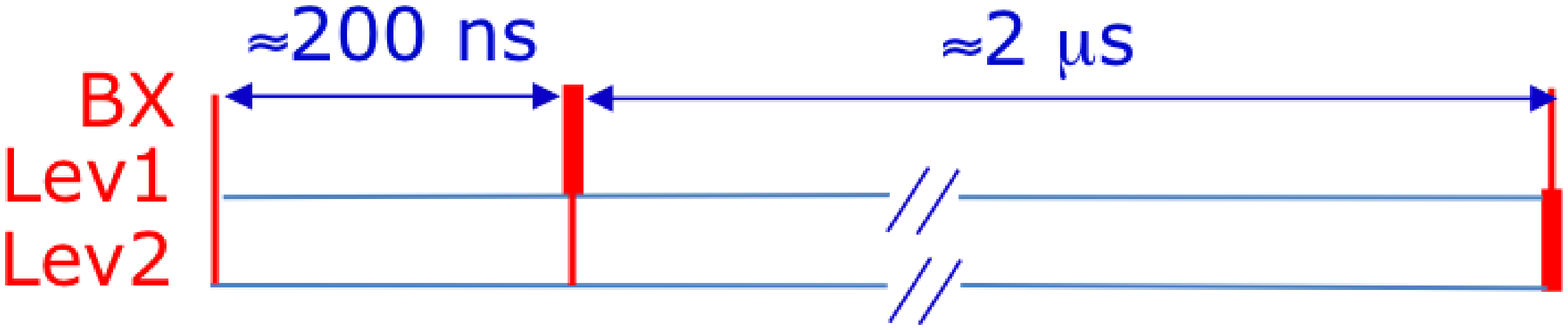}
\caption{KLOE DAQ Timing}
\label{KloeTiming}
\end{figure}
The acquisition procedure is started by Lev1 trigger signal  (fig.~\ref{KloeTiming}). When the GEO boards detect the trigger signal they propagate the signal to the on-detector electronics then loading the GASTONE output registers. After a while the GEO boards activate the download clock and the readout of the (about) 400 boards starts. After 900 ns the readout procedure is completed and, if enabled, the data zero suppression procedure starts. Finally, 1.8 $\mu$s after the Lev1 signal the Lev2  trigger signal can be generated and data transmission from GEO boards to CONCENTRATOR starts.
\subsection{HV and FEE integration}
\subsubsection{FEE Integration}
\label{sec:fee_integration}
\begin{figure}[h!]
\centering
\includegraphics[width= 6 cm]{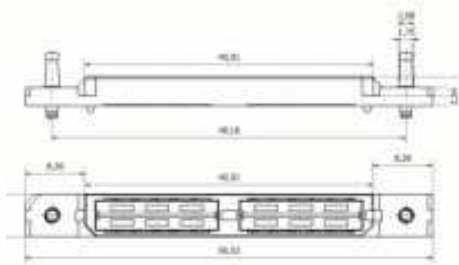}
\caption{Front-end board connector}
\label{connector}
\end{figure}

\begin{figure}[h!]
\centering
\subfigure[]{
\includegraphics[scale=0.60]{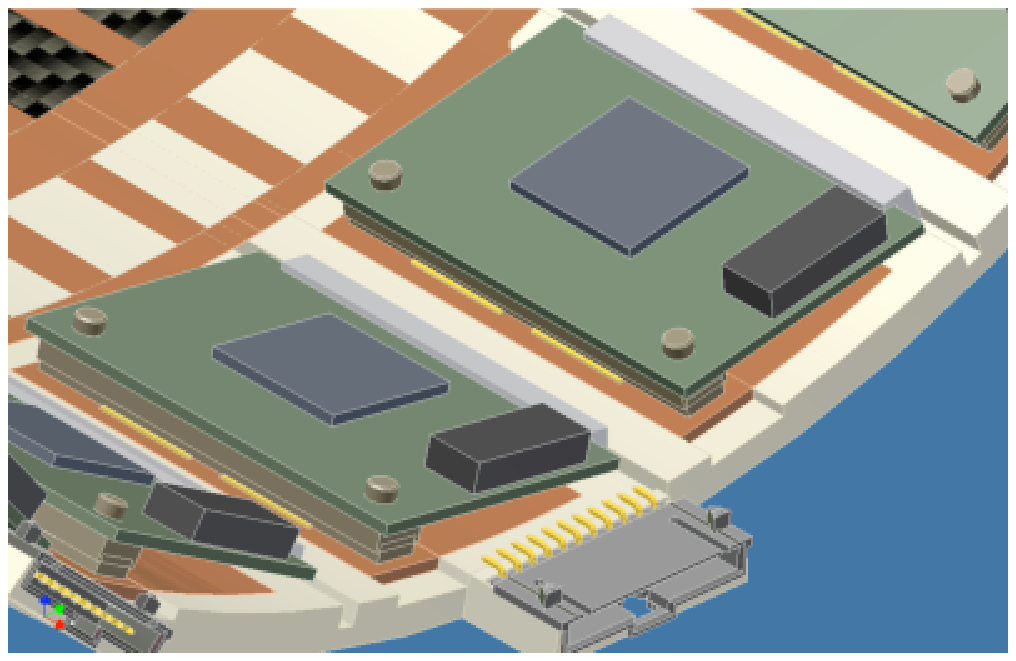}
\label{detectormodel}
}
\subfigure[]{
\includegraphics[scale=0.9]{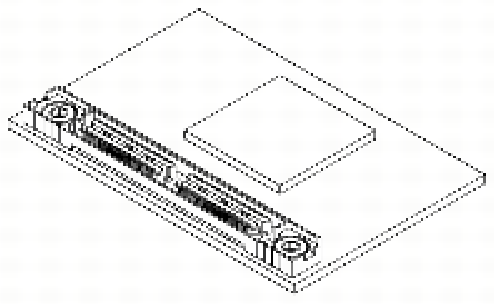}
\label{pcb3dmodel}
}

\caption[Optional caption for list of figures]{Detector 3-D model - front-end pcb insertion detail  \subref{detectormodel}, PCB 3-D  model  \subref{pcb3dmodel} }
\end{figure}

\begin{figure}[h!]
\centering
\includegraphics[width= 6 cm]{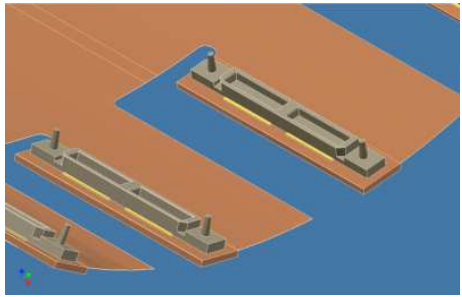}
\caption{Stiffness support and connectors glued into the Kapton foil}
\label{stiffsupp}
\end{figure}

\begin{figure}[h!]
\centering
\subfigure[]{
\includegraphics[scale=0.225]{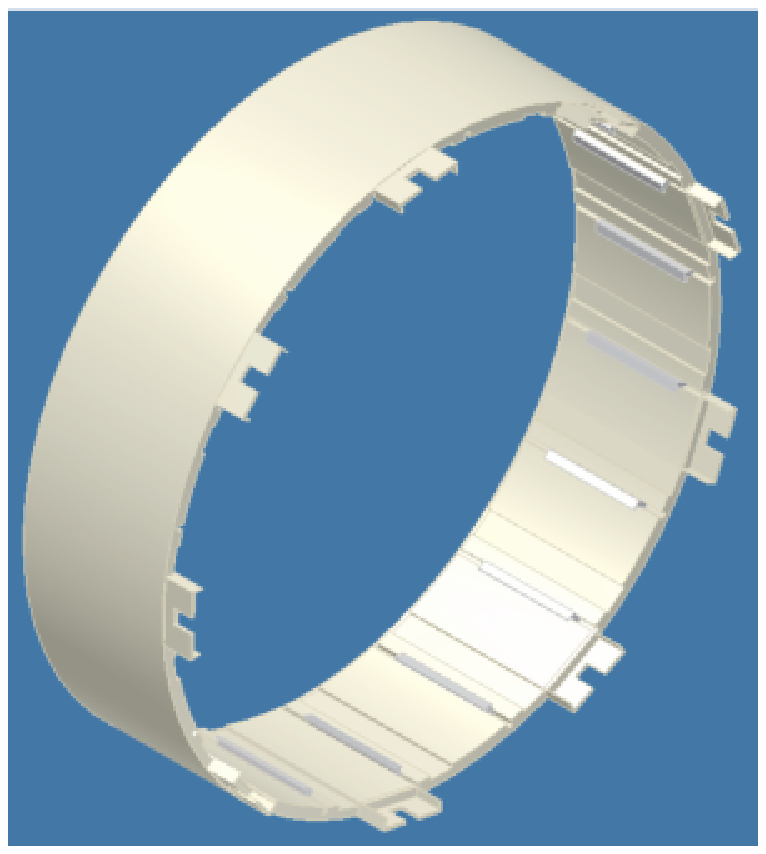}
\label{suppflanges}
}
\subfigure[]{
\includegraphics[scale=0.6]{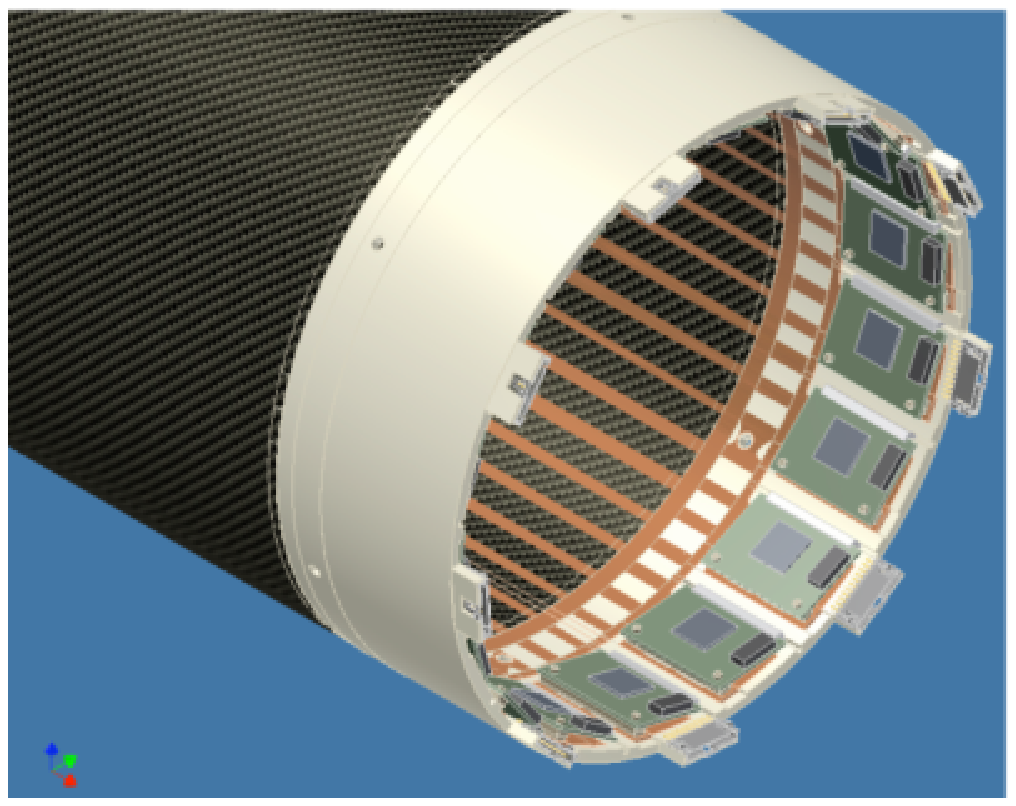}
\label{finass}
}

\caption[Optional caption for list of figures]{Front-end electronics support flanges  \subref{suppflanges}, Detector structure  \subref{finass} }
\end{figure}

\begin{figure}[h]
\centering
\includegraphics[width= 6 cm]{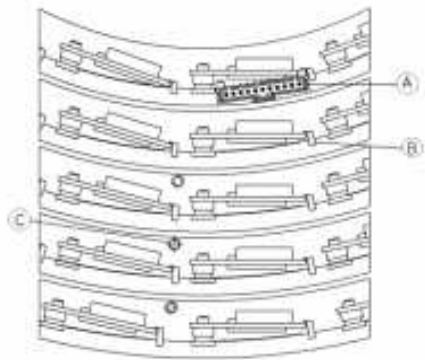}
\caption{Support structure detail : A - HV connectors, B - front-end boards nozzles, C - gas connections}
\label{finassblowup}
\end{figure}

The number of connections between the detector readout plane and the front-end electronics is a function of the layers diameters. As an example, for the innermost layer there are about 3950 readout strips which are connected to the same number of electronics readout channels. \\ \\
The connections density and the cylindrical structure of the readout dictates the usage of a 120 pins/0.5 mm pitch connector (fig.~\ref{connector}) positioned parallel with respect to the detector z axis. The connector has two retention plugs that allow a suitable fixation of the PCB to the mechanical support and the connection to the electrode ground plane. \\ \\
To define the front-end board pcb dimension constraints a 3-D mechanical detector model has been implemented (fig.~\ref{detectormodel}). From the 3-D simulations a front-end board pcb maximum size of 60x40 mm$^2$ has been inferred (fig.~\ref{pcb3dmodel}). \\ \\
The readout signals are extracted from both sides of the cylindrical electrode in such a way to  obtain a symmetrical readout. Finally the electronics channels have been arranged as described in tab.\ref{tab:conxside} allowing the partition of the (cylindrical) readout electrode in three pieces covering 120 degrees each. \\ \\
\begin{table}[ht]
\caption{Front-End and HV connector per side}
\centering
\begin{tabular}{c c c c}
\hline \hline
Layer Number & Radius(mm)  &  FEE Connector per side & HV Connector per side\\[0.5ex]
 &   &  &  for one 120$^o$ sector\\[0.5ex]
1 & 136 & 18 & 4\\
2 & 158 & 21 & 7\\
3 & 180 & 24 & 7\\
4 & 202 & 27 & 7\\
5 & 224 & 27 & 7\\ [1ex]
\hline
\end{tabular}
\label{tab:conxside}
\end{table}
In the shaped far-end electrode protruded part ( see fig.~\ref{stiffsupp}), FR4 stiffness supports will be glued before soldering the front-end board input connectors. The stiffness support will be also used to fix the kapton foil to the cylindrical structure support . \\ \\
For each side of the readout electrode a support flange (fig.~\ref{suppflanges}), implemented by an annular cylinder, has been foreseen. The flange will allow to allocate both the stiffness and the nozzles (located on the opposite side of the connector) used  to hold the connector. Also the supports for the HV connectors (placed at regular angles, taking into account the segmentation of the three GEM detector planes) will be glued to the structure. Finally, the two flanges will be glued to the layer support structure and to the readout electrode \\ \\
Once each layer is built and tested, the detector will be assembled by attaching layers together through suitable structure that will also allow the detector placing on the beam-pipe (fig.~\ref{finass}). \\
In fig.~\ref{finassblowup}  some details concerning the placement of HV connectors, front-end boards and gas connections for the 5 layers are shown.
\subsubsection{HV Integration}
\begin{figure}[h!]
\centering
\subfigure[]{
\includegraphics[scale=0.5]{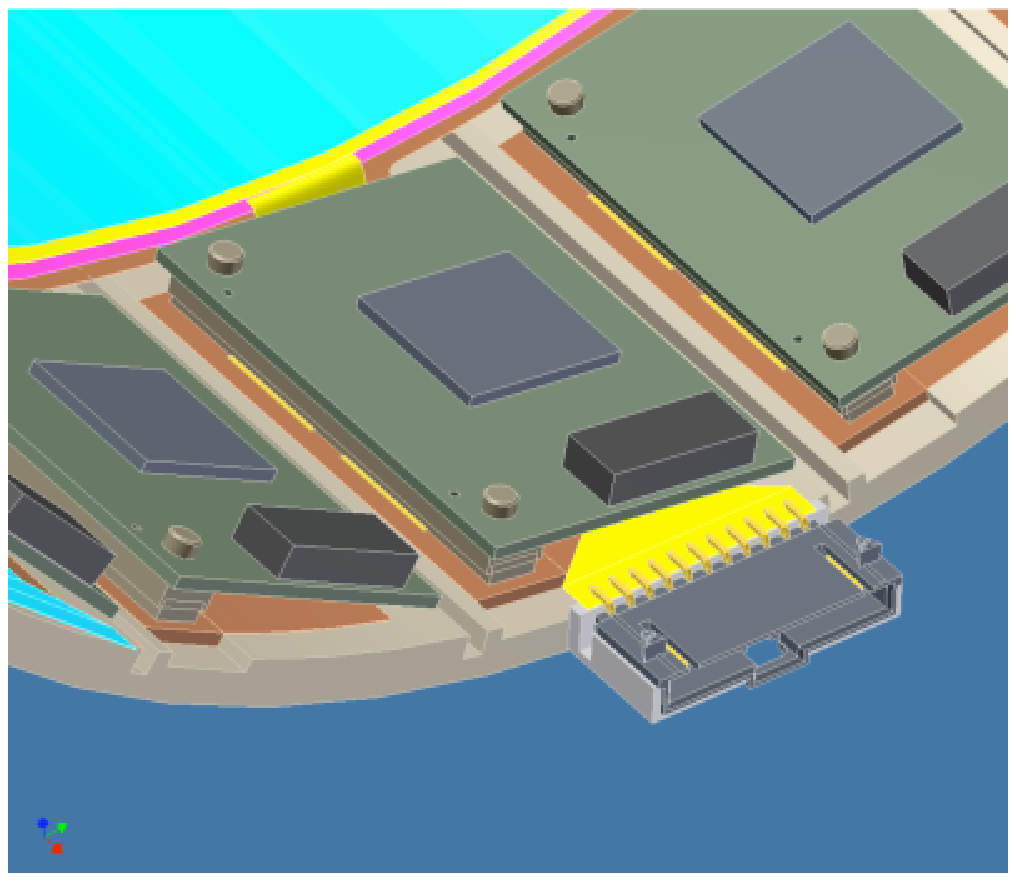}
\label{hvpath1}
}
\subfigure[]{
\includegraphics[scale=0.5]{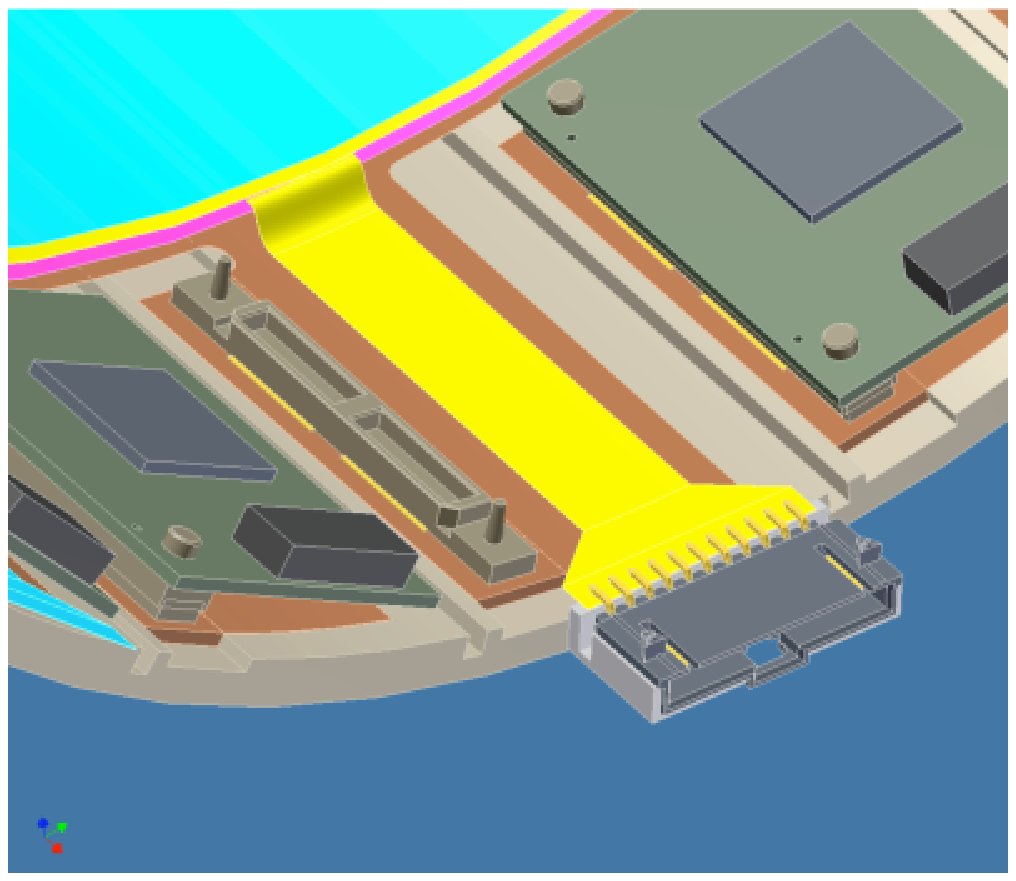}
\label{hvpath2}
}

\caption[Optional caption for list of figures]{HV connection detail  \subref{hvpath1}, HV and front-end board assembling detail  \subref{hvpath2} }
\end{figure}

Each GEM foil has 20 independent HV sections. The relative connections are routed on the same kapton foil into  terminal wings coming out from the detector volume. These wings will be positioned on the front-end support flange, underneath the FEE boards. The HV connectors are soldered and fixed on the same structure, as shown in fig.~\ref{hvpath1} and \ref{hvpath2}.  \\
\subsubsection{Assembling test}
\begin{figure}[!h]
\centering
\includegraphics[width= 9cm]{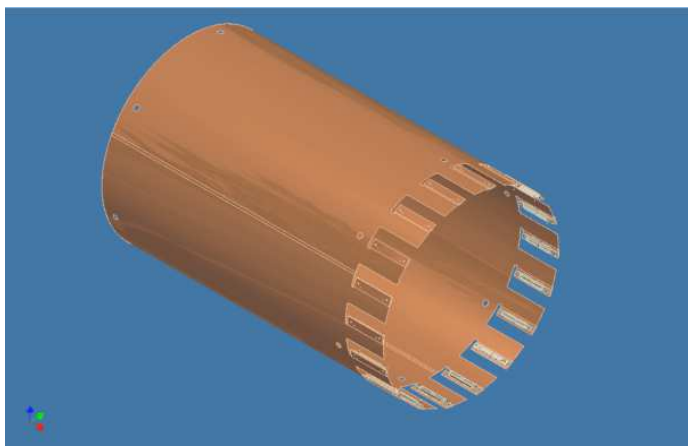}
\caption{Test Layer Assembly}
\label{fig:testlayerass}
\end{figure}

In order to test the assembling procedure a detector layer mock-up will be manufactured, with the same dimensions and the same number of connections of the Layer 1, but with flanges on one side only (fig.~\ref{fig:testlayerass}).

Finally a thermal cycle test has been successfully carried out at 120 $^{\circ}$C to validate the connector soldering and assembling procedure.\\
\begin{figure}[!h]
\centering
\figb CGEM-schedule;13.;
\caption{Time schedule}
\label{fig:schedule}
\end{figure}
%
%
%%%%%%%%%%%%%%%%%%%%%%%%%%%%%%%%%%%%%%%%%%%%%%%%%%%%%%%%%%%%%%%%%%%%%%%%
\newpage
\section{Time schedule and responsibilities sharing}
%%%%%%%%%%%%%%%%%%%%%%%%%%%%%%%%%%%%%%%%%%%%%%%%%%%%%%%%%%%%%%%%%%%%%%%%%
The overall work program and schedule is summarized in fig.~\ref{fig:schedule}.
It is divided into two main parts:
\begin{enumerate}%[label=\alph*)]
\item the finalization of the engineering design of the detector, including tenders,
orders and materials delivery, that takes about 7.5 months;
\item the construction of the five layers, including QC/QA tests of the detectors, and the
successive FEE installation, gas piping, HV connections, that will take less than 15 months.
\end{enumerate}

\par
In the schedule we also reported the sharing of responsibilities among the
INFN groups historically involved in the R\&D phase:
\begin{enumerate}[label=\alph*)]
\item the Bari group will have in charge the production and testing procedure to validate FE electronics together with the design and construction of the support frames.
\item the Rome 1 group is responsible of the detector assembly system tool;
\item the LNF group, beside the overall supervision of the project, is responsible for the
construction and QC tools, the GEM foil design, construction and test of the detector.
\end{enumerate}

Following our schedule, the final design of the IT started in September 2009 and the
detector will be completed and ready for the installation by the end of July 2011.

\newpage
%%%%%%%%%%%%%%%%%%%%%%%%%%%%%%%%%%%%%%%%%%%%%%%%%%%%%%%%%%%%%%%%%%%%%%%%

\end{document}